\documentclass{article}
\PassOptionsToPackage{numbers, compress}{natbib}

\usepackage[main,final]{neurips_2026}
\usepackage[utf8]{inputenc}
\usepackage[T1]{fontenc}
\usepackage{hyperref}
\usepackage{url}
\usepackage{booktabs}
\usepackage{amsfonts}
\usepackage{amsmath}
\usepackage{nicefrac}
\usepackage{microtype}
\usepackage{xcolor}
\usepackage{algorithm}
\usepackage{algpseudocode}
\usepackage{subcaption}
\usepackage{multirow}
\usepackage{enumitem}
\usepackage[flushleft]{threeparttable}

\usepackage{graphicx}

\title{Pixels over Symbols: Sensory Realism Improves Behavioral Alignment in Models of Cognition}

\author{
Hao Yuan Bai$^{1,3}$ \And
Xiaoxuan Lei$^{2,3}$ \And
Zihan Weng$^{1,3}$ \AND
Motahareh Pourrahimi$^{1,3}$ \And
Lucas Gomez$^{1,3}$ \And
Pouya Bashivan$^{2,3}$\\
$^1$ Integrated Program in Neuroscience, McGill University, Montréal, Canada\\
$^2$ Department of Physiology, McGill University, Montréal, Canada\\
$^3$ Mila, Université de Montréal, Montréal, Canada
}

\begin{document}

\maketitle

\begin{abstract}
Computational models of cognition have traditionally relied on abstract symbolic inputs that approximate sensory experiences with simplified vector representations. This simplification reflects historical computational constraints and a long-held assumption that the details of sensory realism are incidental to the study of cognitive processes. Recent advances have enabled the development of sensory-realistic models capable of performing diverse cognitive tasks, yet whether sensory realism meaningfully shapes cognition has not been formally established. We address this gap through a large-scale empirical comparison of abstract and sensory-realistic neural network models trained on a battery of memory-dependent decision-making tasks. Our results show that sensory-realistic models mirror human behavioral response patterns significantly more than their abstract counterparts, though a substantial gap to human internal behavioral consistency remains even for the best tested models. By analyzing the internal dynamics of these systems, we find that the learned recurrent weights are not the main source of this divergence. In fact, frozen abstract-trained weights driven by sensory-realistic inputs recover the human consistency of sensory-realistic models. The resulting dynamics are also within the seed-to-seed variability of sensory-realistic models. Together, our findings argue that across a broad set of visual working-memory tasks, sensory realism accounts for a substantial portion of the model-human behavioral agreement, more than the choice of sequence architecture and comparable to the choice of vision encoder. These results suggest that findings obtained from abstract task models in neuroscience should be more carefully interpreted in light of their potential behavioral limitations.
\end{abstract}

\section{Introduction}
Since their inception, neural network models have been a key tool in understanding computations underpinning perception, memory, and decision-making in the brain \citep{rosenblatt_perceptron_1958,fukushima_neocognitron_1980,hopfield_neural_1982,kohonen_self-organized_1982,malsburg_self-organization_1973}.
In modeling cognition, the stimuli in these models are typically abstract, low-dimensional codes (e.g. one-hot or scalar values) rather than the high-dimensional sensory signals that biological brains receive from the external world. This convention historically rests on two grounds. First, developing cognitive models capable of processing raw sensory data had been computationally infeasible. Training and studying large neural network models at a scale that would enable processing of natural inputs such as high-resolution images was not possible until recently. Second, and perhaps more importantly, it is widely assumed that sensory observations are abstracted when they reach the brain regions implicated in cognitive processing. This viewpoint has been supported by many studies reporting neurons in the primate associational cortices that encode stimuli in a primarily categorical fashion \citep{freedman_categorical_2001,freedman_comparison_2003,swaminathan_preferential_2012}. As a result, most models of cognitive behavior have treated stimuli as idealized signals stripped of their rich perceptual content \citep{mante2013context,chaisangmongkon_computing_2017,yang2019task,song2016training, oreilly_making_2006,lu2022neural}. Beyond these computational and theoretical grounds, the reliance on abstract stimuli also inherits a long experimental tradition in human and animal studies \citep{Fechner1860, Hubel1962, Campbell1968, Newsome1989, IBL2025}, where cognitive paradigms are deliberately built from impoverished stimuli such as dots, gratings, or colored patches, and models are then designed to mirror these paradigms \citep{cueva_low-dimensional_2020, yang2019task}.

Given the long history of abstract models in cognitive neuroscience, many in the field may assume that these models reproduce cognitive behavior as effectively as sensory-realistic models. However, this assumption has not been tested, and it is not known whether sensory-realistic models of cognition are needed at all. Recent advances in developing large-scale neural network models have made it possible to build sensory-realistic cognitive models. Within the past decade, numerous studies have demonstrated the capabilities of these models in capturing behavioral and neural responses during perceptual and language tasks \citep{yamins2014performance, cadieu2014deep, khaligh2014deep, yamins2016using, ratan2021computational,kell2018task, li2023dissecting,wang2021evolving, taleb2024can,caucheteux2022brains, schrimpf2021neural, raugel2025scaling}. 
While several studies have begun applying these models to memory and decision-making settings \citep{bates2024scaling, lei2024geometry, jaffe2025image}, it remains unclear whether sensory realism matters. Importantly, no prior work has systematically asked whether the \textit{format} of the input could modulate the cognitive behavior that emerges, nor whether the internal solutions these models discover depend on the input geometry. The answer may have important implications for neuroscience, as a substantial body of mechanistic claims about circuit dynamics, representational geometry, and the algorithms underlying decision-making is grounded in models trained on abstract inputs. 

We address this gap through a large-scale comparison of abstract-input and sensory-realistic neural network models, evaluated on a battery of memory-dependent decision-making tasks for which we also collected human behavioral data\footnote{https://github.com/BashivanLab/pixels-over-symbols}. We analyze the fine-grained pattern of errors humans and different models commit, and the internal recurrent dynamics that underlie their decisions. We systematically vary the sequence architecture, vision backbone, and model size, and study their effects on behavior. The results converge on the finding that input realism is not an incidental implementation detail but a determinant of the behavior that models produce. Sensory-realistic models track human behavior more closely than abstract-input models, and substitution experiments locate this difference in the inputs rather than in the learned recurrent weights.

Our main contributions are as follows: 
\begin{itemize}[leftmargin=1.2em, nosep]
    \item We show that across a range of memory-dependent decision-making tasks, models trained with sensory-realistic inputs reproduce human behavioral responses significantly more closely than their abstract-input counterparts. Yet a substantial gap remains between the best tested model and human internal behavioral consistency.
    \item We show that pretrained vision-language models, despite approaching or even exceeding human task performance, fall behind most task-trained sensory-realistic models at reproducing the pattern of human responses.
    \item We show that recurrent weights trained on abstract inputs, when driven by sensory-realistic inputs, recover the human consistency of natural models and produce dynamics within the seed-to-seed variability of natural models.
    \item We benchmark how neural network design choices such as sequential architecture, vision backbone, and model size affect human-model behavioral alignment.
\end{itemize}

\section{Related Work} 
\textbf{Sensory-Realistic Models in Neuroscience.}
With parallel GPU computing and the availability of large-scale datasets such as ImageNet \citep{krizhevsky2012imagenet, he2016deep, deng2009imagenet,simonyan_very_2015,szegedy_going_2014}, neural networks have scaled up, allowing their adoption as candidate models of neural processing in the brain. Early work demonstrated that deep convolutional networks trained on natural images can predict neural responses along the primate ventral visual stream at an unprecedented level \citep{yamins_hierarchical_2013,yamins2014performance, cadieu2014deep, khaligh2014deep, yamins2016using, ratan2021computational} while replicating the well-known hierarchy of functional selectivity along the visual pathway \citep{hubel1959receptive, hubel1968receptive,Pasupathy1999,Pasupathy2001,gross_visual_1972,kanwisher_fusiform_1997}. This sensory-realistic modeling approach has since been extended to simulate neural activity across a broad range of modalities and domains, including audition \citep{kell2018task, li2023dissecting}, olfaction \citep{wang2021evolving, taleb2024can}, oculomotor control \citep{pourrahimi2025neural}, and language \citep{caucheteux2022brains, schrimpf2021neural, raugel2025scaling}. 

This shift toward naturalistic inputs has been partially driven by their potential applications. Models operating on natural stimuli are directly applicable to real-world downstream tasks \citep{mehrer2025model}, enable control of neural responses through closed-loop interventions and stimulus optimization \citep{bashivan2019neural, walker_inception_2019}, and provide a framework for explaining large-scale organizational principles in the brain such as cortical topography and functional specialization \citep{margalit2024unifying, dehghani2024credit, qian2026local}. They also generalize more broadly without a predefined input space \citep{ren_how_2024,wang_foundation_2025}. 

\textbf{Models of Cognition.}
The question of whether the format of a representation matters for cognition has a long history in cognitive science. In the classic imagery debate, pictorialists argued that mental images are depictive and preserve the spatial layout of what they represent \citep{kosslyn_visual_1978, kosslyn_image_1980, kosslyn_case_2006}, while descriptivists argued that they are built from abstract, language-like descriptions \citep{pylyshyn_what_1973, pylyshyn_mental_2002}. Here, we ask a related question for neural network models of cognition: whether the format of the input shapes the behavior and internal solutions that these models learn.
Neural networks have also been used to understand neural computations in higher-level cognitive behaviors, spanning a wide range of paradigms including decision-making \citep{yang2019task, sarma_task-specific_2015, masse_circuit_2019, mante2013context, driscoll_flexible_2024}, working memory \citep{oreilly_making_2006, cueva_low-dimensional_2020}, and associative memory \citep{lu2022neural, sucevic_neural_2023}. Much of this work, however, has been constrained to simplified setups where sensory observations and behavioral paradigms are delivered in symbolic form. For example, sensory evidence along different experimental dimensions is typically encoded as a single-channel scalar sequence, while task identities, rules, and contextual cues are supplied as one-hot codes \citep{mante2013context, sussillo2013opening, song2016training, yang2019task, oreilly_making_2006}. More recently, a smaller but growing class of cognitive models has begun incorporating sensory realism, operating directly on high-dimensional natural inputs such as pixels or natural images in tasks of visual working memory and decision-making \citep{bates2024scaling, lei2024geometry, jaffe2025image}.

\section{Methods}
\begin{figure}
\centering
\includegraphics[width=\textwidth]{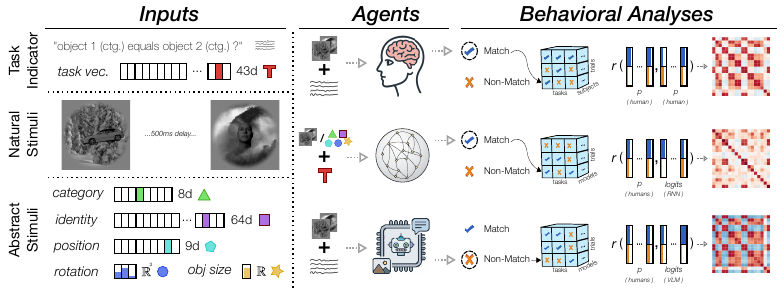}
\caption{\textbf{Schematic of the Experimental Setup.} Models receive both task and stimulus information in their inputs. Task indicator is provided in either textual or one-hot while stimuli are presented in their naturalistic or abstract formats. In the abstract format, information about the primary object in the scene is provided as a combination of one-hot (category, object identity, position) and scalar values. Three groups are studied: humans, neural network models trained on abstract or natural inputs, and pretrained vision-language models. Behavioral responses are collected from each group and used to estimate the choice probabilities for each task trial. These probabilities are then used to calculate behavioral consistency with humans using different metrics.}\label{fig:schematic}
\end{figure}
\subsection{Problem Formulation}
Each task trial consists of $T$ frames, each presenting a stimulus. At the designated response frames, the model must produce a binary decision $\hat{y}_t \in \{0, 1\}$ comparing stimulus information across the frames according to the task rule. In both \texttt{Abstract} and \texttt{Natural} models, task identity is provided as a one-hot vector $\mathbf{k} \in \{0,1\}^{K}$, concatenated to each frame in a trial. For humans and Vision-Language models (VLMs), the task rule is provided in natural language (see \ref{sec:appendix/human_trials} for details and Figs.~\ref{fig:task_1back}, \ref{fig:task_interdms}, \ref{fig:task_ctxdm}, \ref{fig:task_dms} for examples).

\subsection{Task Dataset}
Stimuli are sampled from the HvM dataset \citep{Majaj2015,rajalingham2015comparison}, which contains 5,760 images of objects positioned on naturalistic backgrounds (e.g. mountains, beaches, forests). The dataset contains 8 object categories (cars, planes, boats, fruits, faces, animals, chairs, and tables), each with 8 distinct object identities rendered in multiple viewing conditions. There are 3 variation levels, with each level introducing more extreme viewing conditions, making the objects more difficult to distinguish. Each image is described by discrete features: category $\mathbf{c}$, identity $\mathbf{o}$, and spatial position $\mathbf{p}$ on a $3\times3$ grid, and continuous features: rotation angles $\mathbf{r} \in \mathbb{R}^3$, and scale $\mathbf{s} \in \mathbb{R}$. See \ref{sec:appendix/hvm_dataset} for stimuli examples.

We excluded the 100 images shown to human participants from model training and validation. The remaining 5,660 images are split into training and validation sets at a 3:1 ratio within each object and variation level (4,271 and 1,389 images). As a result, every object appears in training, and validation measures how well models generalize to unseen views of these objects. We chose this split so that the two model types can be compared fairly. Since \texttt{Abstract} models receive object identity as a one-hot vector (Sec.~\ref{sec:model_types}), if some objects were held out from training, the input weights for their identity dimensions would never be updated, and \texttt{Abstract} models would receive no useful identity information for these objects. \texttt{Natural} models do not face this problem, because their pretrained vision encoder produces informative features for images of any object.

We created 43 tasks from four task families: Delayed match-to-sample (DMS), NBack, Interleaved DMS (InterDMS), and contextual decision-making (CtxDMS) using iWISDM \citep{pmlr-v274-lei25a} (see Tables \ref{tab:tasks} and \ref{tab:all_tasks}). DMS tasks are the basic unit tasks, where an observer judges whether a sample stimulus and a test stimulus match on a target feature (position, identity, or category). In NBack tasks, at each step $t$, the observer performs a DMS judgement by comparing the current stimulus with the stimulus on frame $t-n$. In InterDMS tasks, the observer performs two independent DMS subtasks whose frames are interleaved. In CtxDMS tasks, the observer performs two DMS subtasks where the target feature of the second subtask depends on the outcome of the first.

\subsection{Model Types}\label{sec:model_types}
We use three types of models in our analyses (see Fig.~\ref{fig:schematic}): (1) \textbf{\texttt{Abstract} models}: sequence models that receive symbolic vectors in their inputs; (2) \textbf{\texttt{Natural} models}: sequence models that receive a sequence of images; (3) \textbf{Vision-Language Models (VLMs)}: pretrained transformer models that receive a natural-language description of the task and sequence of images in their input prompts.

\paragraph{Natural input.}
\texttt{Natural} input consists of a sequence of images. 
Each image 
$\mathbf{x}_t \in \mathbb{R}^{H\times W \times C}$ is processed by the frozen pretrained vision encoder. 

The encoder maps each image $\mathbf{x}_t$ into a sequence of token embeddings 
$\mathbf{z}_t \in \mathbb{R}^{N\times D_{\mathrm{vision}}}$, where $N = P^2$ is the resulting number of patches from a $P\times P$ grid and $D_{\mathrm{vision}}$ is the embedding dimension. We then apply attention pooling \citep{radford2021learning} on $\mathbf{z}_t$ to obtain a single image embedding $\mathbf{\bar{z}}_t\in \mathbb{R}^D$, and concatenate it with a one-hot task index $\mathbf{k}\in \{0,1\}^{K}$ to produce our pre-projection input vector $\mathbf{\tilde{z}}_t \in \mathbb{R}^{D+K}$. Formally,
\begin{equation*}
    \mathbf{z}_t = g_{\phi} (\mathbf{x}_t),
    \quad \mathbf{\bar{z}}_t = \mathrm{AttentionPool}(\mathbf{z}_t),
    \quad \mathbf{\tilde{z}}_t = (\mathbf{\bar{z}}_t, \mathbf{k}), 
\end{equation*}
where $g$ denotes the vision model with parameters $\phi$.

\paragraph{Abstract input.}
\texttt{Abstract} input consists of a sequence of one-hot and continuous encodings of the object information in the images. Specifically, at each time step $t$,
\begin{equation*}
  \mathbf{\tilde{z}}_t = 
  \bigl[
    \underbrace{\mathbf{c}}_8,\;
    \underbrace{\mathbf{o}}_{64},\;
    \underbrace{\mathbf{p}}_9,\;
    \underbrace{\mathbf{r}}_3,\;
    \underbrace{\mathbf{s}}_1, \;
    \underbrace{\mathbf{k}}_{43}\;
  \bigr],
  \label{eq:encoding}
\end{equation*}
where $\mathbf{c} \in \{0,1\}^8$ is a one-hot category vector,
$\mathbf{o} \in \{0,1\}^{64}$ is a one-hot identity vector (8 identities in 8 categories), $\mathbf{p} \in \{0,1\}^9$ is a one-hot position vector
over the $3\times3$ spatial grid, $\mathbf{r}\in \mathbb{R}^{3}$ is the rotation angle along each axis, $\mathbf{s}\in \mathbb{R}$ is the object scale, and
$\mathbf{k}$ is the same one-hot task index. To remove differences between natural input and abstract input, we propose dimensionality and visual uncertainty matched abstract models (\ref{sec:appendix_dimmatch}).

\paragraph{Model architecture.}
Both \texttt{Natural} and \texttt{Abstract} models project their respective inputs $\mathbf{\tilde{z}}_t$ into $D$ dimensions with a linear layer followed by LayerNorm \citep{ba_layer_2015} and non-linearity (e.g. GELU \citep{hendrycks2016gaussian}):
\begin{equation*}
    \mathbf{z}_{\mathrm{proj}_t} = \mathrm{GELU}(
    \mathrm{LayerNorm}(
    \mathbf{W}_{\mathrm{in}} \mathbf{\tilde{z}}_t
    )).
\end{equation*}
For every time step $t$, $\mathbf{z}_{\mathrm{proj}_t}$ passes through a
sequence model $f_\theta$ followed by a linear output head $\mathbf{W}_{\mathrm{out}}$:
\begin{equation*}
  \mathbf{l}_t = \mathbf{W}_{\mathrm{out}}\,\mathrm{LayerNorm}\bigl(\mathbf{h}_t\bigr) + \mathbf{b},
  \quad \mathbf{h}_t = f_\theta(\mathbf{z}_{\mathrm{proj}_t},\, \mathbf{h}_{t-1}),
\end{equation*}
where $\mathbf{h}_t \in \mathbb{R}^{D}$ is the hidden state at time $t$, and $\mathbf{l}_t \in \mathbb{R}^2$ is the model's class logits.

\subsection{Evaluation Pipeline}
To make the model-human comparisons feasible, we selected 100 stimuli from the HvM dataset and created 100 stimulus pairs that were balanced across all three object features (category, identity, position). Using this fixed set, we generated trials for 12 Amazon Mechanical Turk (MTurk) tasks (see \ref{sec:appendix/human_trials}), with 20 trials for three 1Back tasks and 100 trials for the other nine tasks, totaling 960 task trials. These 12 tasks are a subset of the 43 tasks models were trained on.

The behavioral tasks performed in this study were reviewed and approved by the Research Ethics Board (REB) in accordance with ethical standards for research involving human subjects. We recruited 20 unique participants for each task on Amazon Mechanical Turk (MTurk). Each task page included instructions, example HvM stimuli, and an example trial. Participants then completed the experiment on an interactive website built with PsychoJS (see \ref{sec:appendix/task_description} and \ref{sec:appendix/human_trials} for example trials and details of recruitment).
\footnote{\url{https://rdh1115.github.io/abs_nat_psychopy_mturk/}}

To investigate the alignment between humans and VLMs, we evaluated four state-of-the-art VLMs of different sizes (see \ref{sec:appendix/vlms}). For trained \texttt{Natural} and \texttt{Abstract} models, we evaluated 6 random seeds per architecture on the 12 MTurk tasks. Because model inputs were padded to six frames, we appended delay frames to MTurk trials so that the final action frame always occurred at the sixth frame.
\subsection{Behavioral Analysis}
After evaluating on the 12 MTurk tasks (defined in \ref{sec:appendix/human_trials}), we represent the model responses as matrices $\mathbf{M}\in \mathbb{R}^{18\times100\times2}$ (directly from $\mathbf{l}_t$). The first dimension indexes the 18 elementary task conditions, the second indexes the 100 trial stimulus pairs, and the final dimension indexes the two response logits (match and non-match). The 18 elementary task conditions arise by decomposing compositional tasks into their constituent DMS parts. For example, an InterDMS-ABBA trial is split into two elementary subtasks, InterDMS-AA and InterDMS-BB, each corresponding to a distinct stimulus pair. Applying this decomposition to the 12 MTurk tasks yields 18 elementary task conditions, whose responses are grouped and sorted into the same set of 100 MTurk stimulus pairs for all models, VLMs, and human observers. For humans and VLMs, we obtain binary output matrices $\mathbf{H}, \mathbf{V}\in \left\{0, 1\right\}^{18\times100}$. To quantify how closely model response patterns resemble those of humans, we apply the following behavioral metrics on the above matrices. 

\textbf{Human Consistency.}
Following \citet{rajalingham2018large}, we define human consistency as the Pearson correlation between model and human response vectors across trials (see \ref{sec:appendix/human_cons}). For models, we apply softmax to $\mathbf{M}$ to get the probability of correct class per choice. For humans and VLMs, we use $\mathbf{H}, \mathbf{V}$ respectively and estimate the class probabilities by calculating the fraction of correct choices. For each task, we split the 20 subjects into two groups of 10, treating each group's average as a noisy estimate of the true response probability. We mirror the procedure for trained model logits, partitioning the 6 seeds into groups of 3. We report the mean over splits in Fisher-$z$ space, with the standard deviation over splits as error bars. Algorithm~\ref{alg:split_half} describes this process in detail.

\textbf{Representational Similarity Analysis (RSA).} Using the same set of stimulus pairs on all MTurk tasks allows us to group behavioral patterns and measure how similarly two models organize the task/stimulus space. We performed two distinct RSA analyses: cross-task and cross-stimulus. With cross-task RSA, we compare response patterns across tasks with a fixed set of stimuli, and with cross-stimulus RSA, we compare response patterns across stimuli with a fixed set of tasks. See \ref{sec:appendix/rsa}.

\textbf{Error consistency.}
We additionally compute Cohen's $\kappa$ as a measure of trial-by-trial error
overlap \citep{geirhos2020beyond}:
$
  \kappa = \frac{c_{\text{obs}} - c_{\text{exp}}}{1 - c_{\text{exp}}}
  \label{eq:kappa}
$,
where $c_{\text{obs}}$ is the fraction of trials on which the model and a
human subject give the same response (both correct or both incorrect), and
$c_{\text{exp}} = p_m p_h + (1-p_m)(1-p_h)$ is the expected overlap by chance given
their respective accuracies $p_m$ and $p_h$. 

\subsection{Dynamical Similarity Analysis} 
To compare the recurrent structures learned by different models, we apply Dynamical Similarity Analysis (DSA) \citep{NEURIPS2023_6ac807c9} to the hidden states $\mathbf{S} \in \mathbb{R}^{960\times 6 \times D}$ from our trained models. Unlike $\mathbf{M}$, $\mathbf{S}$ has shape $(960\times 6\times D)$ since model inputs were padded to six frames, and there are 20 trials for the three 1Back tasks plus 100 trials for the other nine tasks. For each task and model seed, we project the states onto their top $\mu$ principal components. We then apply Hankel alternative view of Koopman (HAVOK) \citep{brunton2017chaos} to learn a linear operator $\mathbf{A}$ that advances the delay-embedded state one step. Finally, we compute the pairwise distances between the learned operators for each model seed with Procrustes Analysis on Vector Fields. We report the mean $\pm$ SEM over the 12 tasks and all seed pairs. See \ref{sec:appendix/dsa} for more details.

\subsection{Training}
We train both \texttt{Abstract} and \texttt{Natural} models using cross-entropy loss on batches of random trials from all tasks. A new set of trials is generated for each epoch while balancing for task type to mitigate learning biases. We use the AdamW optimizer \citep{loshchilov2019decoupledweightdecayregularization} with a cosine-decay learning rate scheduler. To compare across different model architectures, we kept model parameters on the same order of magnitude. RNN \citep{10.5555/65669.104449} had $D=256$, while all other sequential modules (LSTM \citep{hochreiter1997long}, GRU \citep{cho2014learning}, Transformer \citep{vaswani2017attention}, BERT \citep{devlin2019bert}) had $D=128$. \texttt{Natural} models additionally use label smoothing of 0.2 (Table~\ref{tab:native-training-config}). The attention pool and the transformer layers use 8 attention heads.

\section{Results}\label{sec:results}
\subsection{\texttt{Natural} models better replicate human behavioral responses}\label{sec:results1}
We first compare human behavioral alignment in \texttt{Abstract} and \texttt{Natural} models across 5 sequential architectures: RNN, GRU, LSTM, BERT (decoder only), and vanilla transformer (decoder only). In this section, we use ResNet-50 \citep{he2016deep} pre-trained with CLIP \citep{radford2021learning} as the vision backbone. \texttt{Abstract} models reached significantly higher validation accuracy compared to \texttt{Natural} ones (0.93-0.99 vs. 0.70-0.81). To control for this difference, we select \texttt{Abstract} checkpoints with validation accuracy closest to 80\%, matching the \texttt{Natural} models' accuracy range on the MTurk dataset. Across the 5 sequential architectures tested, final checkpoints do not have higher human consistency except for RNNs (Fig.~\ref{fig:no_match_global:corr}).
\begin{figure}[ht]
\centering
\begin{minipage}{0.80\textwidth}
    \centering
    \begin{subfigure}{0.48\textwidth}
        \includegraphics[width=\linewidth]{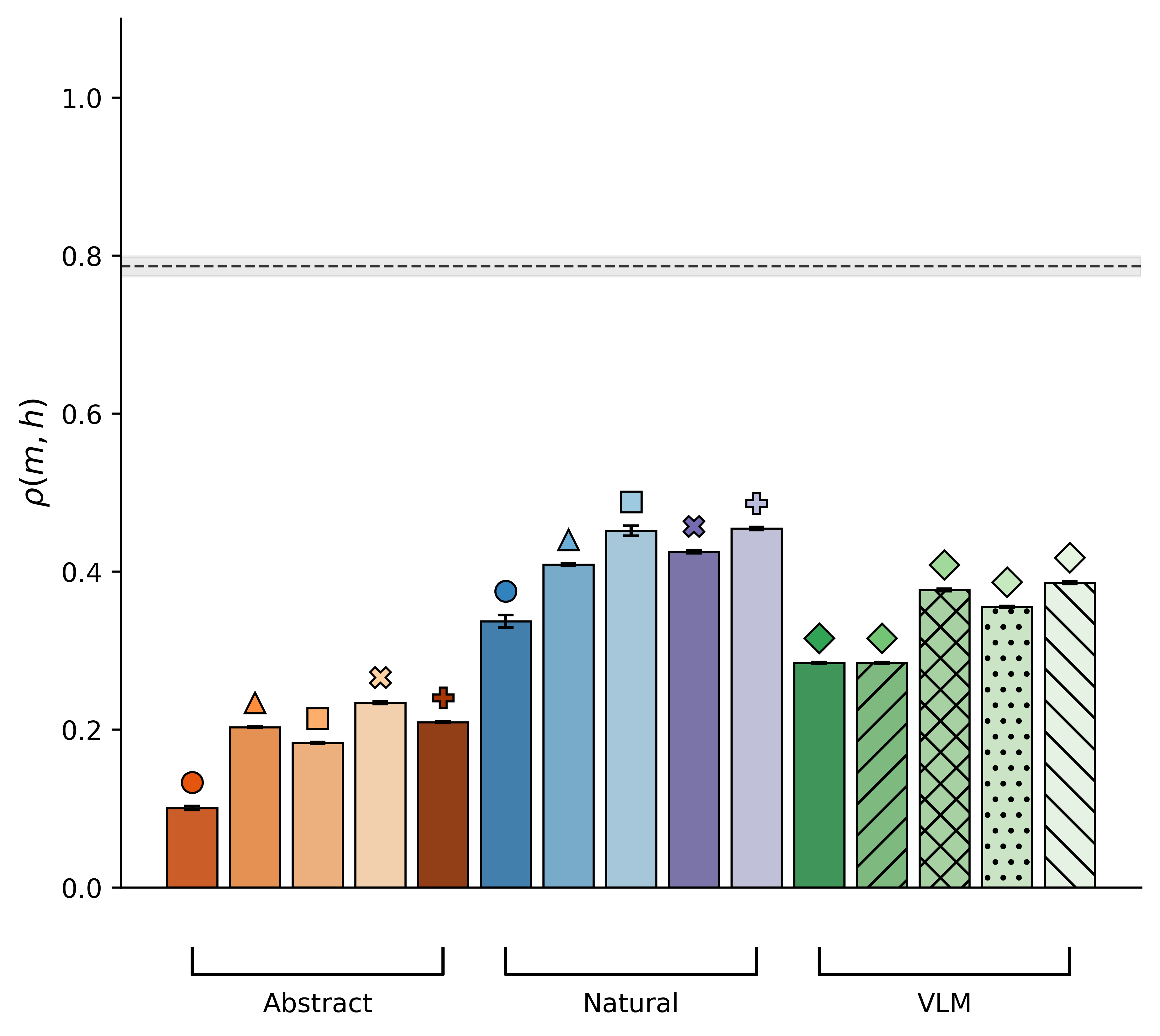}
        \caption{Correlation}\label{fig:seq_type_global:corr}
    \end{subfigure}
    \hfill
    \begin{subfigure}{0.48\textwidth}
        \includegraphics[width=\linewidth]{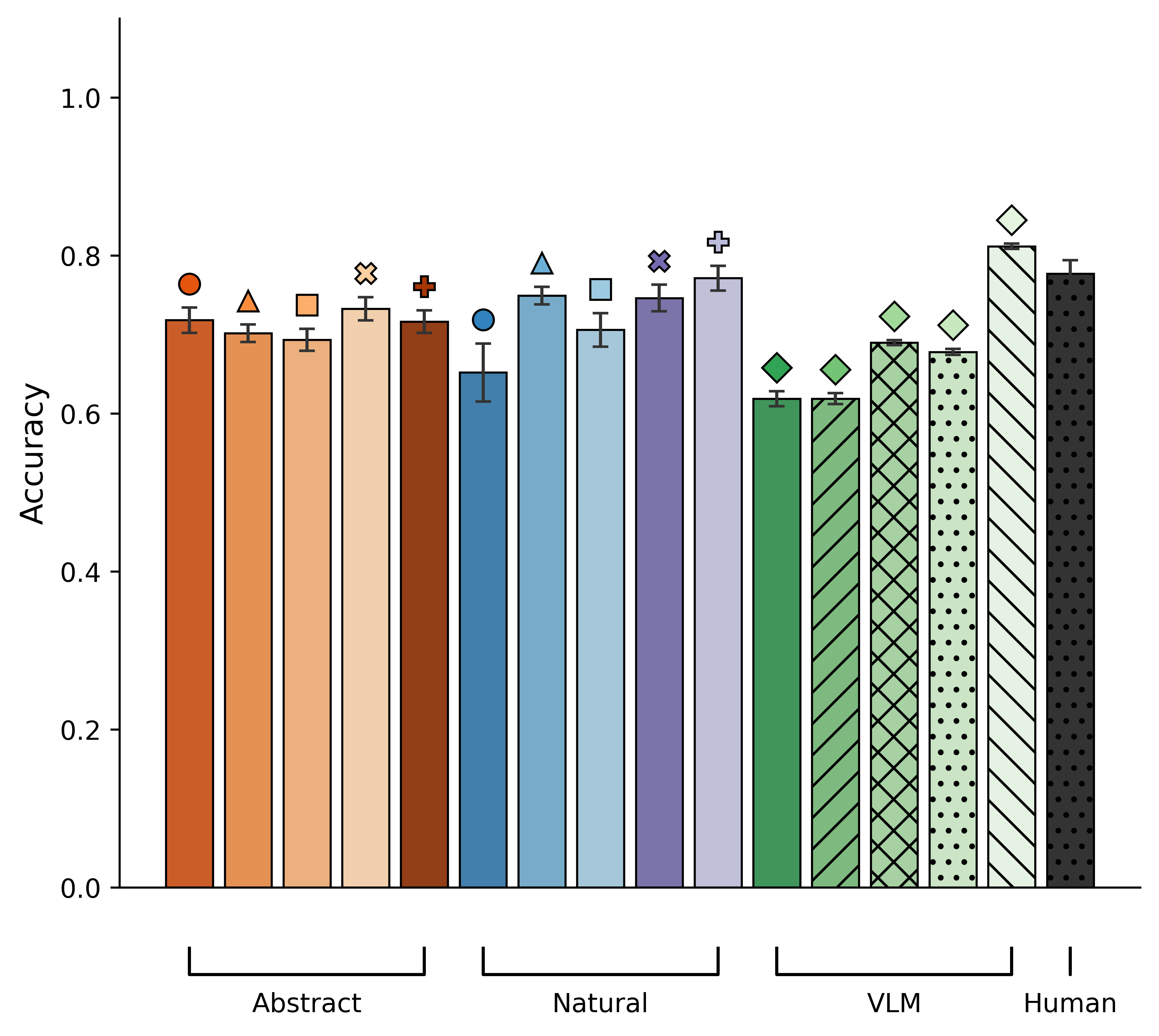}
        \caption{Accuracy}\label{fig:seq_type_global:acc}
    \end{subfigure}
    \end{minipage}
    \begin{minipage}{0.15\textwidth}
    \centering
    \includegraphics[scale=0.5]{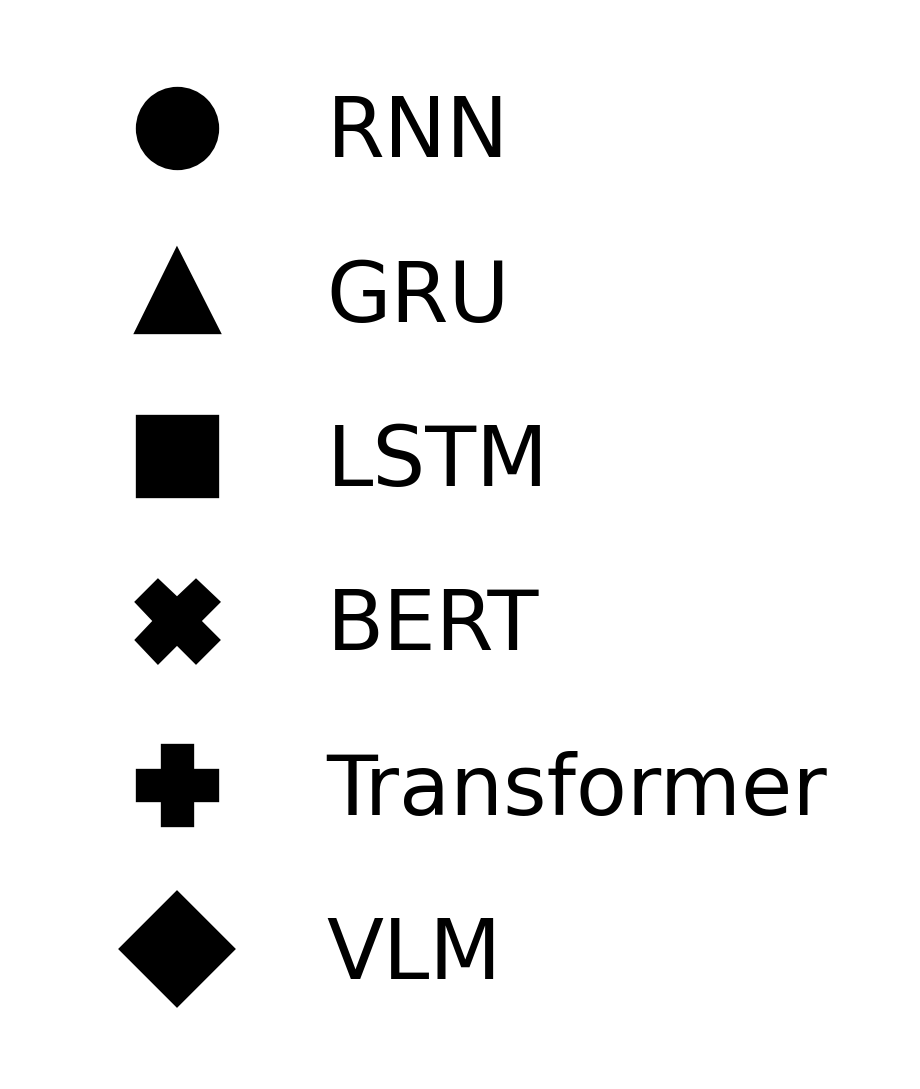}
    \end{minipage}
\caption{
\textbf{Comparison of \texttt{Abstract} and \texttt{Natural} models across sequential architectures.}
(a) Correlation with human behavior. Gray line and zone indicates human internal behavioral consistency, $\rho(h, h)$ in Eq. \ref{eq:hc}. Bars show $\rho(m,h)$ averaged over 20 random split-halves in Fisher-$z$ space; error bars show the standard deviation across split-halves. (b) Task accuracy. Mean and standard deviation reported as error bars, 6 seeds for models and 20 participants for humans.
VLMs (left to right): \texttt{nemotron-nano-12b-v2}, \texttt{nemotron-nano-12b-v2-no-thinking}, \texttt{qwen3-vl-30b}, \texttt{qwen3-vl-235b}, and \texttt{gemini-3.1-pro-preview} (surpasses human-level accuracy).
}\label{fig:seq_type_global}
\end{figure}

Regardless of the sequential architecture, models trained on naturalistic stimuli (both task trained and VLMs) have higher human consistency than \texttt{Abstract} models (Fig.~\ref{fig:seq_type_global:corr}). \texttt{Natural} models exceed their abstract counterparts by 0.199 (95\% CI [0.170, 0.229], paired $t(4)=18.76$, $p=4.8\times10^{-5}$), and do so in all 18 task conditions (paired $t(17)=9.36$, $p=4.1\times10^{-8}$). Regressing image-based visual discriminability out of the human and model responses changes this gap by at most 0.019 (see \ref{sec:appendix_residualization}).

To dissect this discrepancy, we perform feature-level analysis by correlating responses that only compare categories (30\%), identities (33\%), and positions (37\%). \texttt{Natural} models are more human-consistent than \texttt{Abstract} models on every feature. Within the three features, \texttt{Abstract} models come closest on identity trials. VLMs are the most human-consistent on category trials, but fall below \texttt{Natural} models on identity trials (Fig.~\ref{fig:seqsize_feature}). Notably, \texttt{gemini-3.1-pro-preview} exceeds human accuracy on every feature (on identity only marginally, 0.821 vs.\ 0.814). It has the lowest category consistency among the VLMs but the highest position consistency of any model except the \texttt{Natural} LSTM (0.369 vs.\ 0.381). This is in contrast to reports that VLMs struggle on spatial reasoning \citep{weng2025caption, chen2025spatial, wang2024picture, liu2023, kamath2023, shiri2024, pmlr-v274-lei25a}.

We correlate probability vectors across trials within each task family to examine how task paradigm affects human consistency (Fig.~\ref{fig:seq_type_task_type}). \texttt{Natural} LSTMs are the most human-like in InterDMS and DMS trials, Transformers lead on CtxDMS, and four \texttt{Natural} architectures are within 0.015 of each other on NBack.
Among \texttt{Abstract} models, BERT has the highest human consistency in every task family. On DMS trials (DMS-position only) every \texttt{Abstract} model has zero human consistency.

We performed RSA to identify which properties of a model's response structure predict its consistency with humans. Cross-task RSA measures whether responses to a given stimulus pair are similar regardless of task type. We find that \texttt{Natural} models have much higher cross-task RSA alignment values than \texttt{Abstract} models (paired $t(4)=8.13$, $p=1.2\times10^{-3}$; Fig.~\ref{fig:model_task_rsa_alignment}).
Cross-stimulus RSA, in contrast, measures how similarly a model responds to each stimulus pair (across the 12 tasks) relative to all other stimulus pairs. We find that it is also a strong proxy for human consistency ($r=0.962$). Moreover, the human consistency measure was strongly correlated with the $\kappa$ error-consistency metric \citep{geirhos2020beyond} (Fig.~\ref{fig:metric_correlation}). 

Sweeping over a range of vision encoder architectures, we find that vision encoders pretrained with CLIP \citep{radford2021learning} generally have higher human consistency and accuracy than their counterparts trained on ImageNet with supervision (Fig.~\ref{fig:vision_type_global:corr}). Overall, DINOv3 \citep{simeoni2025dinov3} is the most human-consistent vision encoder. As a negative control, a frozen, randomly initialized ResNet-CLIP barely learns the tasks (MTurk accuracy 0.52) and its human consistency drops to almost zero (see \ref{sec:appendix_random_encoder}). We also notice a positive correlation between the linear separability of a vision model's features on the HvM dataset (linear SVM) and its human consistency on category tasks (Fig.~\ref{fig:metric_alignment_cat_cons}). Beyond $h=128$, increasing the hidden dimension raises the \texttt{Natural} LSTM's accuracy but not its human consistency, and the \texttt{Natural} Transformer's consistency decreases with size (Fig.~\ref{fig:vision_type_global}).

To test whether simple properties of the natural input explain this behavioral gap, we matched the dimensionality and the visual difficulty of the \texttt{Abstract} input to those of the \texttt{Natural} input (see \ref{sec:appendix_dimmatch}). Both changes reduced the behavioral alignment of \texttt{Abstract} models. At similar validation accuracy, matching the input dimensionality lowered human consistency from 0.183 to 0.115, and adding noise to mimic perceptual uncertainty lowered it to zero (Fig.~\ref{fig:dimmatch_dose}). This suggests that the gap is not explained by input dimensionality or perceptual uncertainty alone.

\subsection{Why do \texttt{Natural} models better replicate human behavior?}\label{sec:results2}
\begin{figure}[ht] 
    \centering 
    \begin{minipage}{0.8\textwidth} 
    \begin{subfigure}{0.48\textwidth}
    \includegraphics[width=\linewidth]{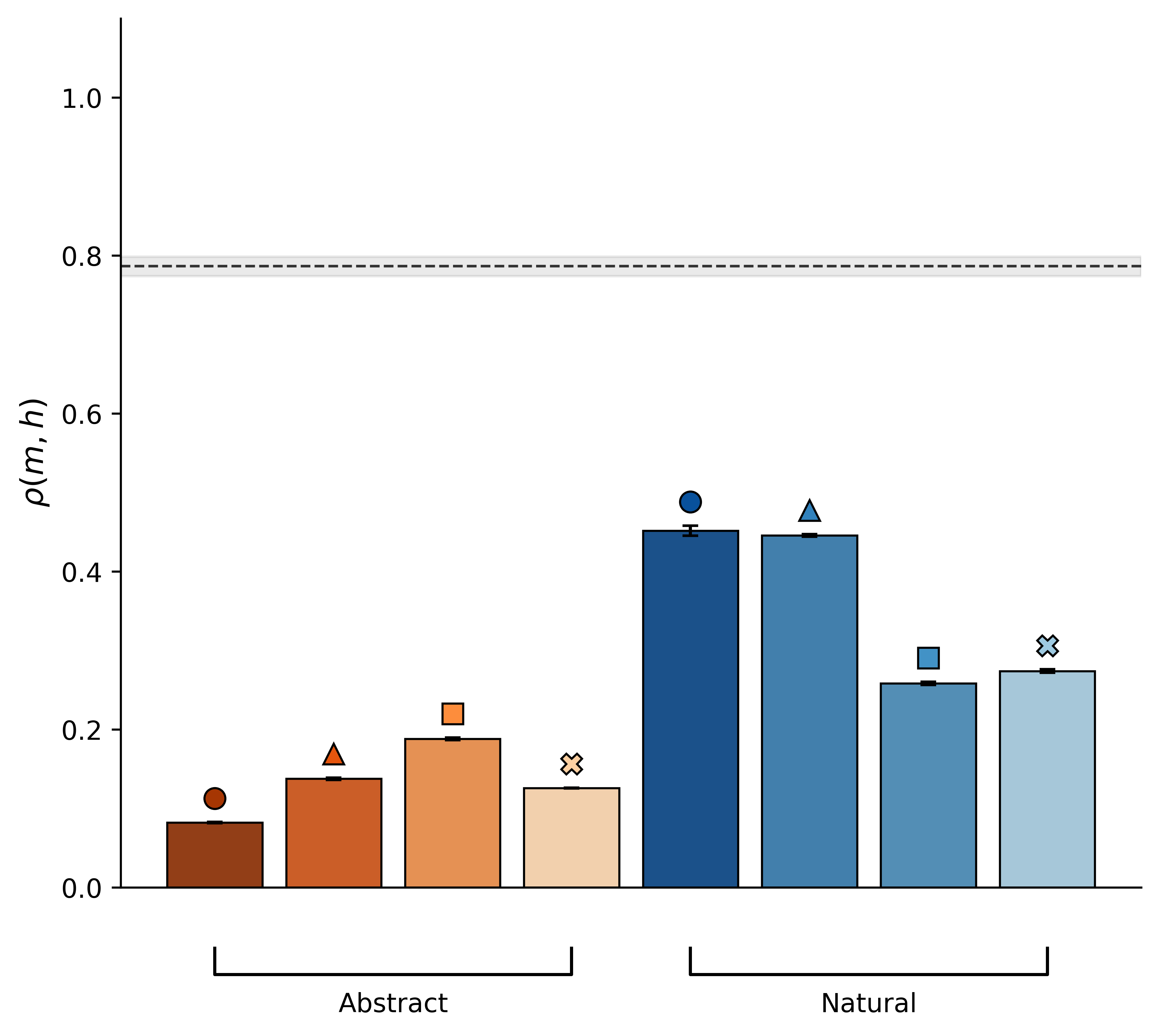}
    \caption{Correlation}\label{fig:transplant_final_global:corr}
    \end{subfigure} 
    \hfill 
    \begin{subfigure}{0.48\textwidth}
    \includegraphics[width=\linewidth]{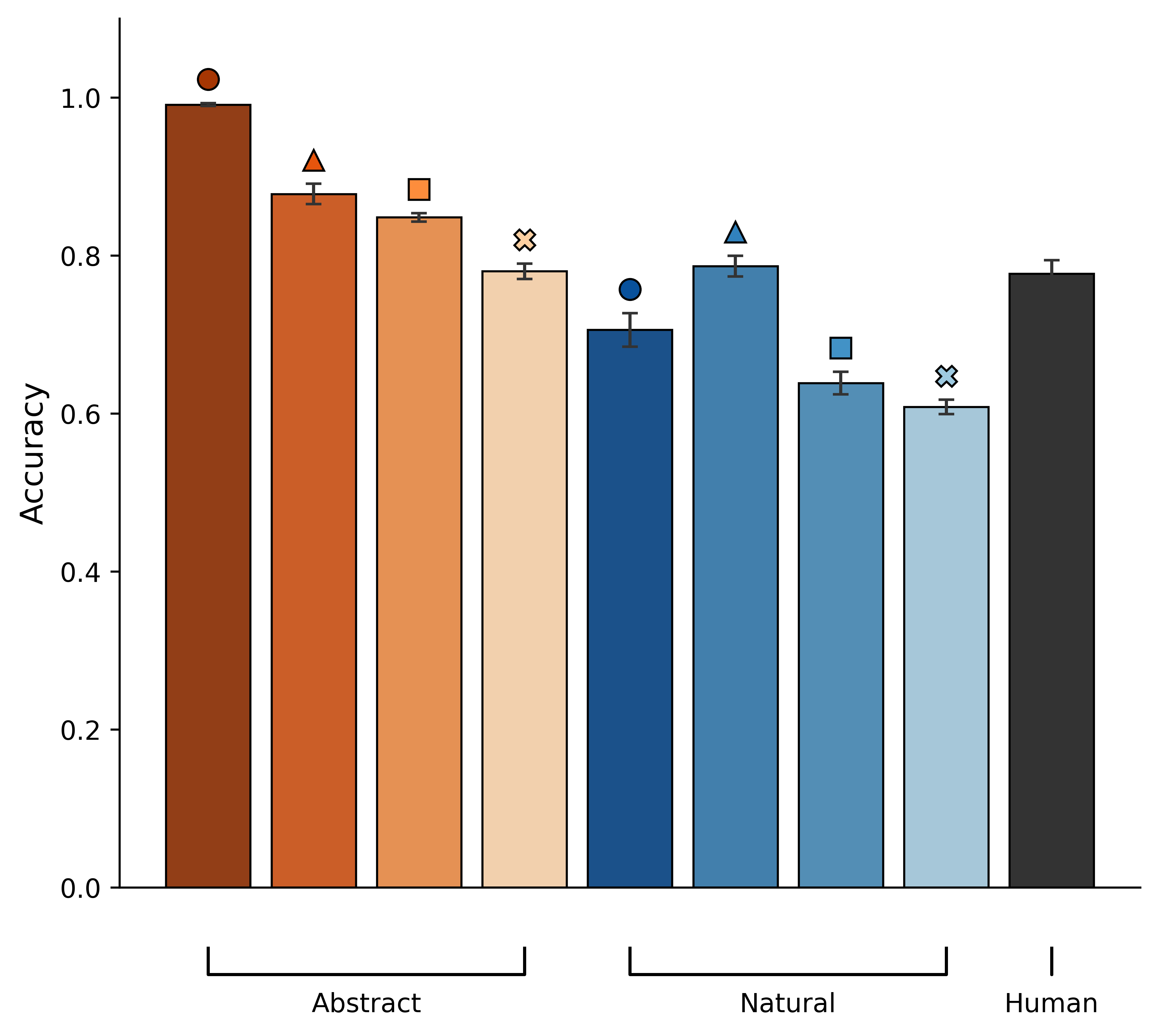}
    \caption{Accuracy}\label{fig:transplant_final_global:acc}
    \end{subfigure}
    \end{minipage} 
    \begin{minipage}{0.15\textwidth}
    \centering
    \includegraphics[scale=0.5]{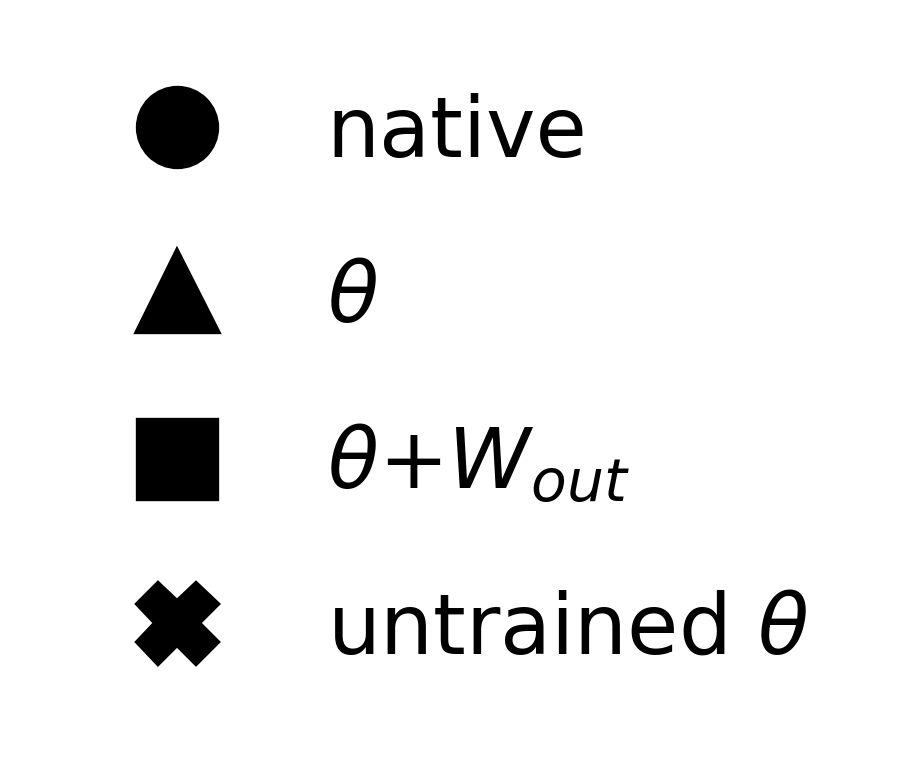}
    \end{minipage}
    \caption{\textbf{Behavioral metrics of \texttt{Abstract} and \texttt{Natural} models with substituted components.}
From left to right, warm-colored bars show $\theta_\mathrm{Abs}$, $\theta_{\mathrm{Nat}\to\mathrm{Abs}}$, $(\theta+\mathbf{W}_{\mathrm{out}})_{\mathrm{Nat}\to\mathrm{Abs}}$ and untrained $\theta$; cold-colored bars show $\theta_\mathrm{Nat}$, $\theta_{\mathrm{Abs}\to\mathrm{Nat}}$, $(\theta+\mathbf{W}_{\mathrm{out}})_{\mathrm{Abs}\to\mathrm{Nat}}$ and untrained $\theta$.
\texttt{Abstract} models use final checkpoints.
(a) Correlation with human behavior.
The gray line and band indicate human internal behavioral consistency, $\rho(h,h)$ in Eq.~\ref{eq:hc}.
Bars show $\rho(m,h)$ averaged over 20 random split-halves in Fisher-$z$ space; error bars show the standard deviation across split-halves.
On \texttt{Abstract} input, substitution improves human consistency only marginally (orange triangle and square vs.\ red circle: 0.138 and 0.188 vs.\ 0.082), barely above the untrained $\theta$ (0.126).
On natural input, $\theta_{\mathrm{Abs}\to\mathrm{Nat}}$ recovers the human consistency of the native model (blue triangle vs.\ circle: 0.446 vs.\ 0.452), well above the untrained $\theta$ (0.274), whereas $(\theta+\mathbf{W}_{\mathrm{out}})_{\mathrm{Abs}\to\mathrm{Nat}}$ reaches only 0.258 (blue square).
(b) Task accuracy; mean and standard deviation over 6 seeds for models and 20 participants for humans.
With an untrained $\theta$, models still reach 0.78 MTurk accuracy on abstract input but only 0.61 on natural input.}
    \label{fig:transplant_final_global}
\end{figure}
Sec.~\ref{sec:results1} shows that \texttt{Natural} models better capture human behavioral patterns, but the source of this gap remains unclear. We reason that input geometry and learned recurrent computations can each contribute to this gap. To disentangle them, we conduct substitution analyses on the LSTM models (ResNet-CLIP for \texttt{Natural}). We freeze modules trained on one modality, and retrain the remaining modules on the other. We denote the sequence modules of natively trained \texttt{Abstract} and \texttt{Natural} models as $\theta_{\mathrm{Abs}}$ and $\theta_{\mathrm{Nat}}$. For example, a $\theta_{\mathrm{Abs} \to \mathrm{Nat}}$ model loads a frozen $\theta_{\mathrm{Abs}}$ and trains the input and output projections on naturalistic input.
In $(\theta + \mathbf{W}_{\mathrm{out}})_{\mathrm{Abs} \to \mathrm{Nat}}$, the output projection is also frozen, so only the input projection is trained. In $(\theta + \mathbf{W}_{\mathrm{in}})_{\mathrm{Abs} \to \mathrm{Nat}}$, only the output projection is trained. Because \texttt{Abstract} and \texttt{Natural} inputs differ in dimensionality, $\theta + \mathbf{W}_{\mathrm{in}}$ substitution uses the dimensionality-matched \texttt{Abstract} models (see \ref{sec:appendix_dimmatch}).
As a baseline, we also retrain the projections with a frozen, randomly initialized $\theta$.
All \texttt{Abstract} models in this section use final checkpoints rather than the accuracy-matched checkpoints. See Fig.~\ref{fig:transplant_acc_matched_global} for accuracy-matched substitution results.

Surprisingly, the $\theta_{\mathrm{Abs} \to \mathrm{Nat}}$ model matches the human consistency of the natively trained \texttt{Natural} model (blue triangle vs. blue circle in Fig.~\ref{fig:transplant_final_global:corr}) and exceeds its task accuracy (Fig.~\ref{fig:transplant_final_global:acc}). In the other direction, $\theta_{\mathrm{Nat} \to \mathrm{Abs}}$ raises human consistency only slightly over the native \texttt{Abstract} model (orange triangle vs. red circle), and barely above an untrained $\theta$ (orange cross).
Freezing $\mathbf{W}_{\mathrm{out}}$ as well further increases consistency on abstract input but lowers it on naturalistic input. $\theta_{\mathrm{Abs} \to \mathrm{Nat}}$ also matches the native \texttt{Natural} model in cross-task and cross-stimulus RSA alignment, while all substitutions on abstract input remain well below it (Fig.~\ref{fig:new_transplant_rsa}).

In dimensionality-matched models, $\theta$ substitutions show the same asymmetry as above: $\theta_{\mathrm{Abs} \to \mathrm{Nat}}$ approaches the native \texttt{Natural} model, while $\theta_{\mathrm{Nat} \to \mathrm{Abs}}$ stays at the level of an untrained $\theta$. Retraining only the input projections ($\theta + \mathbf{W}_{\mathrm{out}}$, square markers) recovers behavioral alignment and retains task accuracy. However, retraining only the output projection drops task accuracy to chance in both directions, and reduces human consistency (diamond markers, Fig.~\ref{fig:transplant_dm_global}). 

In these experiments, human consistency follows the input modality more than the origin of the recurrent weights. For a model receiving naturalistic input, a trained $\theta$ is needed, but whether it was trained on abstract or naturalistic input matters little. For a model receiving abstract input however, an untrained $\theta$ already reaches 0.78 MTurk accuracy, yet no $\theta$ brings its human consistency to the level of \texttt{Natural} models.
This asymmetry is consistent with how the two inputs represent the stimuli. Human errors are reliable at the level of single trials, so they depend on the specific stimuli.
One-hot \texttt{Abstract} inputs make every stimulus equidistant, so no $\theta$ driven by them can reproduce these stimulus-specific errors, regardless of the input it was trained on.
In contrast, a pretrained vision encoder represents stimuli visually with task-specific features mixed with background and viewpoint in the same feature space.
A $\theta$ capable of solving the task inherits the geometry of this representation in its errors.

\subsection{Under natural inputs, \texttt{Abstract} recurrent weights produce dynamics similar to \texttt{Natural} models}\label{sec:results3}
\begin{figure}[ht]
  \centering
  \begin{minipage}{0.75\linewidth}
    \centering
    \begin{subfigure}{0.45\linewidth}
      \includegraphics[width=\linewidth]{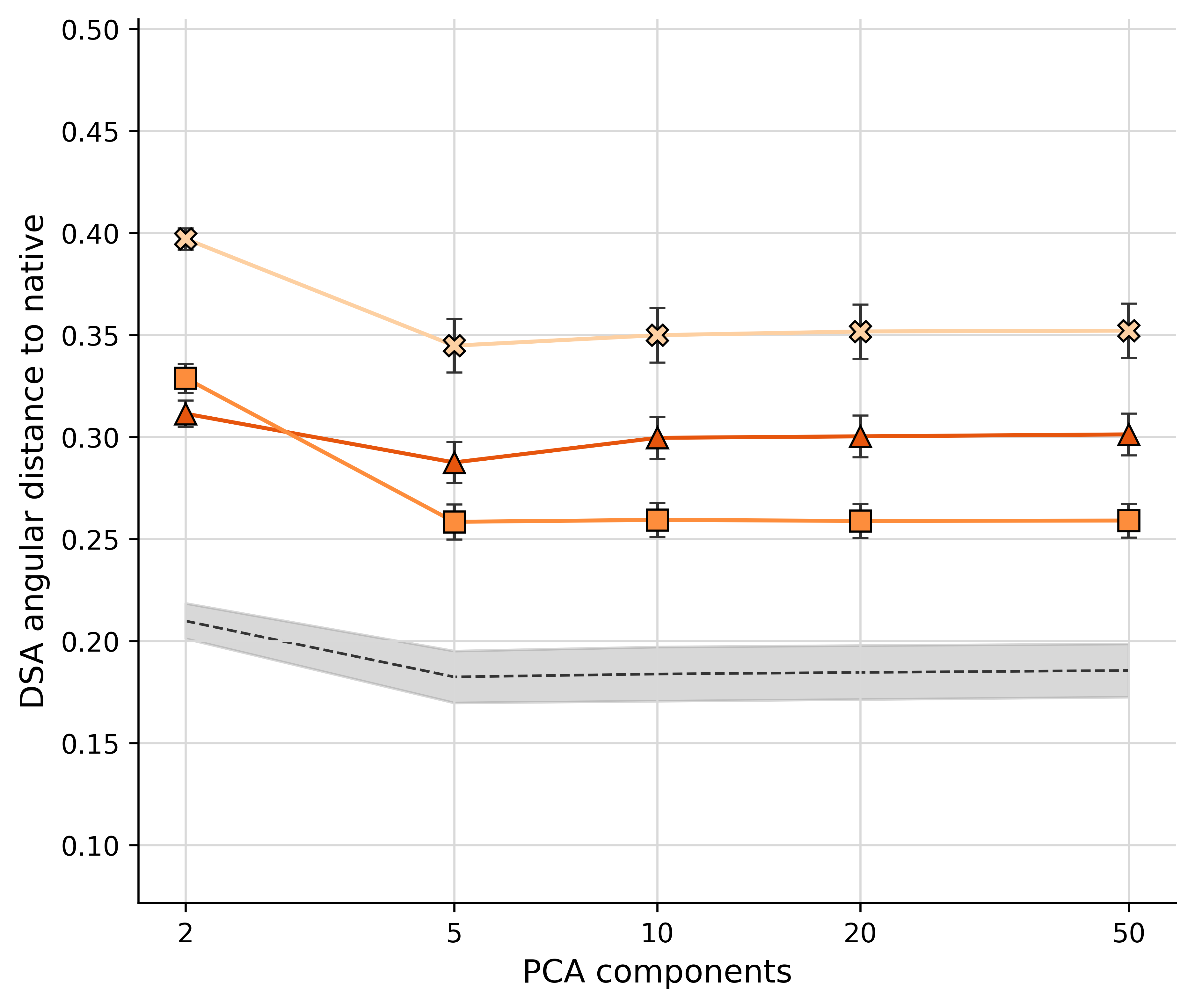}
      \caption{Abstract $\mathbf{h}_t$}\label{fig:dsa:ab}
    \end{subfigure}\hfill
    \begin{subfigure}{0.45\linewidth}
      \includegraphics[width=\linewidth]{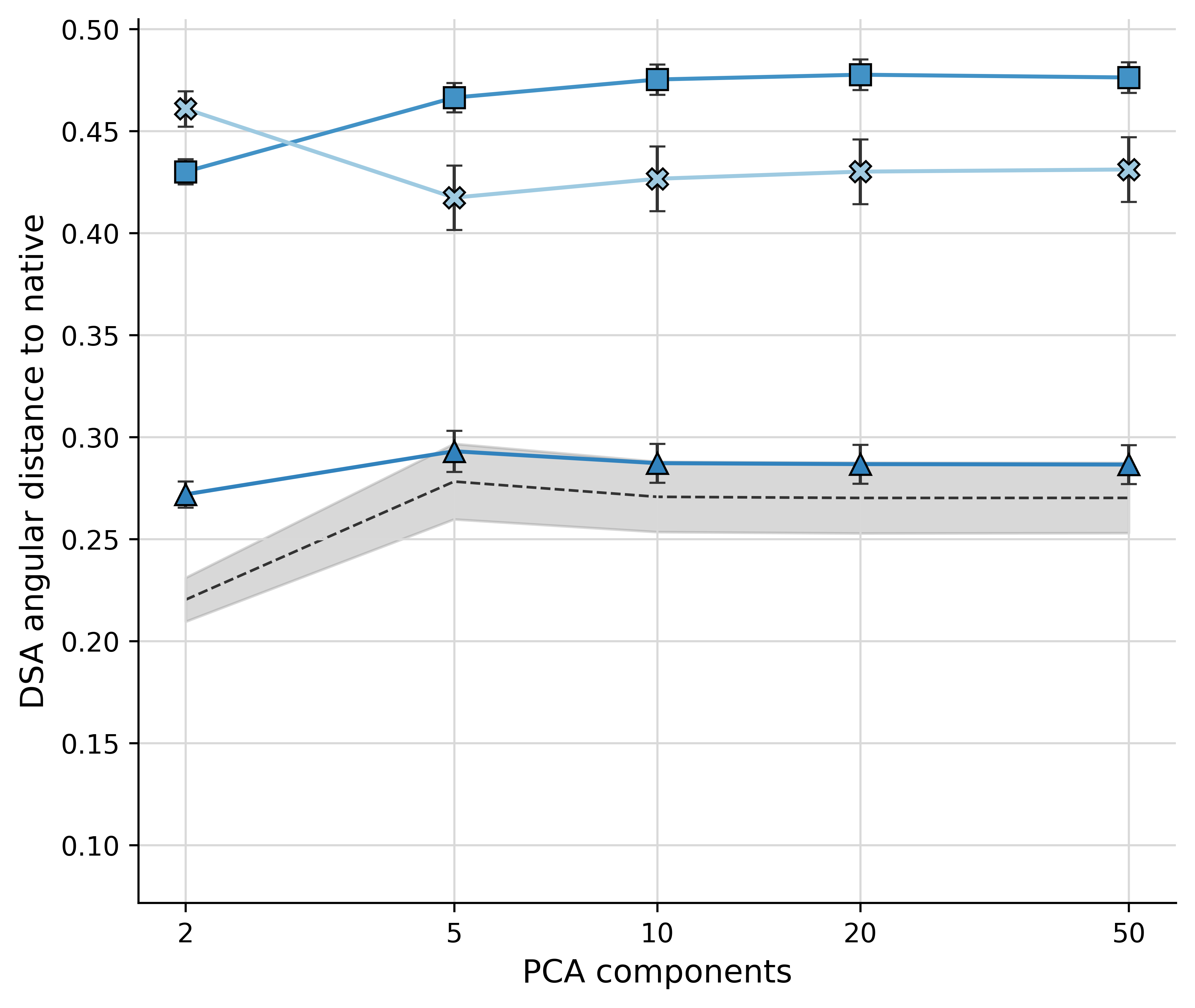}
    \caption{Natural $\mathbf{h}_t$}\label{fig:dsa:nat}
    \end{subfigure}
  \end{minipage}
  \begin{minipage}{0.2\textwidth}
    \centering
    \includegraphics[scale=0.5]{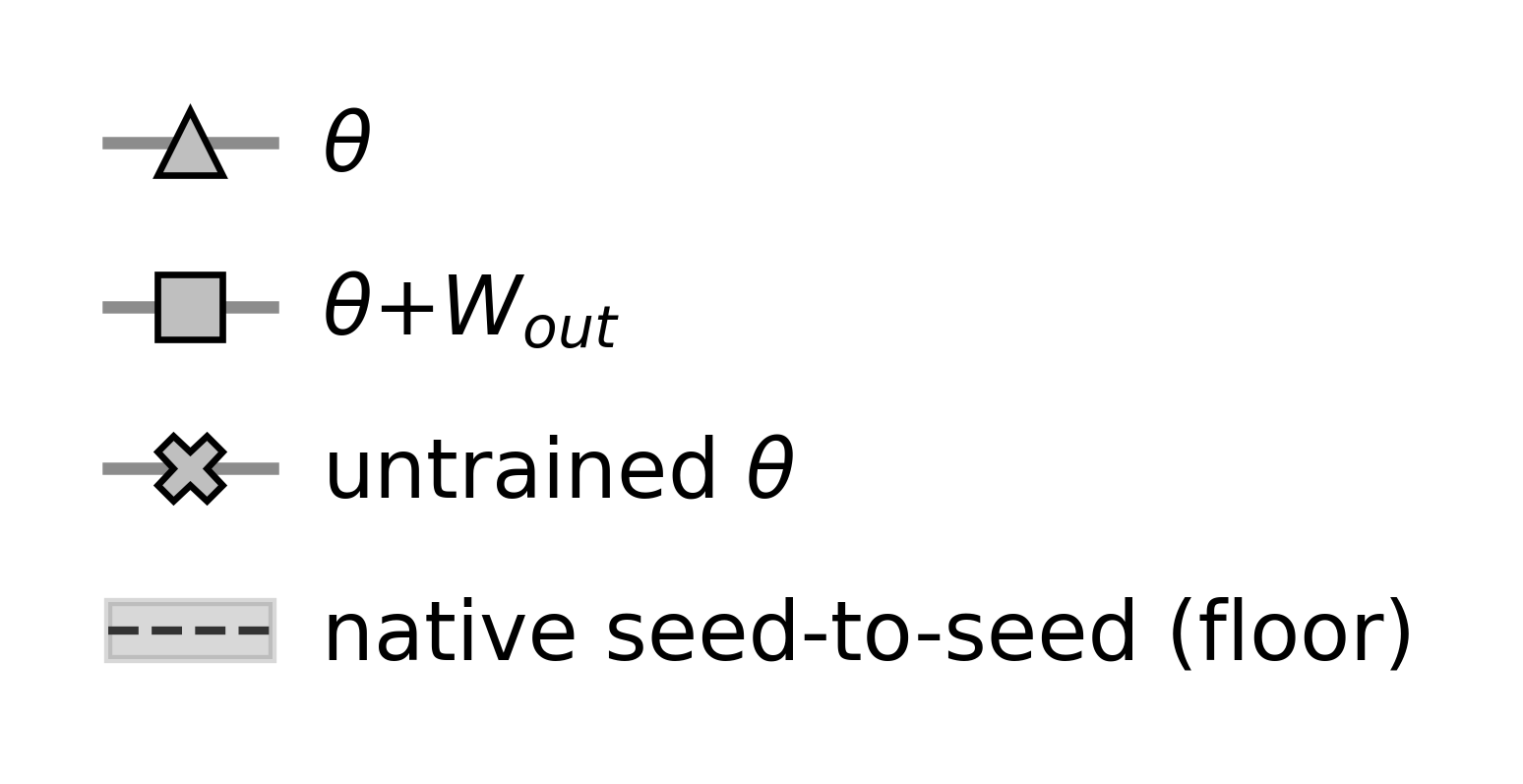}
    \end{minipage}
  \caption{
  \textbf{Dynamic distances of retrained models to native models.}
    (a). In models that receive abstract input,
    $\theta$ and $\theta+\mathbf{W}_{\mathrm{out}}$ substitution lie above the noise floor at every dimension but well below the untrained $\theta$.
    (b). In models that receive ResNet embedding input,
    from 5 dimensions on, $\theta$ substitution is at the noise floor, while $\theta+\mathbf{W}_{\mathrm{out}}$ and the untrained $\theta$ sit far above it, with $\theta+\mathbf{W}_{\mathrm{out}}$ farthest.
    Distances are between sequence-core hidden states.
    Gray bands and dashed lines indicate native models' seed-to-seed distances, which is the noise floor.
    Error bars indicate standard error of mean over tasks and seed pairs.}\label{fig:dsa}
\end{figure}
Neural networks trained on symbolic inputs have been widely used to model cognitive behaviors, particularly to inform how the neural dynamics of these networks (and potentially the brain) solve decision-making tasks. Our previous experiments showed that while \texttt{Abstract} models poorly replicate human behavior, their consistency improves when the projections are retrained on natural inputs. Two mechanisms are possible. Either the two models converge to similar recurrent computations, allowing $\theta$ trained on abstract input to transfer to natural input, or the retrained input projection compensates for differences in the recurrent computation, reaching human-like behavior through different neural dynamics. To distinguish between these two mechanisms, we used DSA to measure the distance between the models' internal dynamics.

Fig.~\ref{fig:dsa:nat} supports the first mechanism.
When the \texttt{Abstract} $\theta$ is frozen inside a model that receives ResNet embeddings and only the input and output projections are retrained, its dynamics lie within the seed-to-seed variability of native \texttt{Natural} models, for $\mu \ge 5$ and most other DSA settings (see \ref{sec:appendix/dsa}).
The retrained projection alone does not produce this match in dynamics, since an untrained $\theta$ stays far from the native models.
The dynamical similarity is also not inherited from the input, since the projected input $\mathbf{z}_{\mathrm{proj}_t}$ of this model remains far from that of native \texttt{Natural} models (Fig.~\ref{fig:dsa_pre:nat}).
This suggests that, when driven by natural inputs, the trained \texttt{Abstract} $\theta$ produces dynamics similar to those of \texttt{Natural} models, which may support the human-like behavior of this model. Freezing the readout as well breaks the match: the dynamics of $(\theta+\mathbf{W}_{\mathrm{out}})_{\mathrm{Abs}\to\mathrm{Nat}}$ stay above the seed-to-seed variability of native \texttt{Natural} models in every DSA setting we tested.

The match of $\theta_{\mathrm{Abs}\to\mathrm{Nat}}$ also holds in dimensionality-matched \texttt{Abstract} models, where, unlike with one-hot inputs, $(\theta+\mathbf{W}_{\mathrm{out}})_{\mathrm{Abs}\to\mathrm{Nat}}$ also stays within the seed-to-seed variability of native \texttt{Natural} models (Fig.~\ref{fig:dsa_dm:nat_seq}). This parallels the behavioral alignment of $(\theta+\mathbf{W}_{\mathrm{out}})_{\mathrm{Abs}\to\mathrm{Nat}}$ in dimensionality-matched models (Fig.~\ref{fig:transplant_dm_global}).
Substituting $\theta+\mathbf{W}_{\mathrm{in}}$ instead places the model far above the floor with either input, and the models perform at chance.
A trained $\theta$ thus needs an input projection adapted to the new modality.

In models receiving abstract input, substitution recovers native dynamics partly. \texttt{Natural} $\theta$ driven by abstract inputs stays above the seed-to-seed variability of native \texttt{Abstract} models at every dimension (Fig.~\ref{fig:dsa:ab}, Fig.~\ref{fig:dsa_dm:ab_seq}), even when the projected input matches the native models (Fig.~\ref{fig:dsa_pre:ab}, Fig.~\ref{fig:dsa_dm:ab_pre}).

Together, these results show that recurrent weights learned from symbolic inputs can work with sensory-realistic inputs and, given an input projection trained on them, produce dynamics that we cannot distinguish from those of models trained on natural inputs. This is consistent with the view that the behavioral gap between \texttt{Abstract} and \texttt{Natural} models depends more on their input representations than on their recurrent weights.
\section{Discussion and Limitations}
Through substitution analyses, we further showed that most of this improvement can be recovered by freezing the recurrent weights of an abstract sequence model and retraining only its input and output projections on naturalistic stimuli. This suggests that task input matters more for behavioral alignment than the input on which recurrent weights are trained. However, our experiments do not identify which properties of input representation are responsible.

Current models explained only a portion of the reliable structure in human behavior; the best-performing models accounted for 64\% of the behavioral consistency observed across human participants. The remaining gap likely reflects factors absent from the present framework such as reward-history-dependent adaptation \citep{botvinick2019reinforcement}, metacognitive monitoring \citep{fleming2012neural}, individual differences in learning strategy \citep{collins2018within}, and the linguistic and conceptual priors accumulated through a lifetime of naturalistic experience \citep{lake2017building, tenenbaum2011grow}. Closing this gap might require models that integrate sensory realism with richer learning histories, more structured inductive biases, and more flexible forms of adaptation.

Notably, large VLMs reproduced a substantial fraction of human behavioral consistency despite receiving no task-specific training, suggesting that broad multimodal pretraining produces representations that transfer non-trivially to structured cognitive demands \citep{mirchandani2023large, lampinen2022tell}. 
Fine-tuning such models on larger collections of compositional cognitive tasks may provide a practical route toward closing the remaining behavioral gap. 
More broadly, future work should examine whether in-context adaptation \citep{brown2020language, garg2022can}, continual learning \citep{hadsell2020embracing}, or reinforcement learning from environmental feedback \citep{wang2018prefrontal} better approximate the conditions under which human-like cognitive behavior is acquired.

\textbf{Limitations.} Several limitations should be addressed in future work. We derived the human behavioral consistency from a relatively small participant pool of 20 per task. 
While this sample size is sufficient for reliable group-level estimates, larger and more demographically diverse cohorts would yield more stable measures of human behavioral consistency and enable analyses of individual differences in model-human alignment.

Our comparison also changes several factors at once. \texttt{Natural} models use vision encoders pretrained on large image datasets, and with a frozen random encoder they do not learn the tasks (Appendix~\ref{sec:appendix_random_encoder}), so we cannot separate the contribution of pretraining from image input. Training \texttt{Abstract} models on inputs that gradually move toward the vision encoder space of each would directly test this account. Also, our tasks are binary match/non-match judgments over a single stimulus set, so the line between perceptual and cognitive contributions is not sharp. 

Although we attempted to eliminate extrinsic differences between abstract and natural input by matching their dimensionality and visual difficulty (see \ref{sec:appendix_dimmatch}), it remains unclear what intrinsic geometric properties of sensory realism drive behavioral alignment. Our dimensionality-matched control uses random up-sampling projections that keep the one-hot encoding in an equidistant simplex, whereas natural images are known to occupy a low-dimensional manifold within a high-dimensional pixel space \citep{arjovsky2017towards, goodfellow2014generative}.

Future work could address this with class-conditional optimal transport, morphing each one-hot encoding toward the empirical distribution of natural embeddings of its category. This would replace our binary comparison with a dose-response curve of behavioral alignment against sensory realism.

Furthermore, behavioral alignment alone cannot determine whether models and humans rely on similar underlying computations. 
Future work should couple behavioral metrics with comparisons of brain and model neural dynamics. Fixed-point analyses could reveal whether naturalistic training reshapes the attractor structures underlying task performance \citep{sussillo2013opening, golub2018fixedpointfinder}, while representational similarity analysis, centered kernel alignment, and linear probing could identify where and how task-relevant variables emerge and align with neural population responses \citep{kriegeskorte2008representational, kornblith2019similarity, alain2016understanding}. 
Together, these analyses would clarify whether sensory realism merely improves behavioral fit or also steers models toward more human-like computational mechanisms.

\begin{ack}
We thank the Amazon Mechanical Turk participants who took part in our experiments, and the reviewers and area chair, whose feedback improved this paper.
This research was supported by the NSERC Discovery grant RGPIN-2021-03035 and the CIHR Project Grant PJT-191957.
P.B. was supported by the FRQ-S Research Scholars Junior 1 grant 310924 and the William Dawson Scholar award.
X.L. was supported by a CAMBAM fellowship, and L.G. by a Masters Excellence Scholarship from UNIQUE (Unifying Neuroscience and Artificial Intelligence in Québec).
This research was enabled in part by support provided by Calcul Québec (\url{calculquebec.ca}) and the Digital Research Alliance of Canada (\url{alliancecan.ca}), and by funding from the Canada Foundation for Innovation (project 42730).
The funders had no role in study design, data collection and analysis, the decision to publish, or the preparation of the manuscript.
The authors declare no competing interests.
\end{ack}
\bibliographystyle{unsrtnat}
\bibliography{ref}

\newpage
\appendix
\section{Appendix}\label{sec:appendix}
\subsection{Tasks}
Here, we detail our task space, task generation pipeline, sensory-realistic task stimuli, and human task evaluation.
\subsubsection{Task Description}\label{sec:appendix/task_description}
\begin{table}[h]
  \centering
  \caption{Summary of the 43 DMS tasks.}
  \label{tab:tasks}
  \resizebox{\linewidth}{!}{%
  \begin{tabular}{llcc}
    \toprule
    Family & Description & \# Tasks & Task IDs \\
    \midrule
    \textbf{NBack}   & $n$-back task for each feature & 9  & 1--9 \\
    \textbf{InterDMS}& Two interleaved DMS sub-tasks (patterns: AABB, ABAB, ABBA) & 27 & 10--36 \\
    \textbf{CtxDMS}  & Contextual DMS: first comparison gates the feature to use in the second comparison & 4  & 37--40 \\
    \textbf{DMS}     & Simple delayed match-to-sample & 3  & 41--43 \\
    \bottomrule
  \end{tabular}
  }
\end{table}
We define 43 DMS tasks that compare three stimulus features: category, identity, and position. The tasks are organized by their structure into four families (Table~\ref{tab:tasks}). For a complete breakdown of each task, see Table~\ref{tab:all_tasks}.

\begin{figure}
\centering
\includegraphics[width=0.8\textwidth]{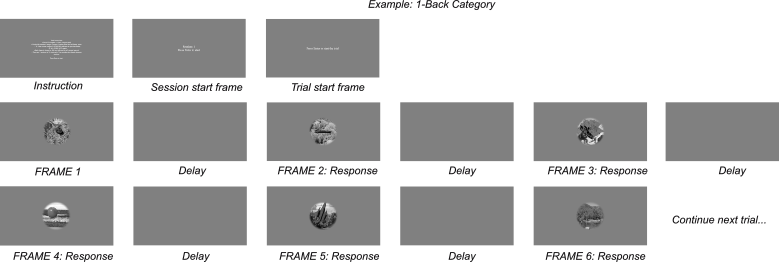}
\caption{An example trial of 1-Back Category task }\label{fig:task_1back}
\end{figure}

\paragraph{NBack (tasks 1--9).}
A sequence of six frames is shown. At each step $t$, the model must report whether
frame $t$ matches frame $t-n$ on the target feature, yielding up to $6 - n$ decision
frames per trial (see Fig.~\ref{fig:task_1back}).

\begin{figure}
\centering
\includegraphics[width=0.8\textwidth]{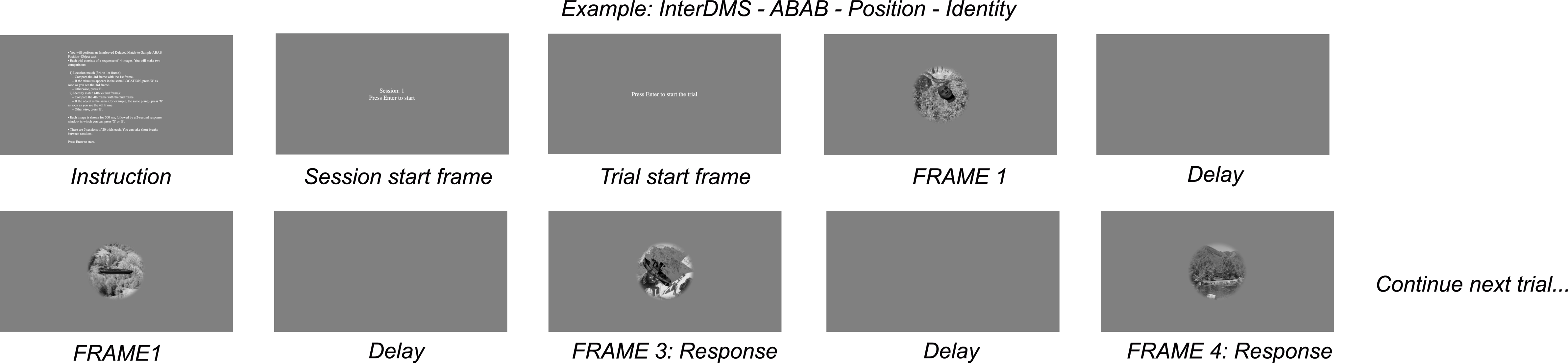}
\caption{An example trial of InterDMS - ABAB - Position - Identity task }\label{fig:task_interdms}
\end{figure}

\paragraph{InterDMS (tasks 10--36).}
Two independent DMS sub-tasks (see Fig.~\ref{fig:task_interdms}), one for feature $f_1$ and one for feature $f_2$, are
interleaved across four stimulus frames according to one of three temporal patterns:
\textbf{AABB} (both $f_1$ frames precede both $f_2$ frames),
\textbf{ABAB} (frames alternate between the two sub-tasks), and
\textbf{ABBA} (the $f_2$ sub-task is nested inside the $f_1$ sub-task).

\begin{figure}
\centering
\includegraphics[width=0.8\textwidth]{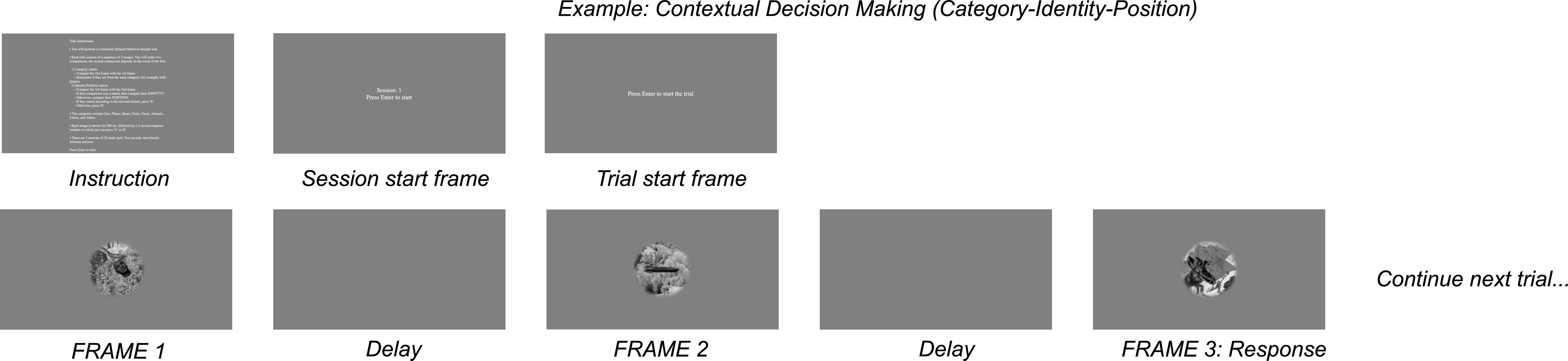}
\caption{An example trial of CtxDMS - Category - Identity - Position task }\label{fig:task_ctxdm}
\end{figure}

\paragraph{CtxDMS (tasks 37--40).}
Three frames are presented. The first comparison (frames 1 and 2) tests a \emph{context}
feature $f_0$. Based on whether the context match is positive or negative, the second
comparison (frames 2 and 3) tests one of two possible \emph{test} features
($f_1$ or $f_2$), requiring the model to flexibly route its decision (see Fig.~\ref{fig:task_ctxdm}).

\begin{figure}
\centering
\includegraphics[width=0.8\textwidth]{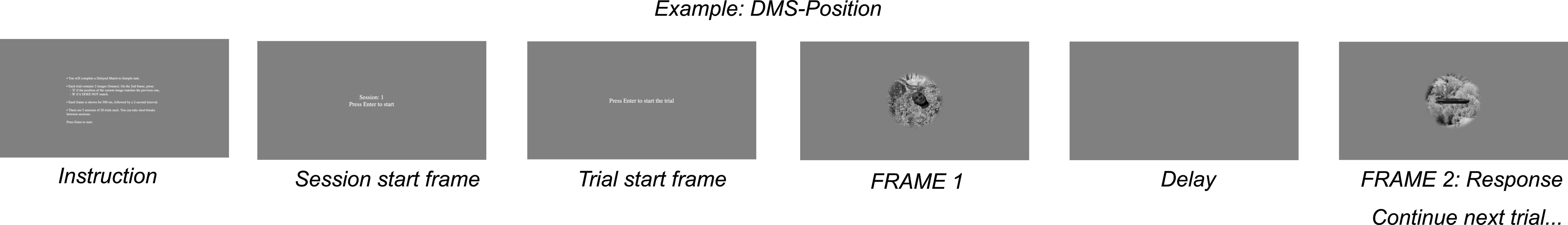}
\caption{An example trial of DMS-position task }\label{fig:task_dms}
\end{figure}

\paragraph{DMS (tasks 41--43).}
Two frames are presented. The model observes a \emph{sample} stimulus and must
judge whether the immediately following \emph{test} stimulus matches on the target
feature (position, identity, or category, see Fig.~\ref{fig:task_dms}).

\subsubsection{HvM Dataset}\label{sec:appendix/hvm_dataset}
At training time, trials are generated on-the-fly. For each frame, the appropriate feature constraints are applied as a Boolean mask over all eligible HvM rows, and a stimulus is sampled uniformly from the resulting candidate set. With 50\% probability, the target feature value is preserved from the reference frame (match); otherwise a new value is drawn that differs on that feature (non-match), while all unconstrained features vary freely.

The image below illustrates the three variation levels in the High-Variation Model (HvM) dataset \citep{rajalingham2015comparison} used for the behavioral and computational experiments.

\begin{figure}
    \centering 
    \includegraphics[width=0.6\textwidth]{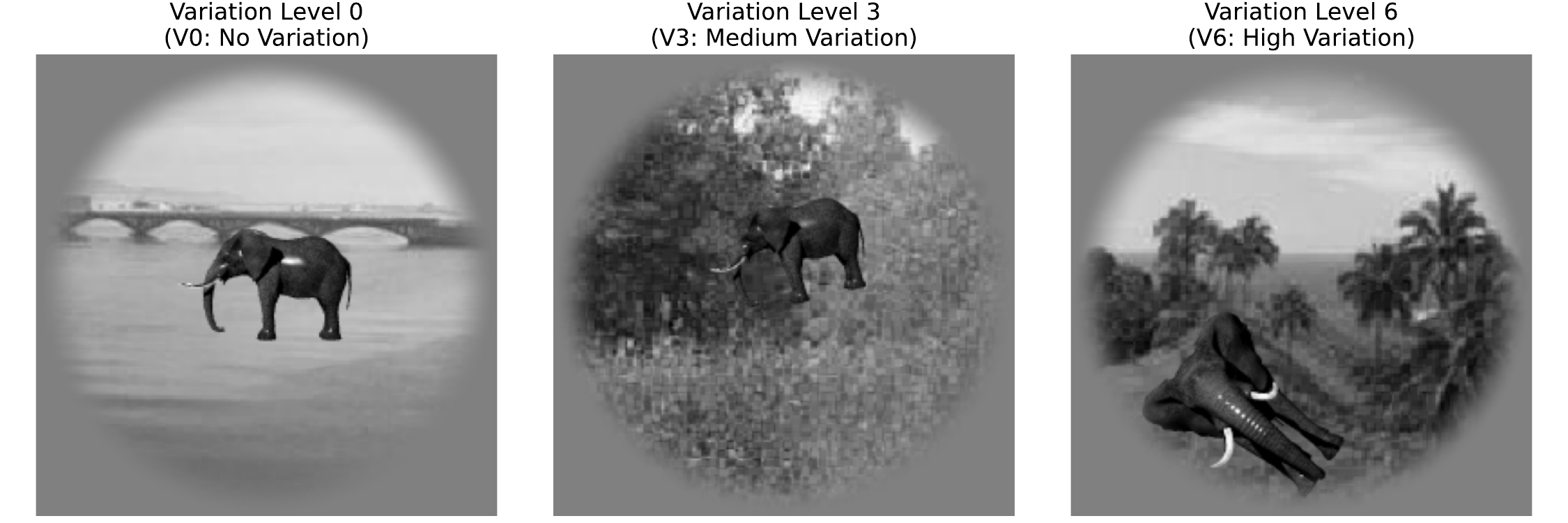}
    \caption{\textbf{Examples of the HvM stimuli seen by sensory-realistic models at 3 variation levels.} Variation 0 (var-0): No variation. Objects are centered, with a fixed neutral view angle and fixed scale. Variation 3 (var-3): Medium variation. Objects undergo moderate translation, rotation, and scaling. Variation 6 (var-6): High variation. Objects undergo large translation, rotation, and scaling, presenting the most challenging visual condition.
}\label{fig:hvm_examples} 
\end{figure}

\paragraph{Variation-level balancing.}
Variation level 6 contains more images per object than levels 0 and 3, and would otherwise be sampled at a disproportionately high rate. To counteract this and reduce sampling bias toward variation level 6, we assign sampling weights of 0.5 to all var-6 candidates relative to 1.0 for other variation levels, making var-6 images approximately half as likely to be drawn.
\paragraph{Task generation.}
Each task is defined by a fixed logic that determines which stimulus pairs are compared and along which feature dimension. For instance, \texttt{InterDMS-ABBA-position-category} compares the position of the first and last stimuli (response on frame 4), and also the category of the second and third stimuli (response on frame 3). The predefined task logic allows us to sample the ground truth action first (match vs. no-match), and then find stimuli that satisfy the task conditions. This generation procedure guarantees balanced action labels and ensures the diversity of stimulus features.

\subsubsection{Human Trials}\label{sec:appendix/human_trials}
The 12 MTurk task IDs are: $\{1,2,3,19,23,27,28,32,36,37,40,41\}$. On the interactive website for human trial collection, each frame was shown for a fixed duration, and participants respond via their keyboard (\textit{x} = match; \textit{b} = non-match) at the response frames. Before the main session, for each task, participants completed a learning session (2 trials for 1Back, 3 for DMS, 5 for InterDMS, and 8 for CtxDMS tasks; 11 early CtxDMS participants received 5) with per-frame on-screen task instructions and response feedback.

From 16 objects (2 per category), we randomly selected 100 images: 6 per object plus 1 more for 4 of the objects. Of these, 95 are high-variation images (31 from variation 3 and 64 from variation 6) and 5 are variation-0 images (centered, unscaled) from 5 unique objects. To enable direct comparison across all 12 tasks, we construct a single set of 100 unique
stimulus \emph{pairs} that are jointly balanced for category, identity, and position match distributions (50\% match / 50\% non-match for each feature). NBack tasks additionally require
temporally chained pairs; for those tasks we greedily construct chains that approximately maintain feature balance.

To enforce quality control, participants whose accuracy on the main session fell below 65\% (approximately
$3\sigma$ above chance level of 50\%) were rejected and excluded from analysis. For task~40
(CtxDMS: category-identity-position), which is the most demanding variant, the threshold
was relaxed to 60\% due to its high difficulty (16\% of recruited workers met the standard threshold). MTurk workers who passed this quality control received \$5 per task.

\subsection{Additional Results}\label{sec:appendix_add_results}
\subsubsection{Human consistency}\label{sec:appendix/human_cons}
\begin{algorithm}
\caption{Split-Half Consistency Evaluation}
\label{alg:split_half}
\begin{algorithmic}[0]

\Require Model logits $\mathbf{M}$, human binary responses $\mathbf{H}$
\Ensure Consistency scores

\State $\mathcal{R} \gets \left[ \ \right]$

\For{each split indices $(S_h^a, S_h^b)$ and $(S_m^a, S_m^b)$}
    \State $\rho(h, h):$
    \State $\mathbf{h}_a \gets \mathrm{mean}(\mathbf{H}_{S_h^a})$
    \State $\mathbf{h}_b \gets \mathrm{mean}(\mathbf{H}_{S_h^b})$
    \State $\rho_{hh} \gets \mathrm{corr}(\mathbf{h}_a, \mathbf{h}_b)$
    \State
    
    \State $\rho(m, m):$
    \State $\mathbf{m}_a \gets \mathrm{mean}(\mathbf{M}_{S_m^a})$
    \State $\mathbf{m}_b \gets \mathrm{mean}(\mathbf{M}_{S_m^b})$
    \State $\rho_{mm} \gets \mathrm{corr}(\mathbf{m}_a, \mathbf{m}_b)$

    \State
    \State $\rho(m, h):$
    \State $\rho_1 \gets \mathrm{corr}(\mathbf{m}_a, \mathbf{h}_a)$
    \State $\rho_2 \gets \mathrm{corr}(\mathbf{m}_a, \mathbf{h}_b)$
    \State $\rho_3 \gets \mathrm{corr}(\mathbf{m}_b, \mathbf{h}_a)$
    \State $\rho_4 \gets \mathrm{corr}(\mathbf{m}_b, \mathbf{h}_b)$
    \State $\rho_{mh} \gets \mathrm{FisherZMean}(\rho_1, \rho_2, \rho_3, \rho_4)$
    \State
    
    \State $\tilde{\rho}(m, h):$
    \State $\tilde{\rho}_{mh} \gets \frac{\rho_{mh}}{\sqrt{\rho_{mm} \cdot \rho_{hh}}}$

    \State
    \State $\mathcal{R}.\mathrm{append}(\rho_{hh}, \rho_{mm}, \rho_{mh}, \tilde{\rho}_{mh})$
\EndFor

\State \textbf{return} $\mathrm{Aggregate}(\mathcal{R})$

\end{algorithmic}
\end{algorithm}

\citet{rajalingham2018large} define a noise-corrected human consistency as:
\begin{equation}
  \widetilde{\rho}(m, h)
  = \frac{\rho(m,h)}{\sqrt{\rho(m,m)\cdot\rho(h,h)}},
  \label{eq:hc}
\end{equation}
where $\rho(m, h)$ is the Pearson correlation between model and human response vectors, $\rho(m, m)$ correlation between model split-halves, and $\rho(h, h)$ is the Pearson correlation between human split-halves. We report the uncorrected $\rho(m,h)$ throughout, clipped at zero after averaging over split-halves ($\max(0,\rho(m,h))$, so a reported 0 means $\rho(m,h)\le 0$), and show $\rho(h,h)$ as its ceiling. $\widetilde{\rho}(m,h)$ is not used. Statistical tests and the residualization analysis use the unclipped values.

We perform 20 random split-halves, compute $\rho(m, h), \rho(m, m), \rho(h, h)$, take the Fisher z-transform of each correlation, and average across splits. Since we are primarily interested in the behavioral alignment between models and humans and less interested in the model's internal reliability across seeds, we visualize the uncorrected $\rho(m, h)$ directly in Sec.~\ref{sec:results}, and show $\rho(h, h)$ as the noise-correction ceiling. 

For significance tests, we correlate each seed’s response probabilities with the mean human response, average over seeds within a model, and pair over the five sequence architectures or the 18 task conditions.
\begin{figure}
\centering

\begin{subfigure}{0.3\textwidth}
    \includegraphics[width=\linewidth]{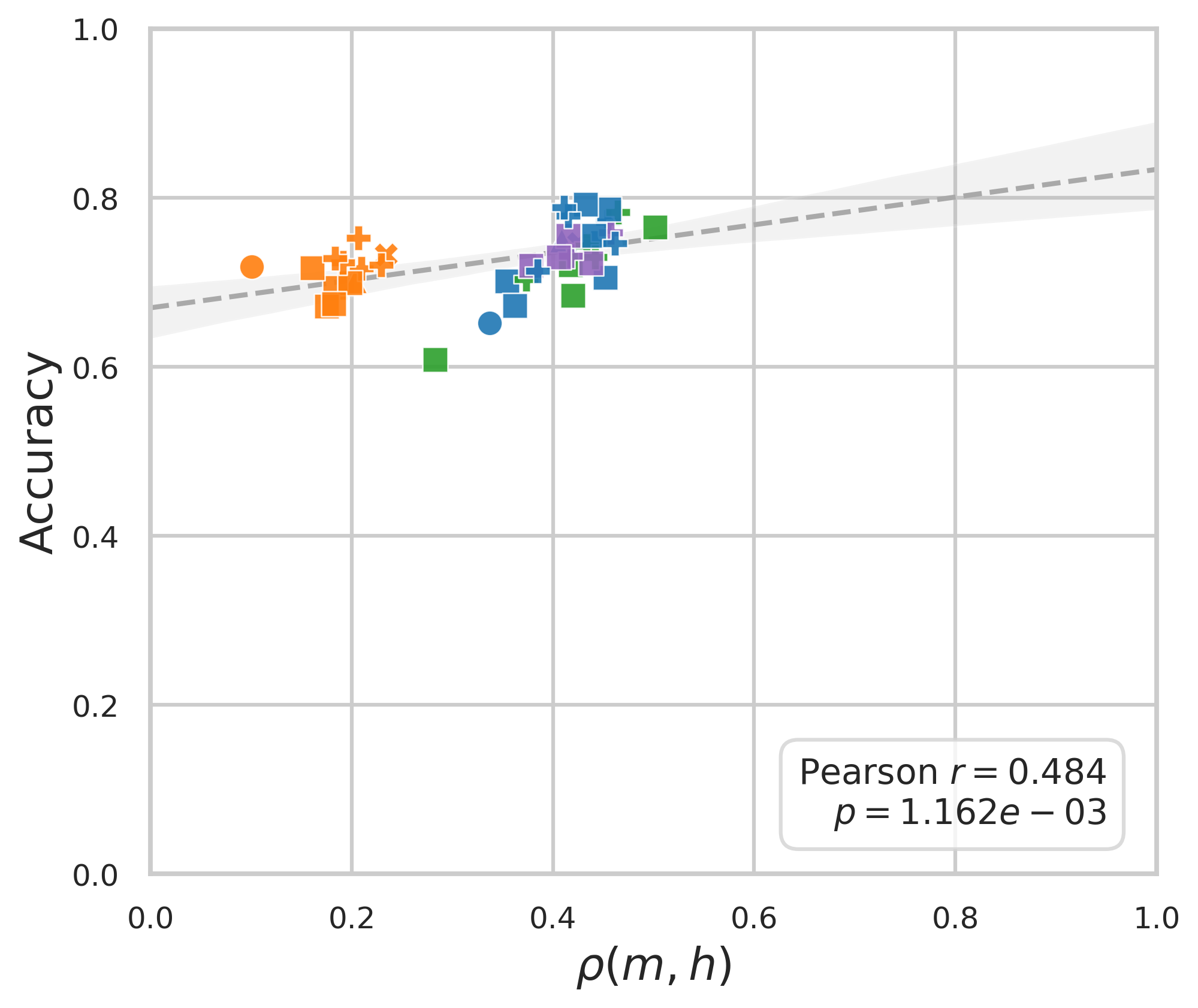}
    \caption{$\rho,$ Accuracy}
\end{subfigure}
\hfill
\begin{subfigure}{0.3\textwidth}
    \includegraphics[width=\linewidth]{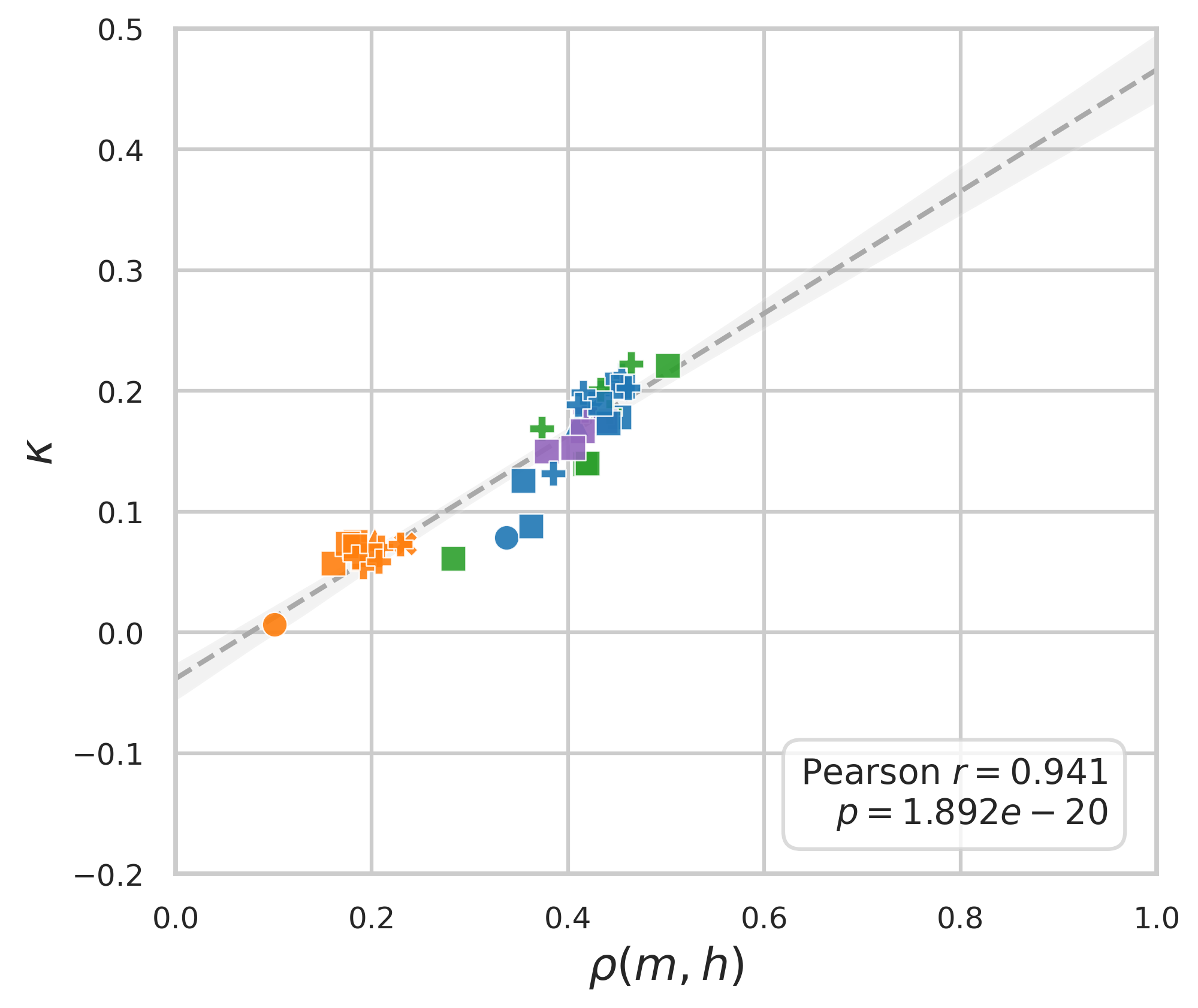}
    \caption{$\rho,\kappa$}
\end{subfigure}
\hfill
\begin{subfigure}{0.3\textwidth}
    \includegraphics[width=\linewidth]{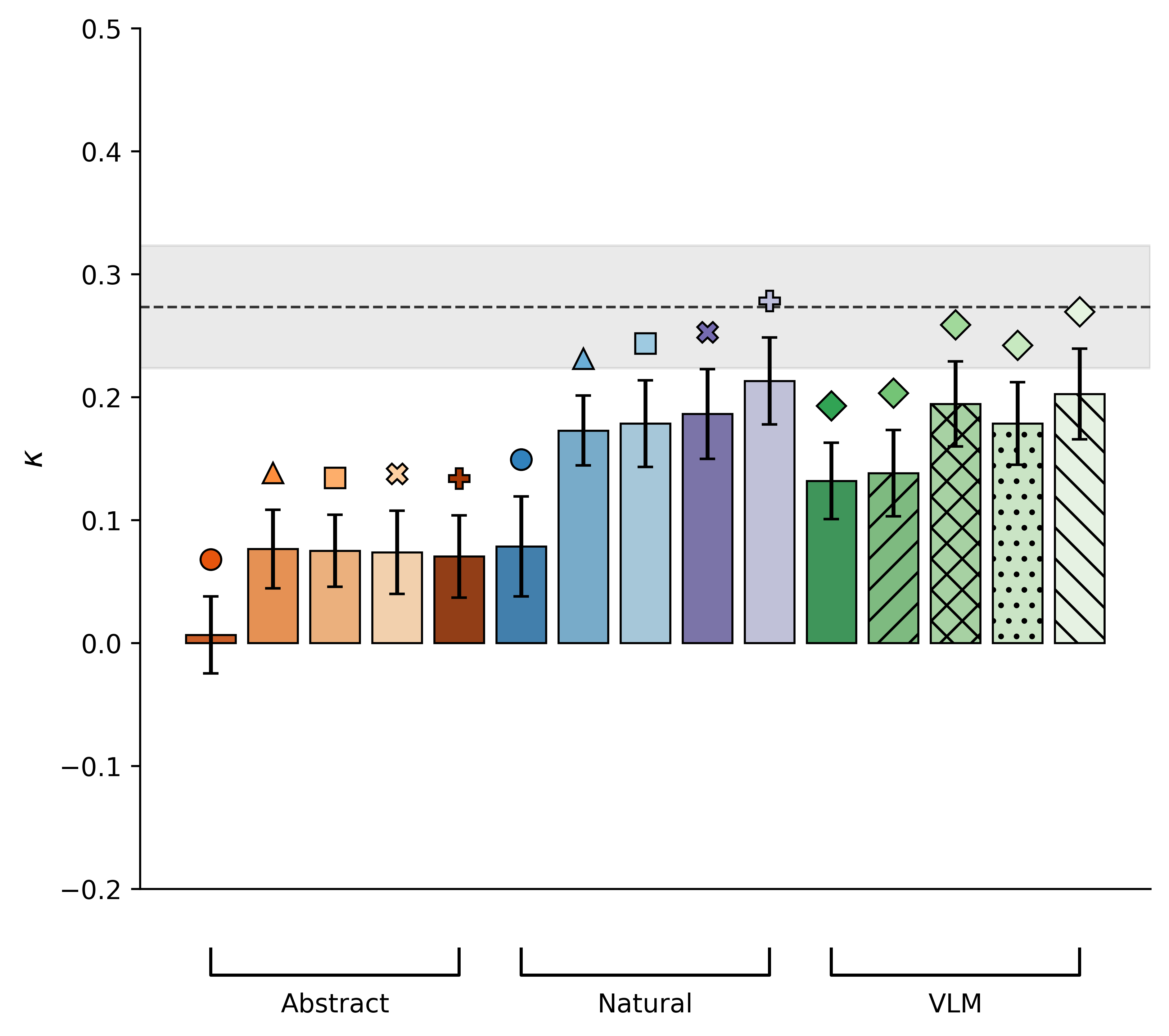}
    \caption{$\kappa$}
\end{subfigure}
\caption{\textbf{Correlation of behavioral metrics.} 
Across models, human consistency $\rho(m,h)$ is almost linearly related to error consistency $\kappa$ (Pearson $r=0.941$) and much less to accuracy ($r=0.484$), which motivates measuring model behavior beyond accuracy.}
\label{fig:metric_correlation}
\end{figure}

\begin{figure}
    \centering
\begin{subfigure}{0.48\textwidth}
    \includegraphics[width=\linewidth]{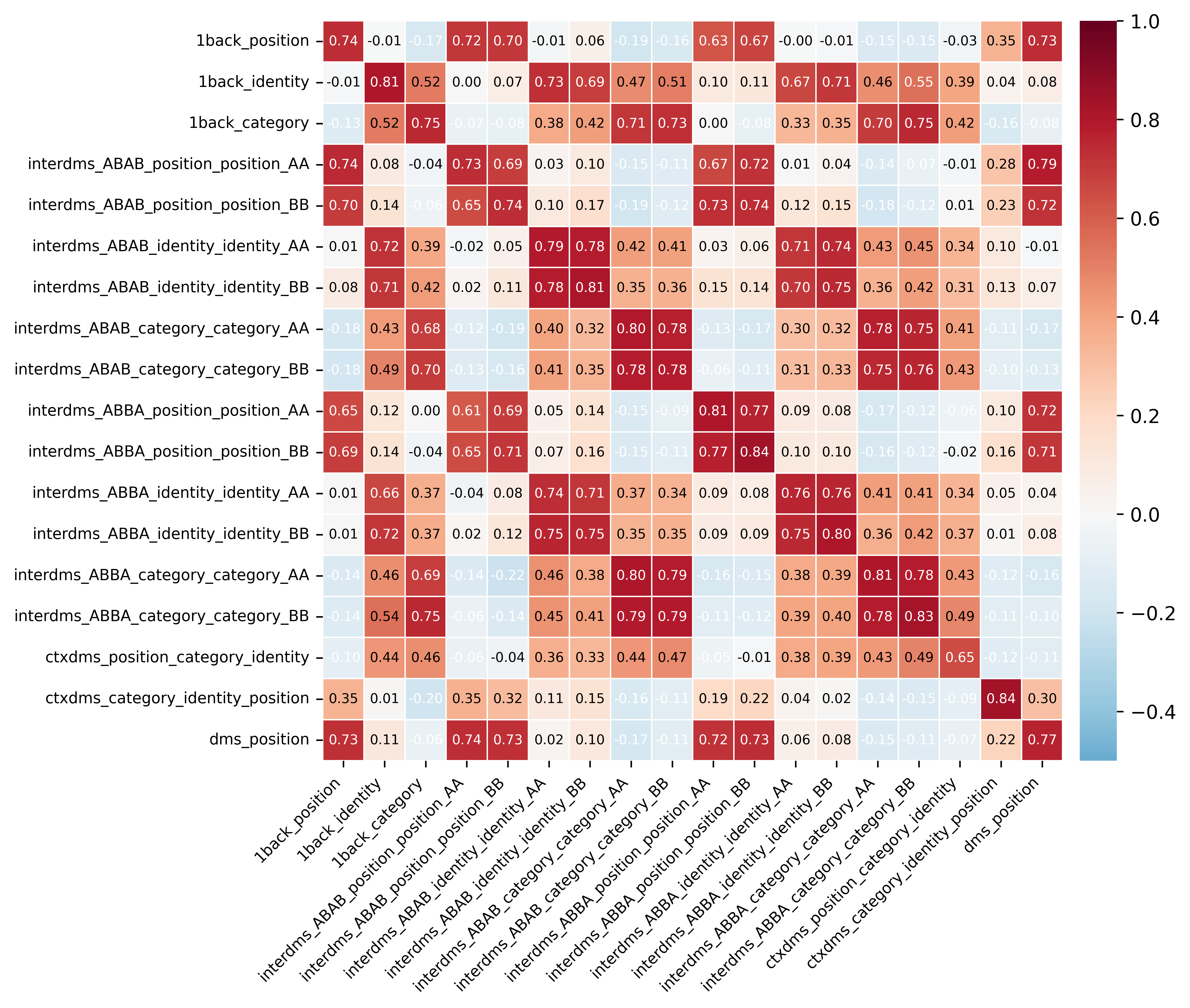}
    \caption{}\label{fig:human_ct}
\end{subfigure}
\hfill
\begin{subfigure}{0.48\textwidth}
    \includegraphics[width=\linewidth]{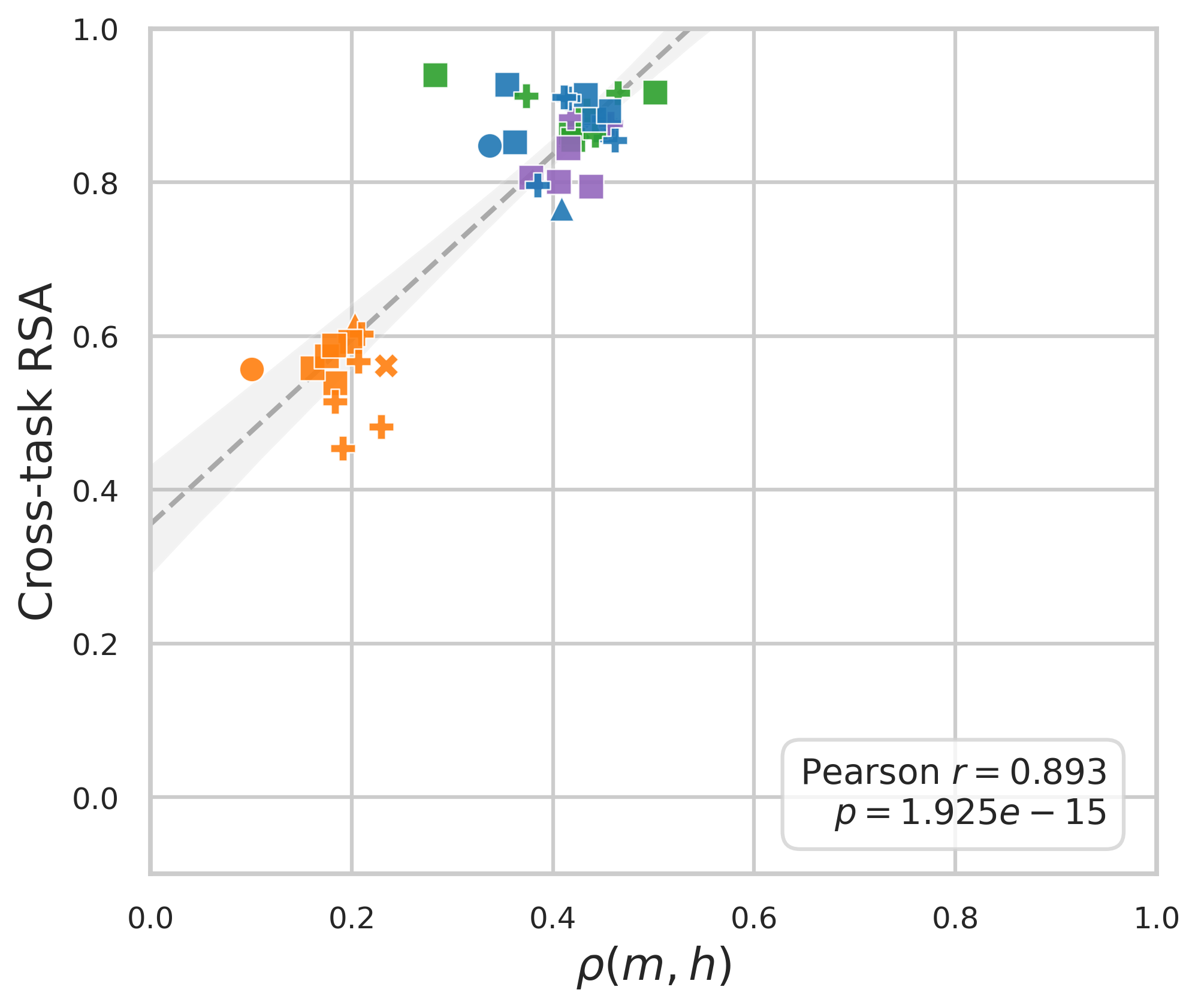}
    \caption{}\label{fig:model_task_rsa_alignment}
\end{subfigure}
\vspace{0.5em}
\begin{subfigure}{0.48\textwidth}
    \includegraphics[width=\linewidth]{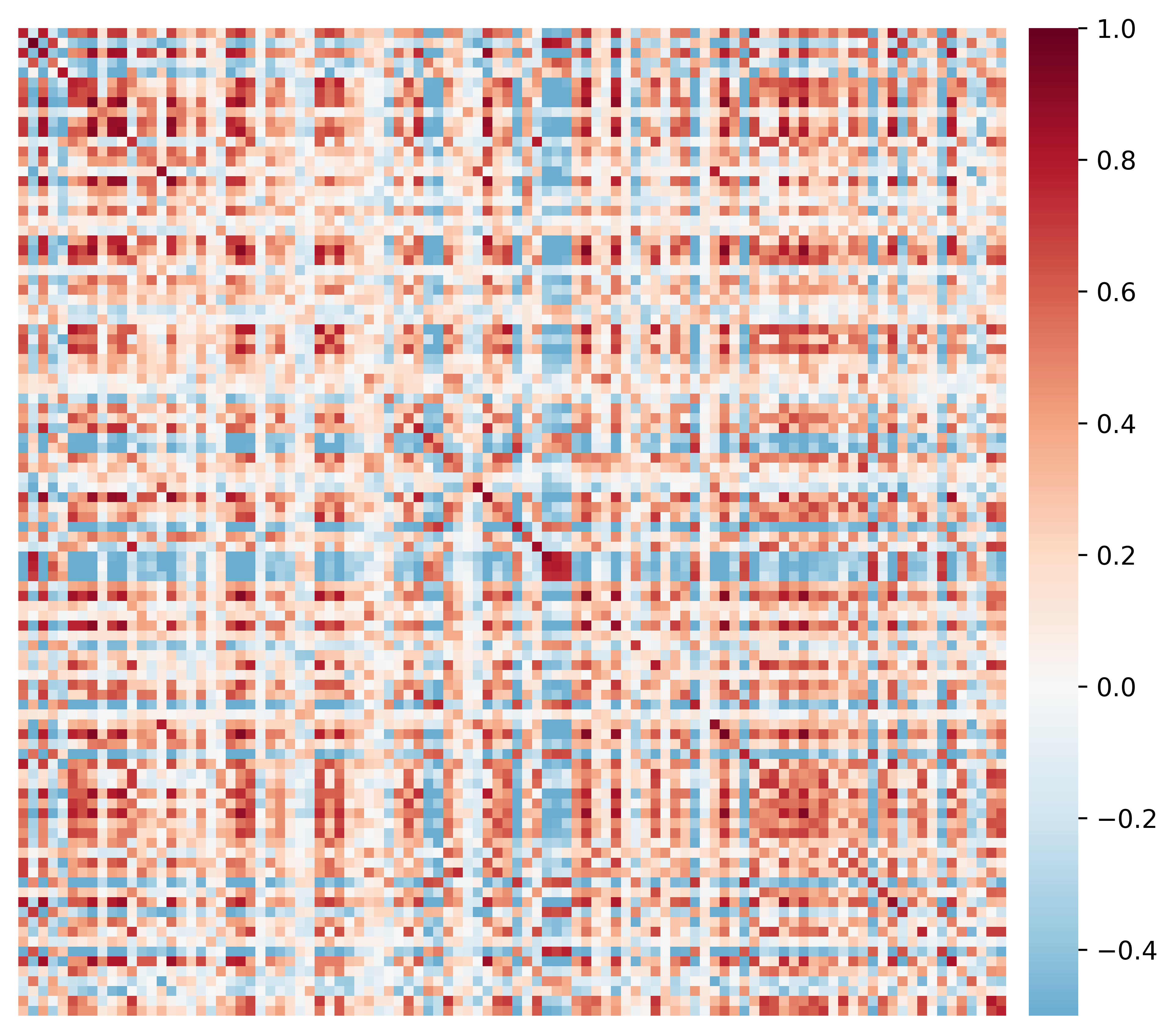}
    \caption{}\label{fig:human_cs}
\end{subfigure}
\hfill
\begin{subfigure}{0.48\textwidth}
    \includegraphics[width=\linewidth]{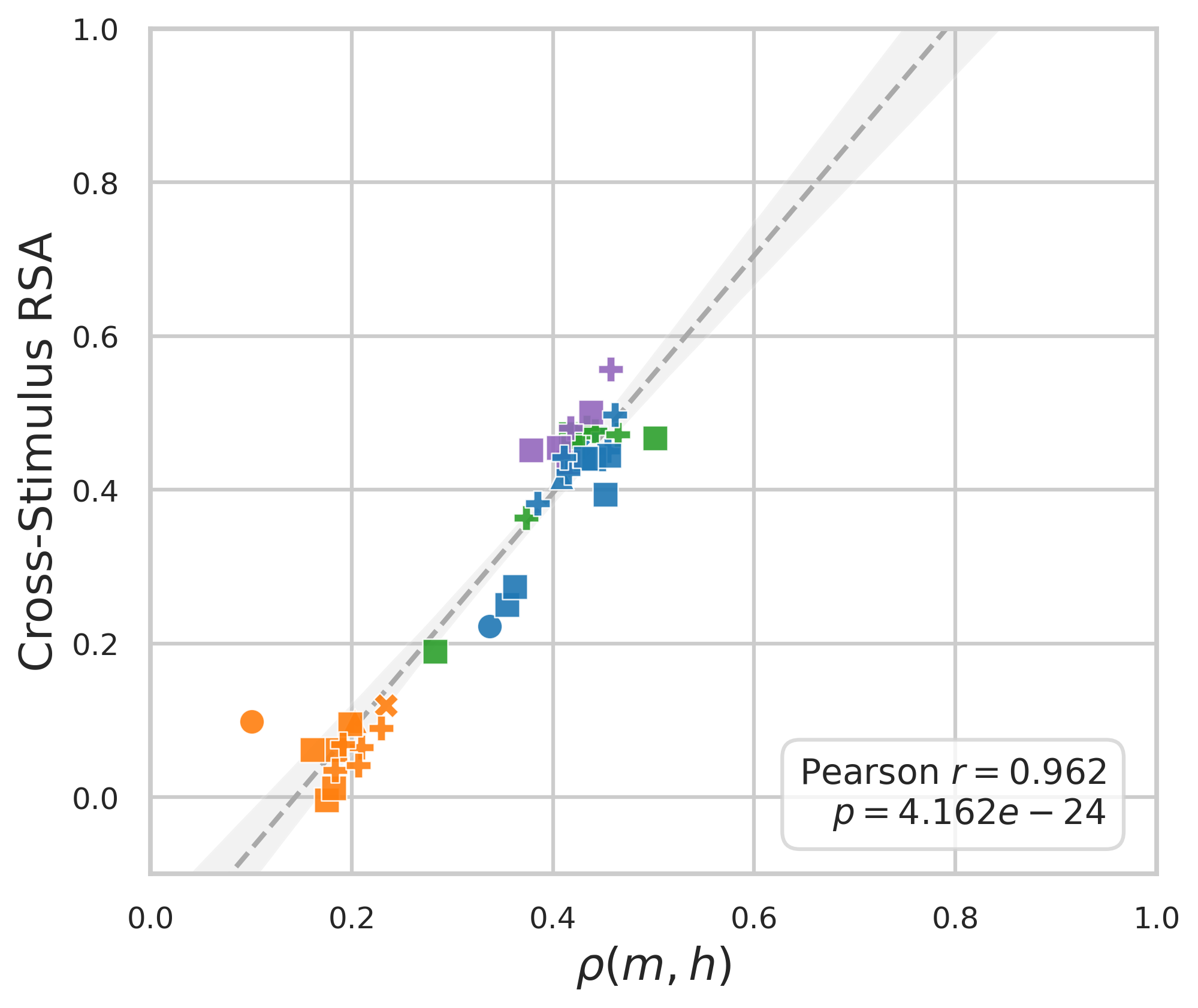}
    \caption{}\label{fig:model_stimuli_rsa_alignment}
\end{subfigure}
\caption{\textbf{Cross-task and cross-stimulus RSA.} (a) Human cross-task RSA matrix shows clusters of high correlation in tasks using the same stimuli features, e.g. 1back position and InterDMS position. Since identity tasks are related to category tasks, we also see correlation between the response vectors in these cross-task comparisons. (b) Correlation between cross-task RSA and human consistency. Abstract models (orange) have much lower alignment values than natural models (blue, green, and purple colors match the colors used in Fig.~\ref{fig:vision_type_global}). (c) Human cross-stimulus RSA matrix. If two stimuli are similar in category, identity, and position, then they would have high correlation across tasks. (d) Correlation between cross-stimulus RSA and human consistency. Cross-stimuli RSA alignment predicts human consistency better than cross-task RSA.}
\end{figure}
\begin{figure}
\centering

\begin{subfigure}{0.3\textwidth}
    \includegraphics[width=\linewidth]{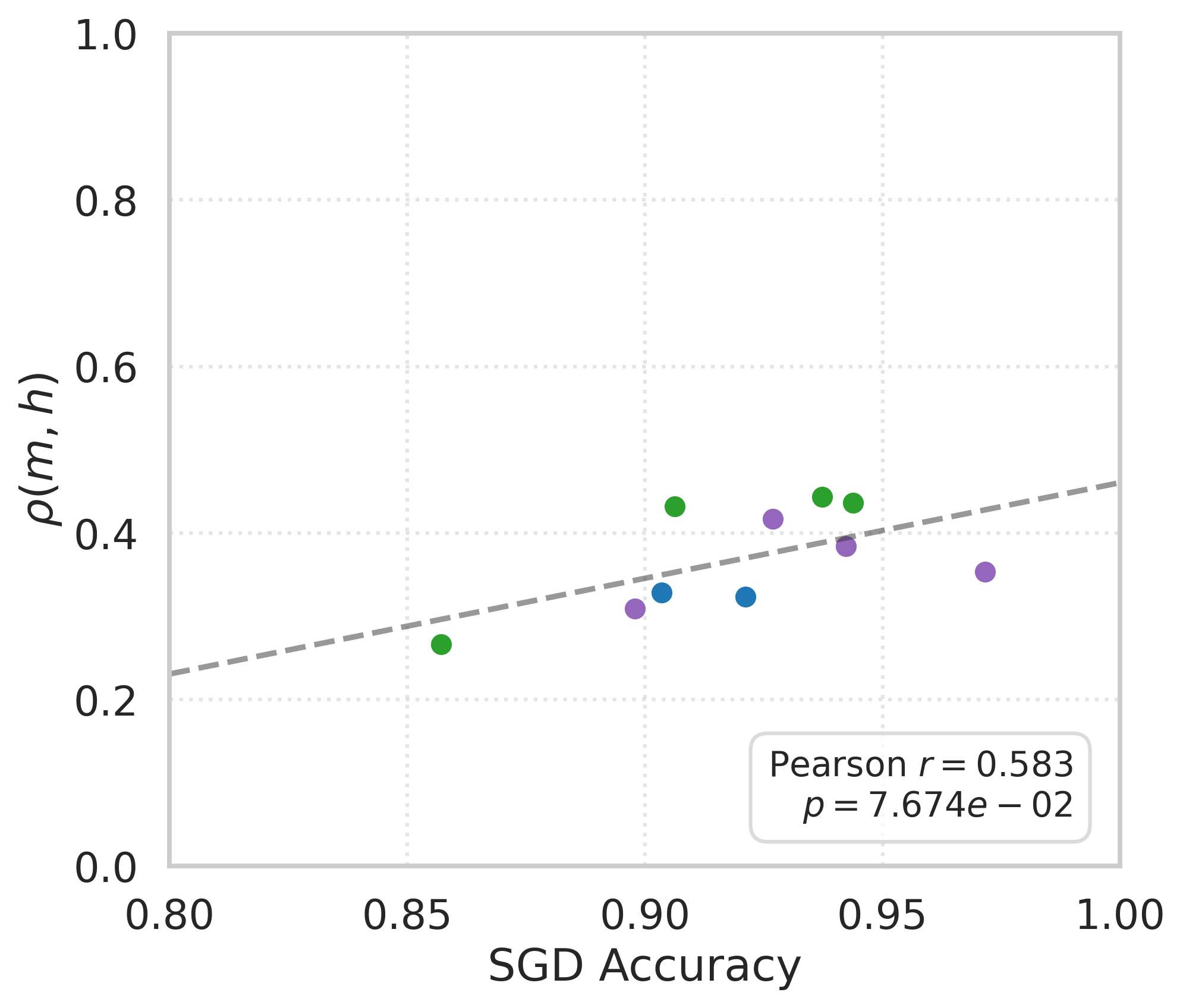}
    \caption{CAT, $\rho$}
    \label{fig:metric_alignment_cat_cons}
\end{subfigure}
\hfill
\begin{subfigure}{0.3\textwidth}
    \includegraphics[width=\linewidth]{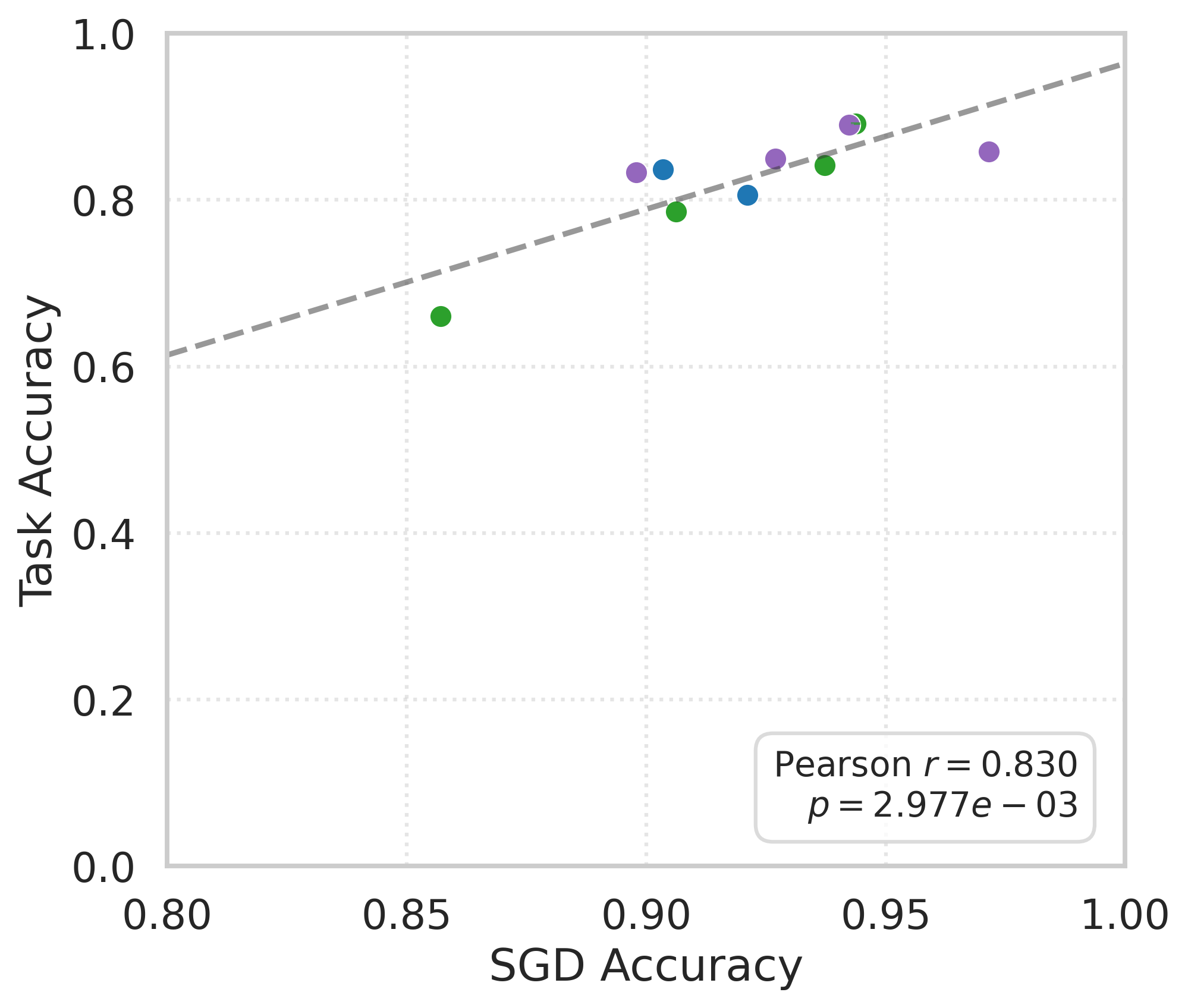}
    \caption{CAT, Acc}
    \label{fig:metric_alignment_cat_acc}
\end{subfigure}
\begin{subfigure}{0.3\textwidth}
    \includegraphics[width=\linewidth]{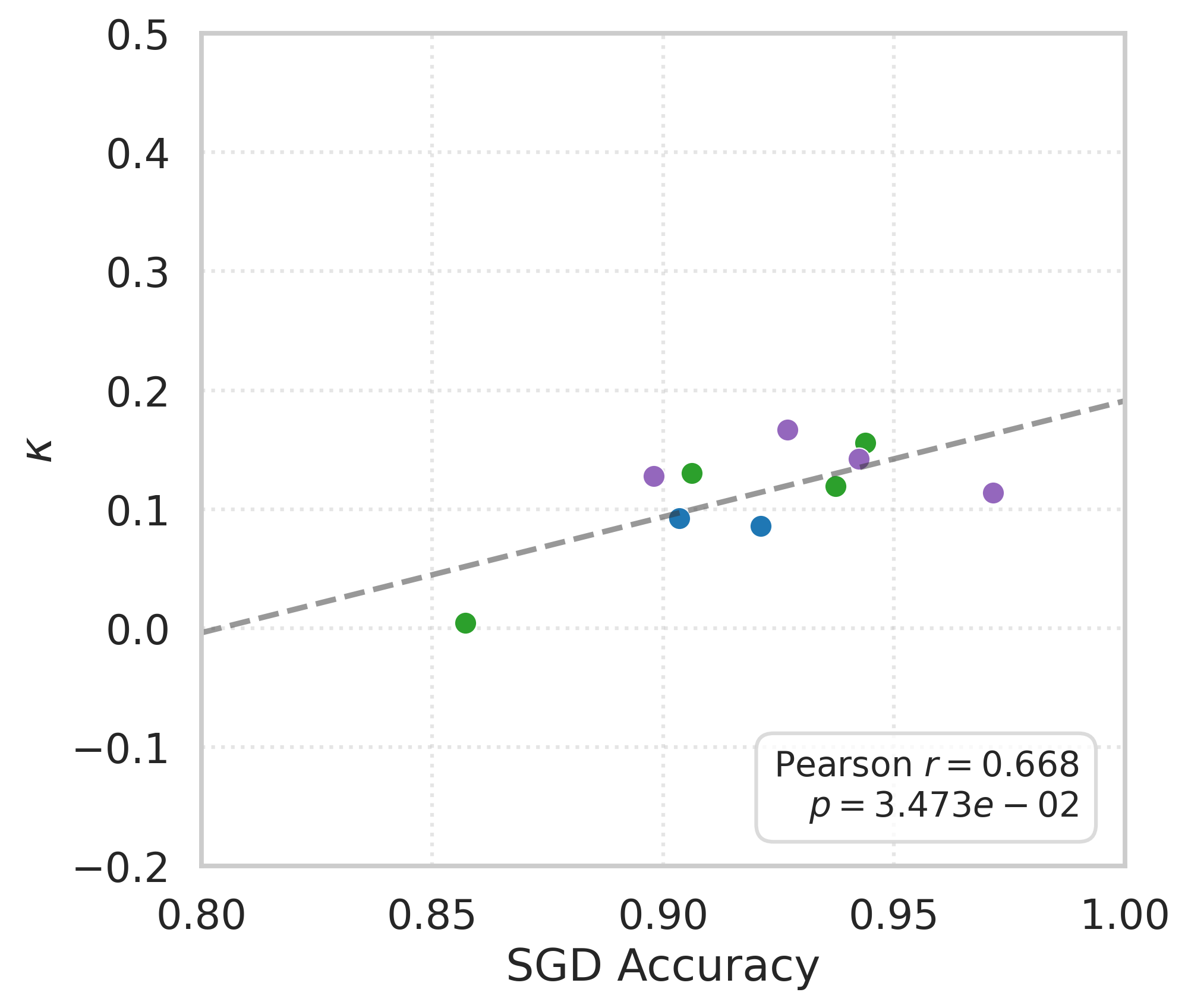}
    \caption{CAT, $\kappa$}
    \label{fig:metric_alignment_cat_ec}
\end{subfigure}
\vspace{0.5em}

\begin{subfigure}{0.3\textwidth}
    \includegraphics[width=\linewidth]{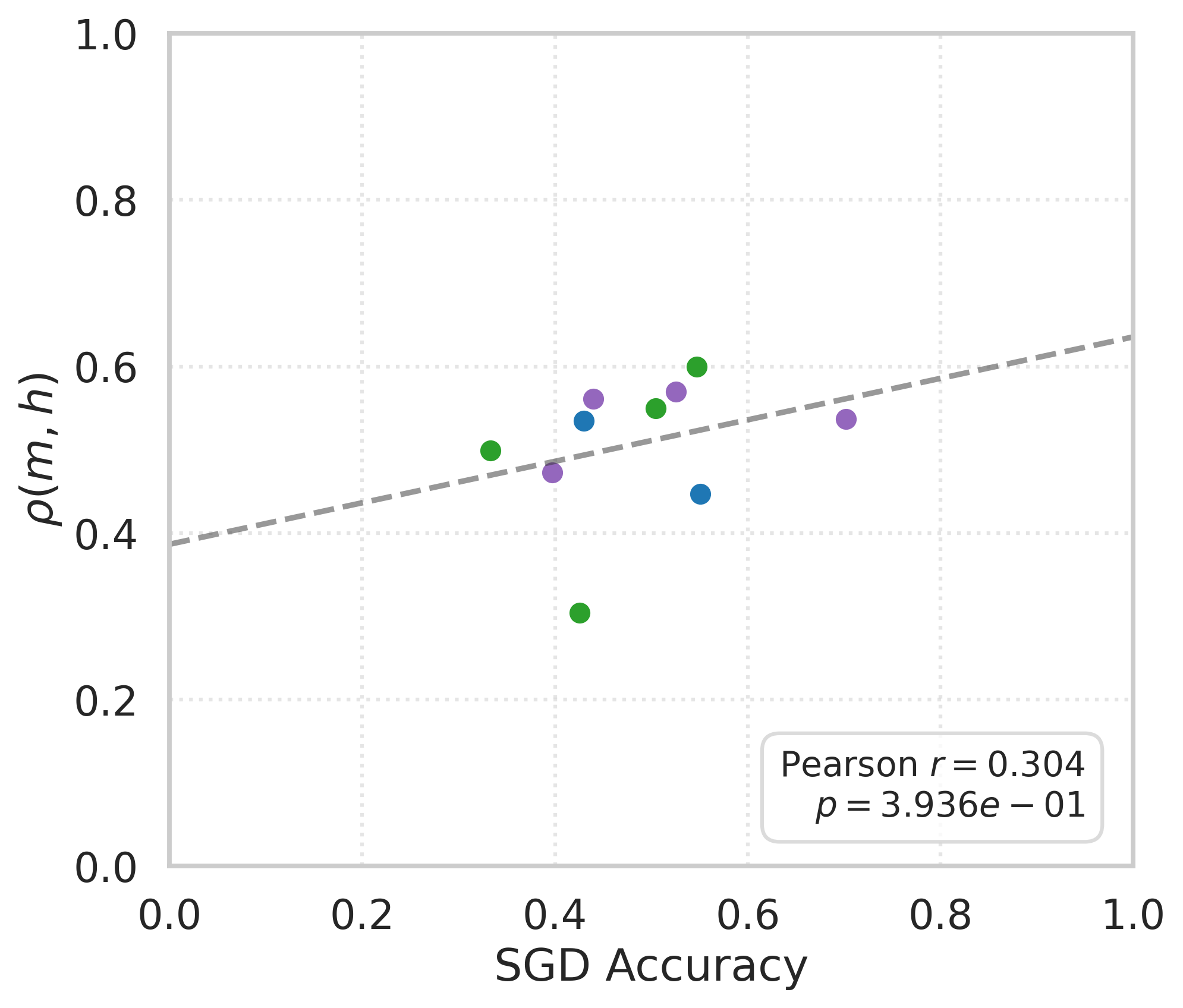}
    \caption{ID, $\rho$}
    \label{fig:metric_alignment_id_cons}
\end{subfigure}
\hfill
\begin{subfigure}{0.3\textwidth}
    \includegraphics[width=\linewidth]{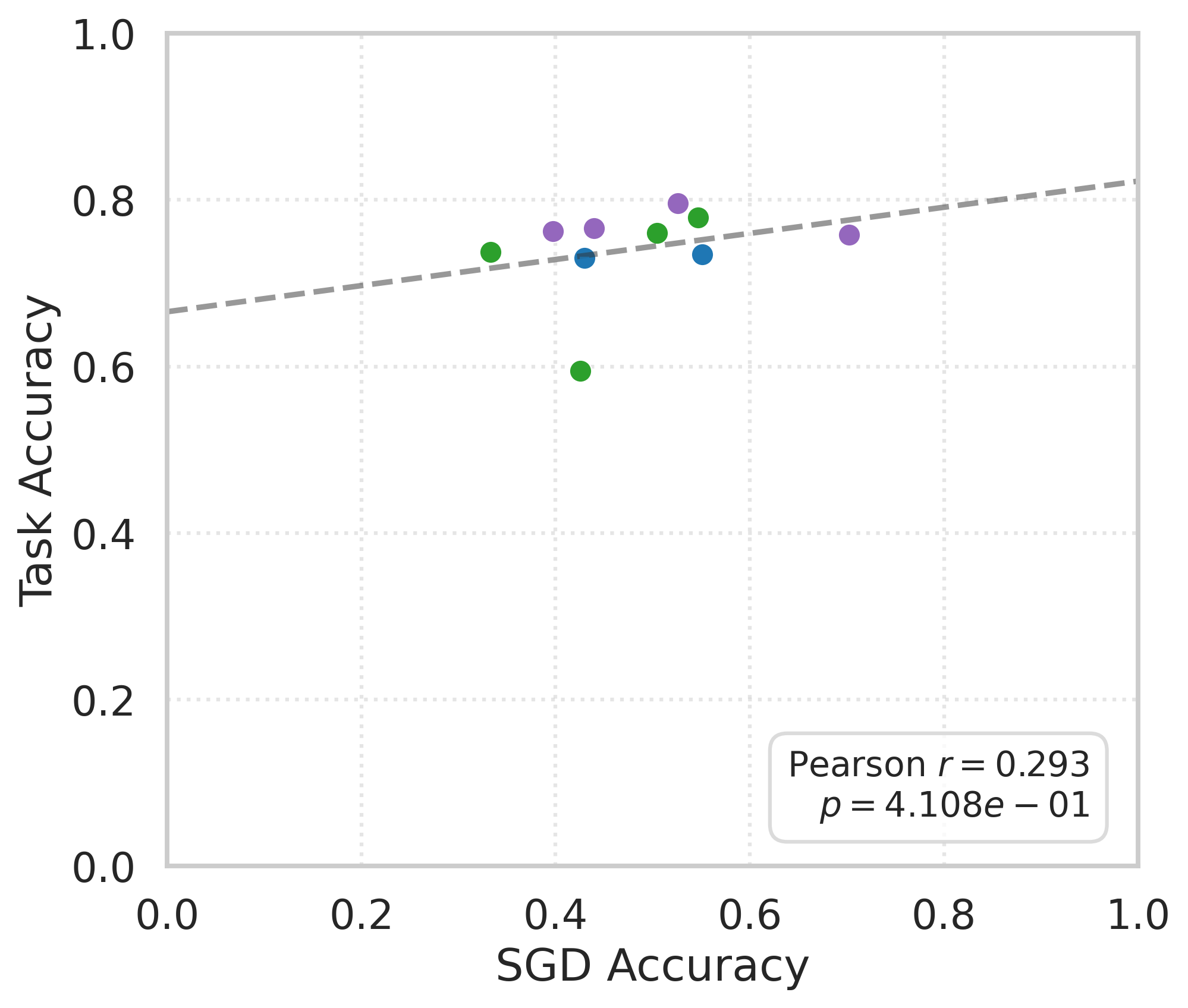}
    \caption{ID, Acc}
    \label{fig:metric_alignment_id_acc}
\end{subfigure}
\begin{subfigure}{0.3\textwidth}
    \includegraphics[width=\linewidth]{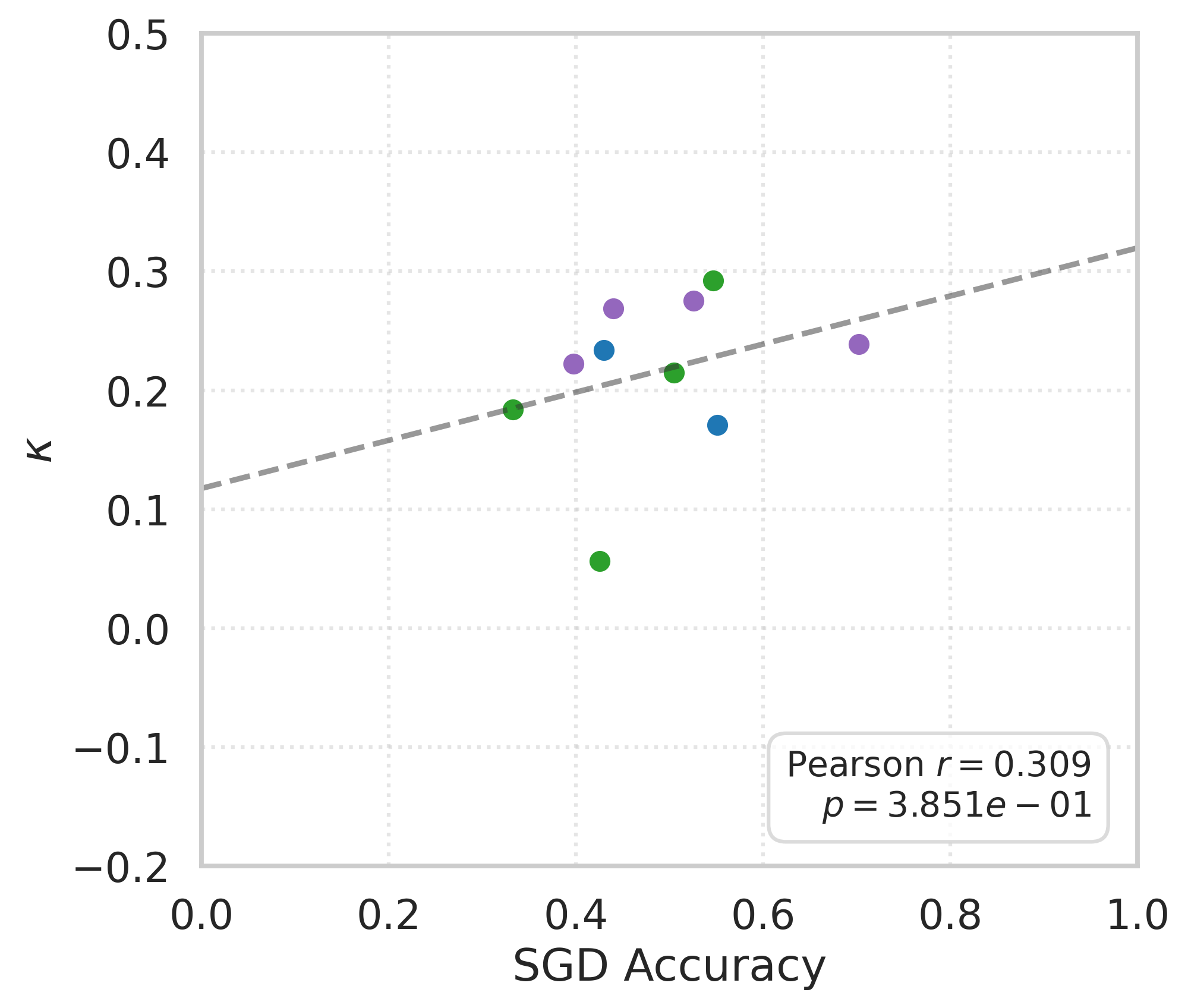}
    \caption{ID, $\kappa$}
    \label{fig:metric_alignment_id_ec}
\end{subfigure}
\vspace{0.5em}

\vspace{0.5em}

\begin{subfigure}{0.3\textwidth}
    \includegraphics[width=\linewidth]{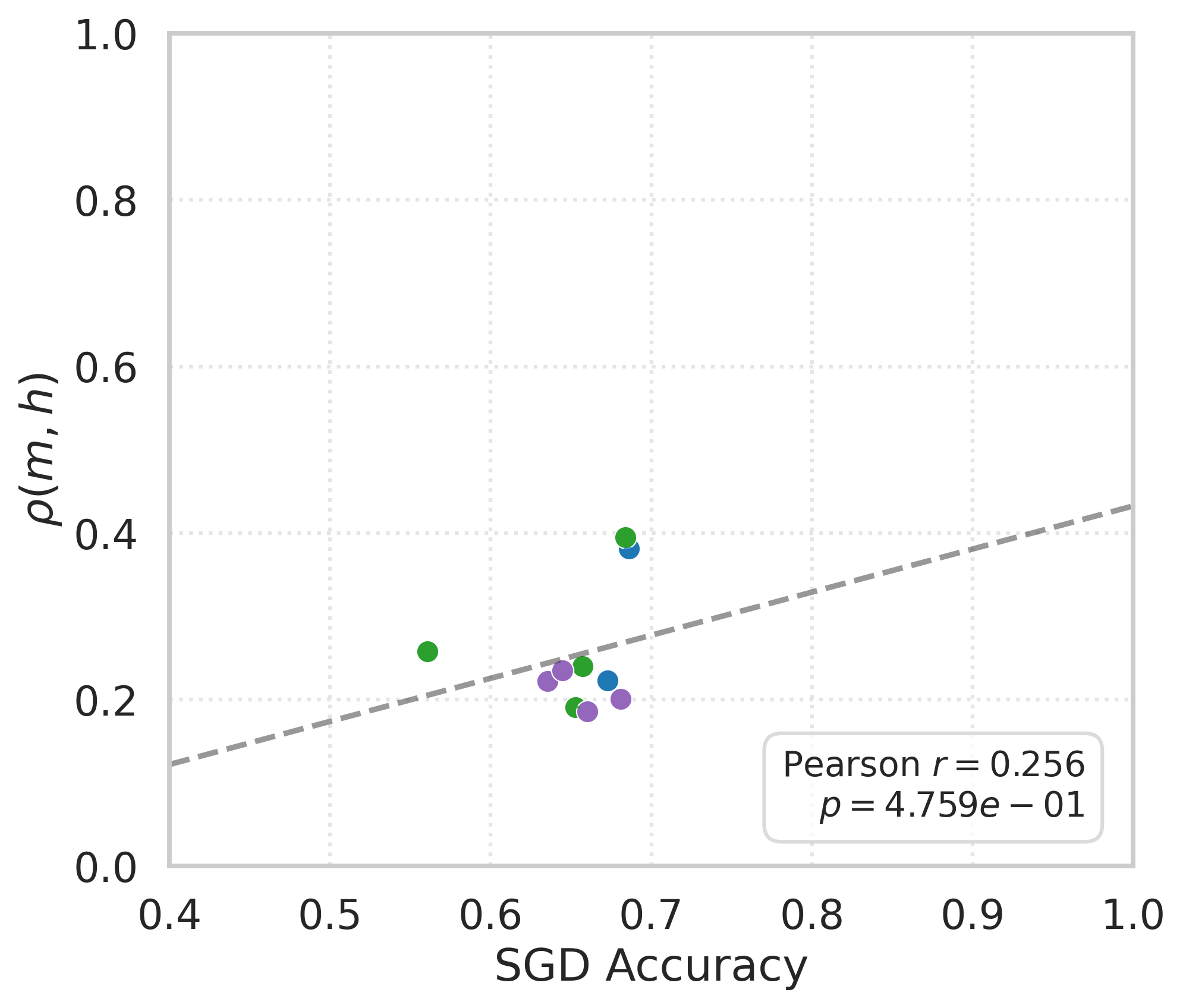}
    \caption{POS, $\rho$}
    \label{fig:metric_alignment_pos_cons}
\end{subfigure}
\hfill
\begin{subfigure}{0.3\textwidth}
    \includegraphics[width=\linewidth]{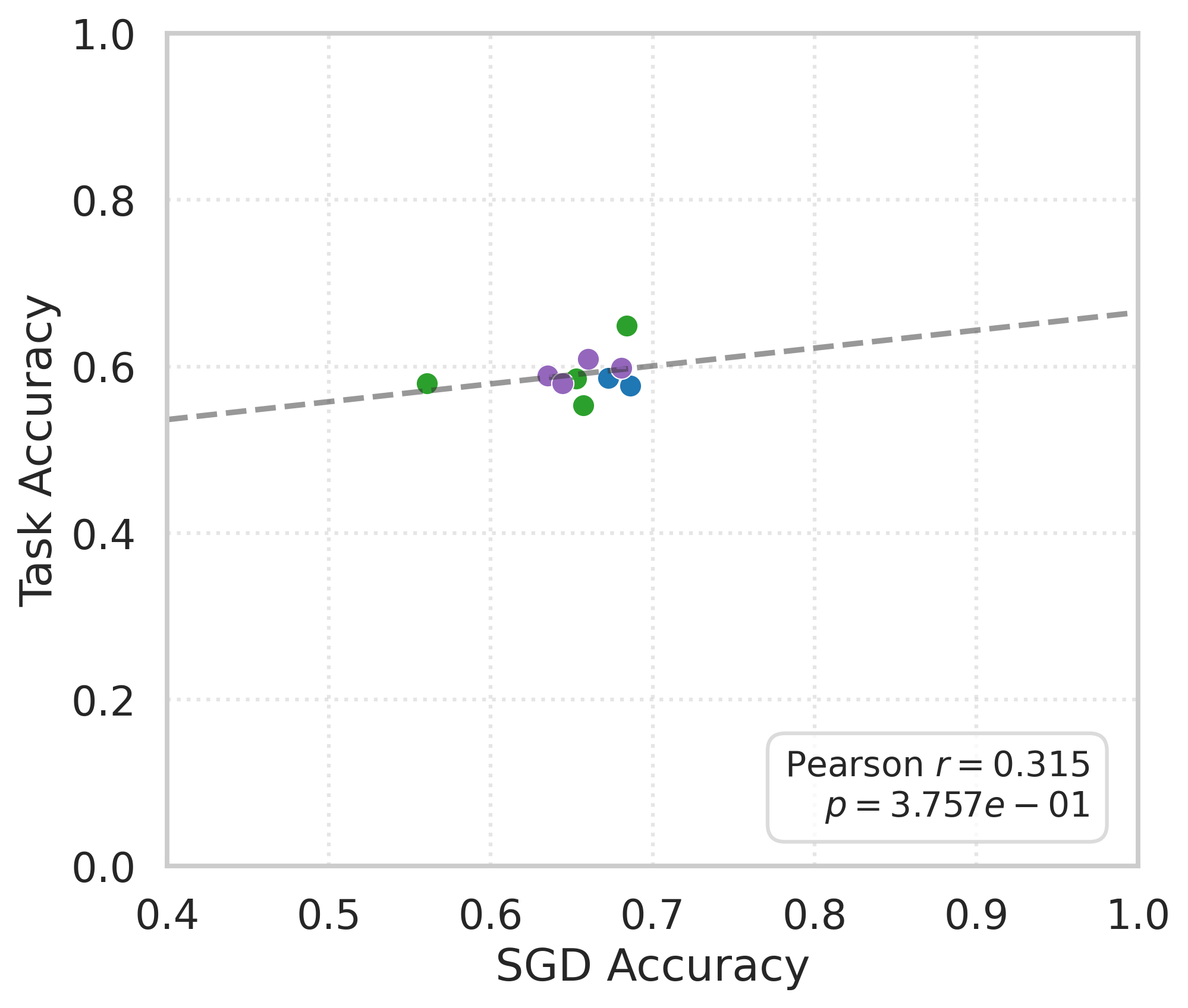}
    \caption{POS, Acc}
    \label{fig:metric_alignment_pos_acc}
\end{subfigure}
\begin{subfigure}{0.3\textwidth}
    \includegraphics[width=\linewidth]{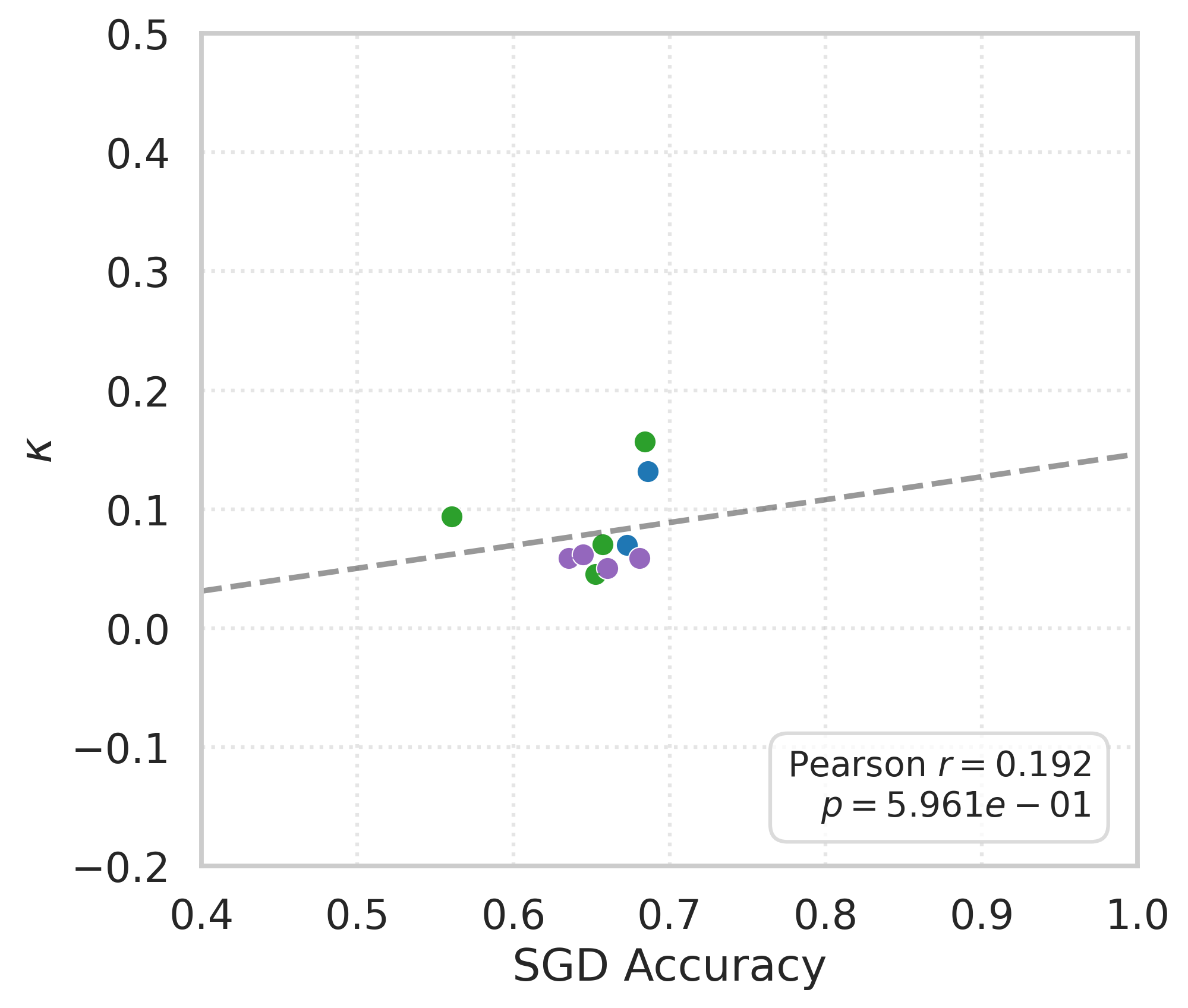}
    \caption{POS, $\kappa$}
    \label{fig:metric_alignment_pos_ec}
\end{subfigure}
\caption{\textbf{Feature-level linear-SVM accuracy and behavioral metric correlation.}}
\label{fig:metric_alignment}
\end{figure}

\subsubsection{Accuracy matched vs. final \texttt{Abstract} model}\label{sec:add_result_acc_match}
\begin{figure}
\centering
\begin{minipage}{0.8\textwidth}
    \centering
    \begin{subfigure}{0.48\textwidth}
        \includegraphics[width=\linewidth]{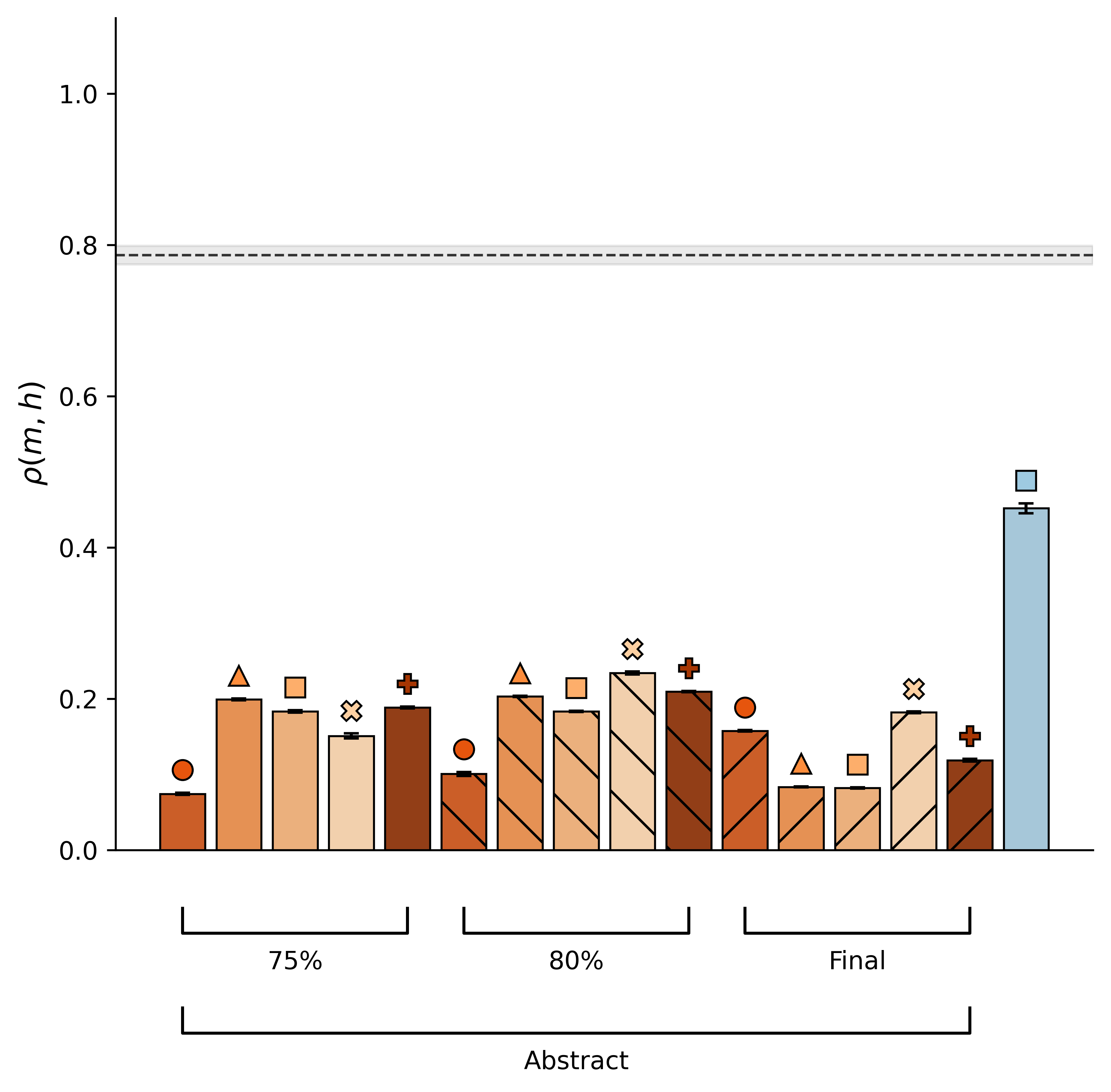}
        \caption{Correlation}\label{fig:no_match_global:corr}
    \end{subfigure}
    \hfill
    \begin{subfigure}{0.48\textwidth}
        \includegraphics[width=\linewidth]{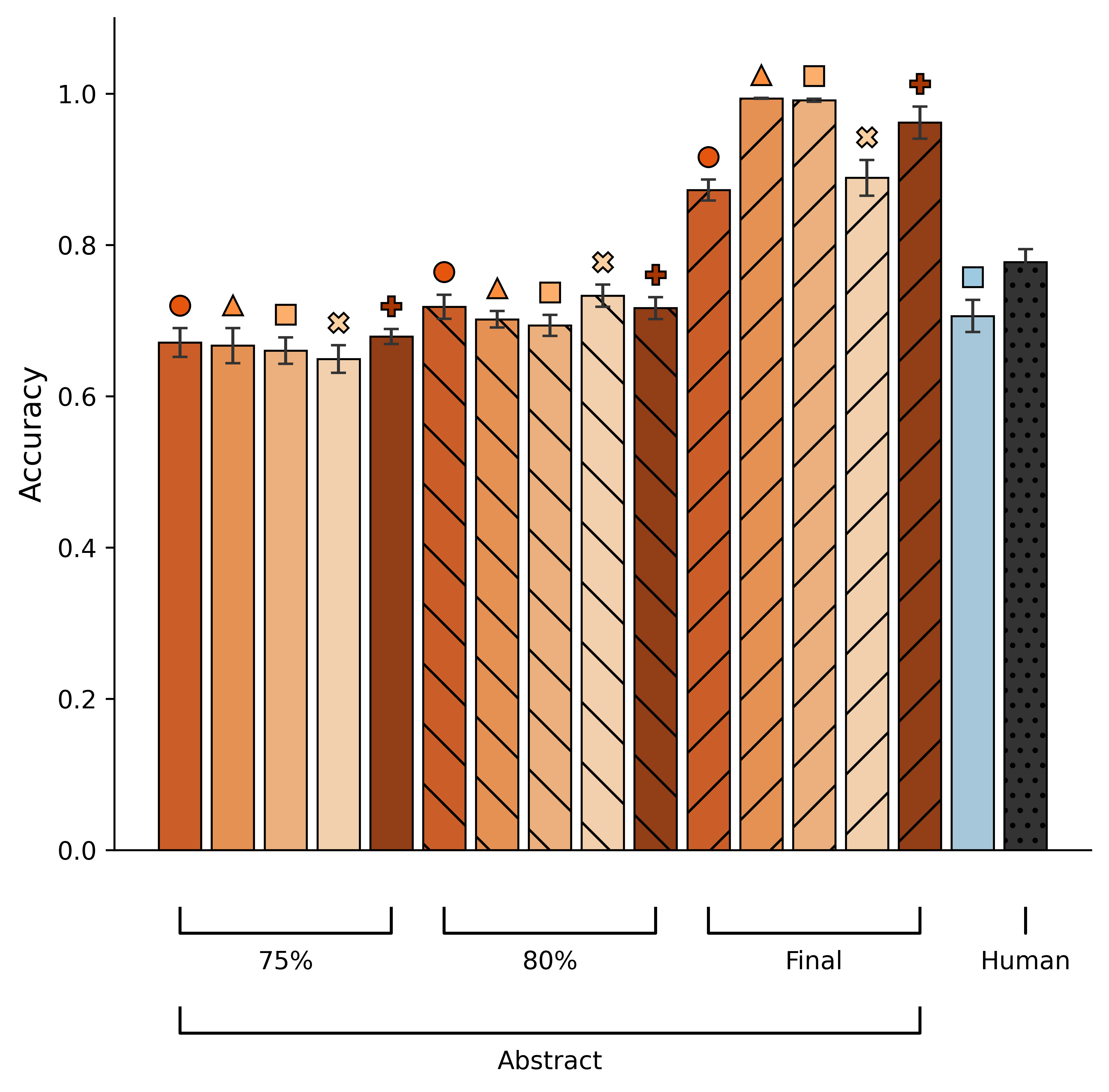}
        \caption{Accuracy}\label{fig:no_match_global:acc}
    \end{subfigure}
\end{minipage}
\begin{minipage}{0.15\textwidth}
    \centering
    \includegraphics[scale=0.5]{plots/seq_type/legend.png}
    \end{minipage}
\caption{\textbf{Comparison of \texttt{Abstract} models with matched accuracy and final accuracy.}
(a) Correlation with human behavior. (b) Task accuracy. Empty orange bars are model checkpoints with 75\% validation accuracy, orange bars with backward slashes are 80\% matched accuracy, and orange bars with forward slashes are final model checkpoints. The blue bar is the ResNet-CLIP LSTM natural model. Although accuracy on the MTurk trials increases (right figure), the gap in behavioral alignment persists (left figure).
}\label{fig:no_match_global}
\end{figure}

When training \texttt{Abstract} models, we realized their validation accuracies quickly approached 100\% and their logits were almost deterministic. Since human consistency is computed on a model's per-trial confidence, comparing humans against the final model checkpoints is unfair. To account for this, we performed the analyses in Sec.~\ref{sec:results1} using intermediate model checkpoints with around 80\% validation accuracy (most model seeds reached this in the first training epoch). In Fig.~\ref{fig:no_match_global}, we show that regardless of the selected model checkpoint, the behavioral gap still exists between abstract and natural models. In sections~\ref{sec:results2} and~\ref{sec:results3} however, we performed analyses with the final model checkpoints. This is because the \texttt{Abstract} models have not yet formed stable recurrent dynamics in the first epoch. Using DSA, we find that their dynamics are close to that of a random $\theta$.

\begin{figure}[ht] 
    \centering 
    \begin{minipage}{0.8\textwidth} 
    \begin{subfigure}{0.48\textwidth}
    \includegraphics[width=\linewidth]{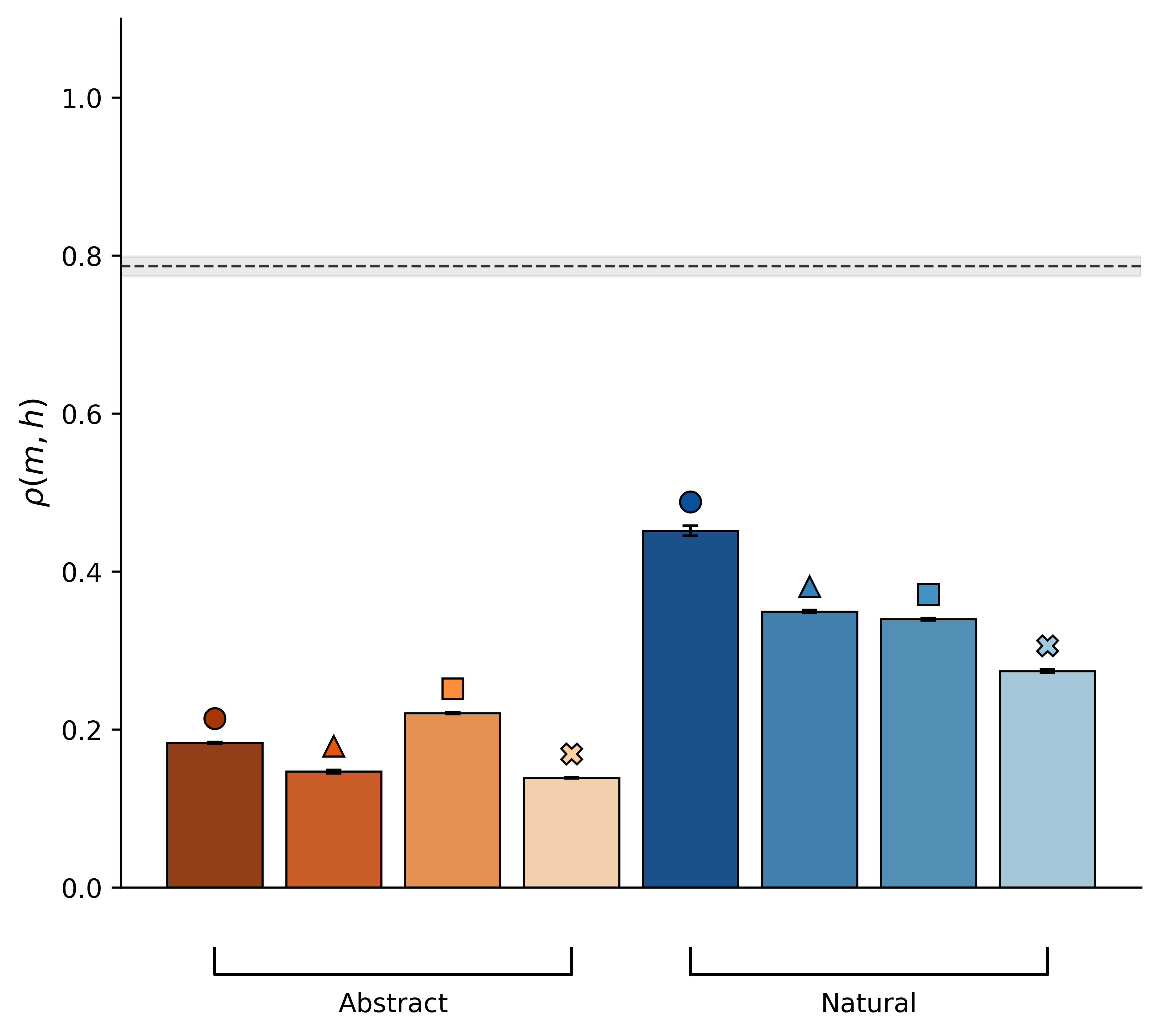}
    \caption{Correlation}\label{fig:transplant_acc_matched_global:corr}
    \end{subfigure} 
    \hfill 
    \begin{subfigure}{0.48\textwidth}
    \includegraphics[width=\linewidth]{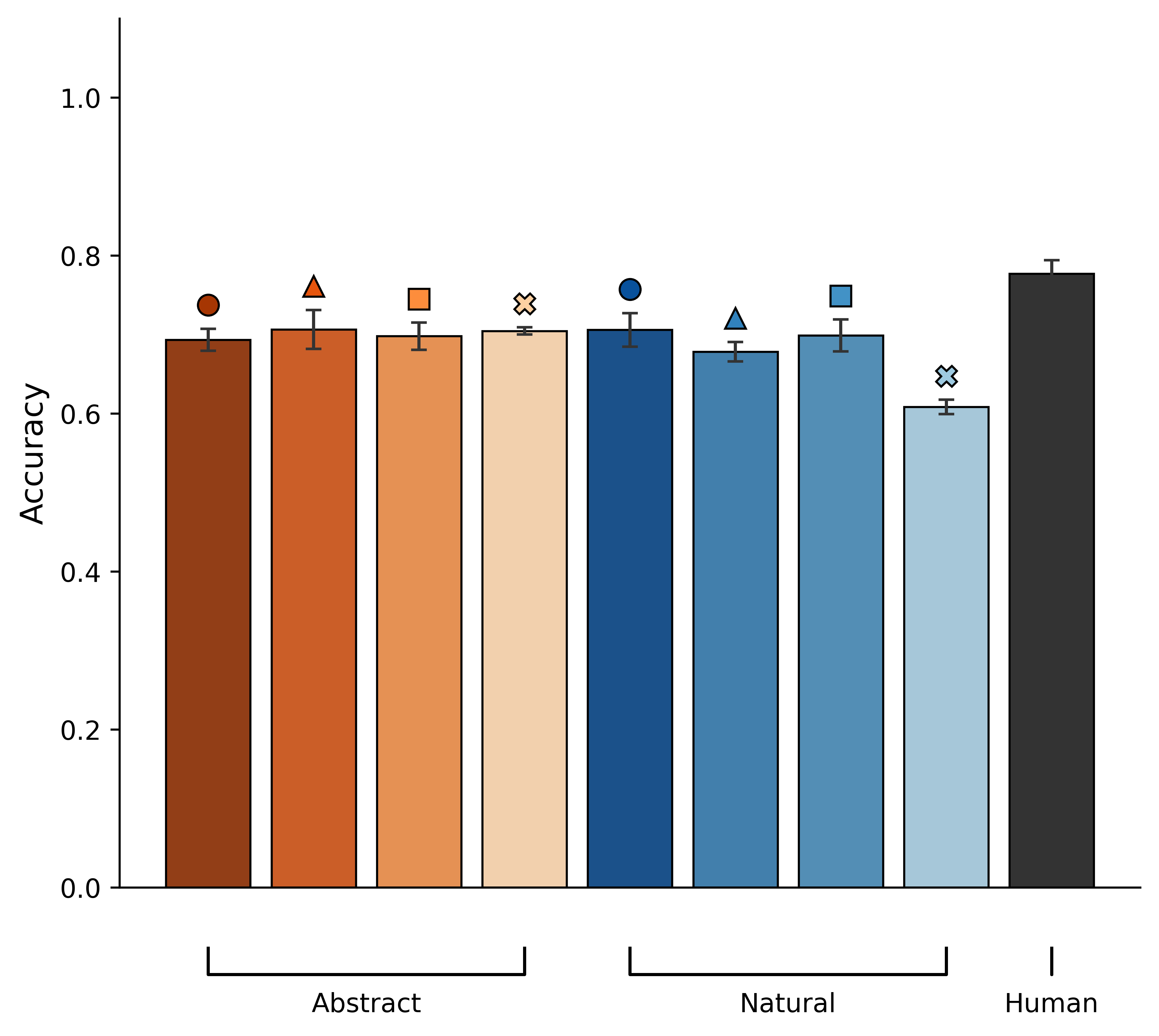}
    \caption{Accuracy}\label{fig:transplant_acc_matched_global:acc}
    \end{subfigure}
    \end{minipage} 
    \begin{minipage}{0.15\textwidth}
    \centering
    \includegraphics[scale=0.5]{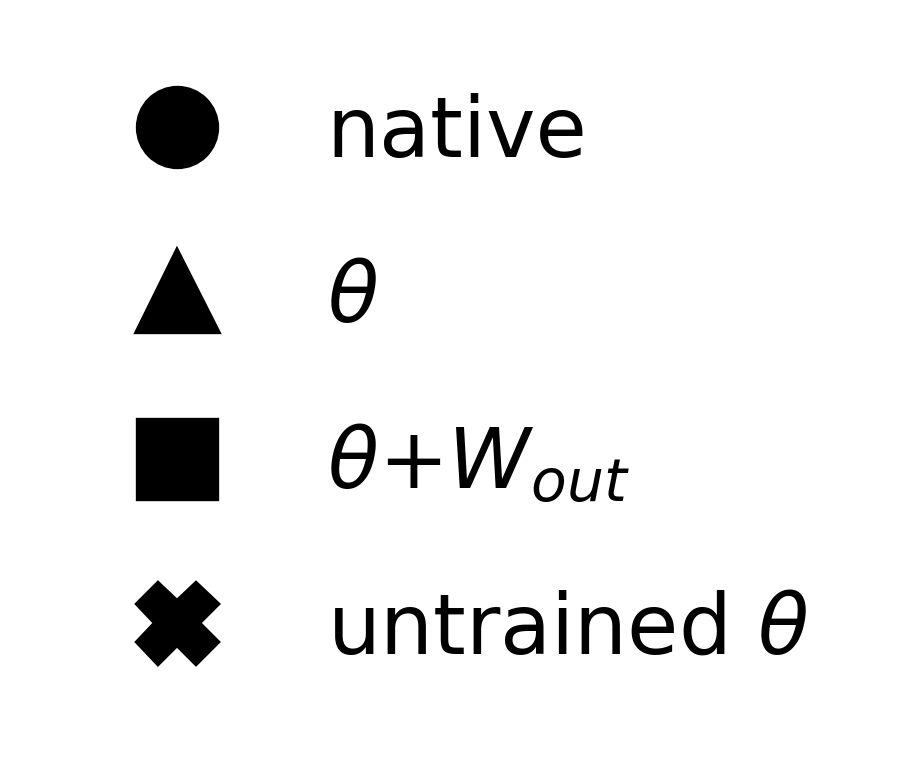}
    \end{minipage}
    \caption{\textbf{Behavioral metrics of accuracy-matched \texttt{Abstract} and \texttt{Natural} models with substituted components.}
Same conditions and colors as Fig.~\ref{fig:transplant_final_global}, but with accuracy-matched \texttt{Abstract} checkpoints.
Models receiving abstract input (orange) are evaluated at the checkpoint closest to 80\% validation accuracy, and the $\mathrm{Abs} \to \mathrm{Nat}$ substitutions (blue) load the sequence module of the 80\%-matched \texttt{Abstract} model.
(a) Human consistency $\rho(m,h)$. (b) Task accuracy on the MTurk trials.
All trained models fall within 0.03 of each other in accuracy, except the untrained-$\theta$ \texttt{Natural} baseline.
With accuracy controlled, the input still dominates behavior alignment. Every model receiving naturalistic input, including the untrained-$\theta$ baseline, is more consistent with humans than every model receiving abstract input.
A natural $\theta$ barely changes abstract-input models: $\theta_{\mathrm{Nat} \to \mathrm{Abs}}$ falls slightly below the native \texttt{Abstract} model (0.147 vs.\ 0.183), to the level of an untrained $\theta$. Keeping the natural readout as well raises consistency only to 0.221, below every model receiving naturalistic input.
The 80\%-matched abstract $\theta$ supports higher consistency than an untrained $\theta$ but lower than $\theta_{\mathrm{Nat}}$, whereas the fully trained module in Fig.~\ref{fig:transplant_final_global} matches $\theta_{\mathrm{Nat}}$, so the contribution of an abstract sequence module grows with its training.}\label{fig:transplant_acc_matched_global}
\end{figure}
\subsubsection{Dimensionality and visual difficulty matching}\label{sec:appendix_dimmatch}
To identify the key properties of sensory realism that drive human behavioral alignment, we ran additional experiments, isolating the effect of extrinsic and intrinsic properties.
\begin{enumerate}
    \item To assess the effect of dimensionality, we up-projected the one-hot embedding into the same number of dimensions as the vision encoder $\mathbb{R}^{N\times D_{\mathrm{vision}}}$. For instance, ResNet CLIP produces $N=49$ image tokens of dimension $D_{\mathrm{vision}}=2048$, so our pipeline applies a fixed random projection $\mathbf{W}_{\mathrm{up}}\in\mathbb{R}^{85\times49\times2048}$, an independent map for each of the 49 tokens. The resulting abstract embedding thus matches $z_t$ in the number of dimensions. This also has the benefit of connecting the up-projected abstract input to the identical attention pool mechanism in natural models (which produces $\bar{z}_t\in\mathbb{R}^{128}$), and allows both models to share the same input projection ($\mathbf{W}_{\mathrm{in}}\in\mathbb{R}^{171\times128}$).
    \item To match the perceptual uncertainty of sensory-realistic stimuli, we add Gaussian noise to the up-projected tokens of every stimulus frame, resampled at each presentation, with standard deviation $\sigma = s\,(1 + k(\bar r - \langle \bar r\rangle))$, where $\bar r$ is the RMS of the stimulus's normalized rotation and scale, $k=0.5$, and $s$ is the noise level. These models are evaluated at their final checkpoints, since the noise level affects their accuracy, and each trial's response is averaged over 20 seeded noise draws.
\end{enumerate}
\begin{figure}
\centering
    \begin{subfigure}{0.35\textwidth}
        \includegraphics[width=\linewidth]{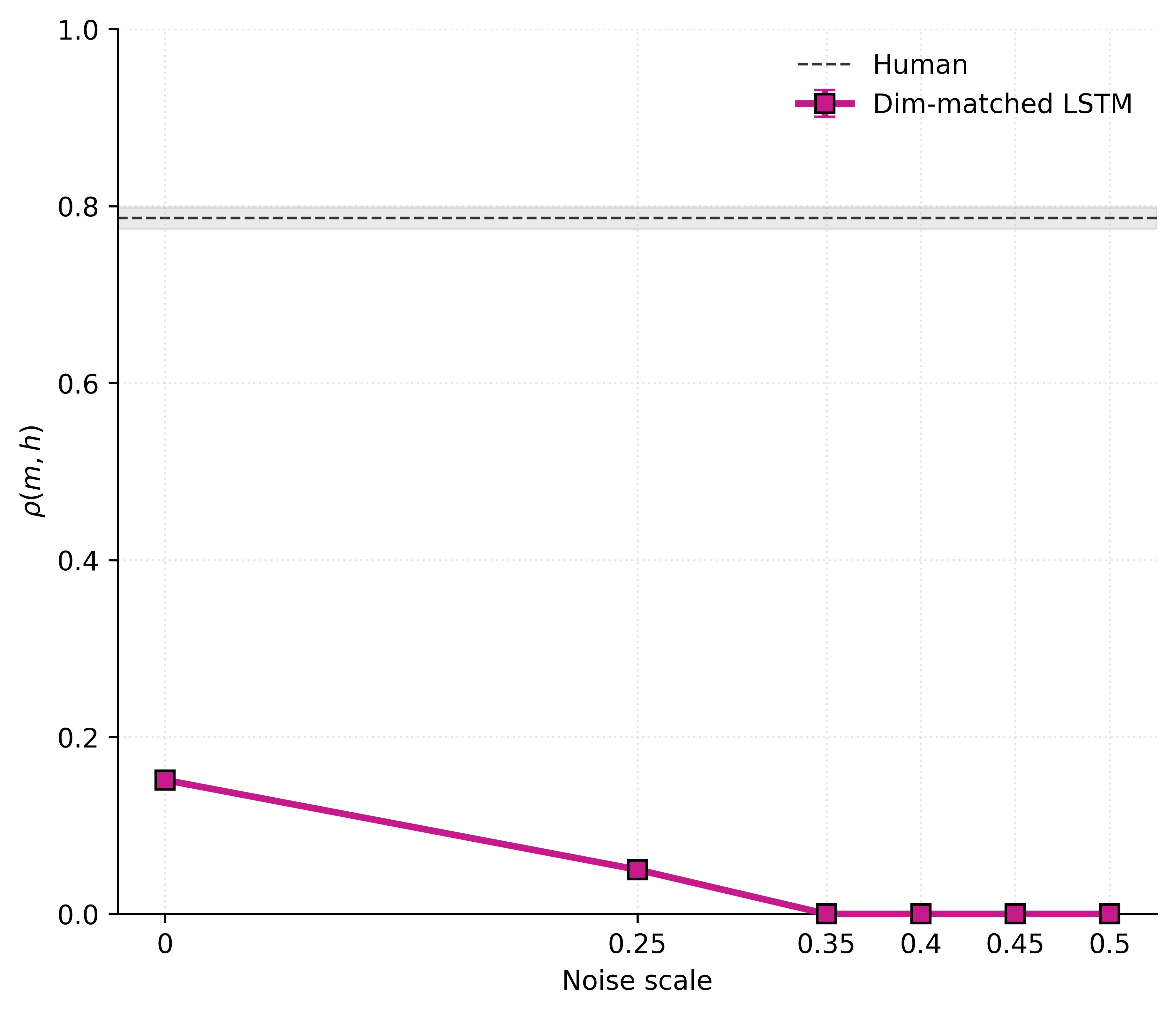}
        \caption{Correlation}\label{fig:dimmatch_dose:corr}
    \end{subfigure}
    \begin{subfigure}{0.35\textwidth}
        \includegraphics[width=\linewidth]{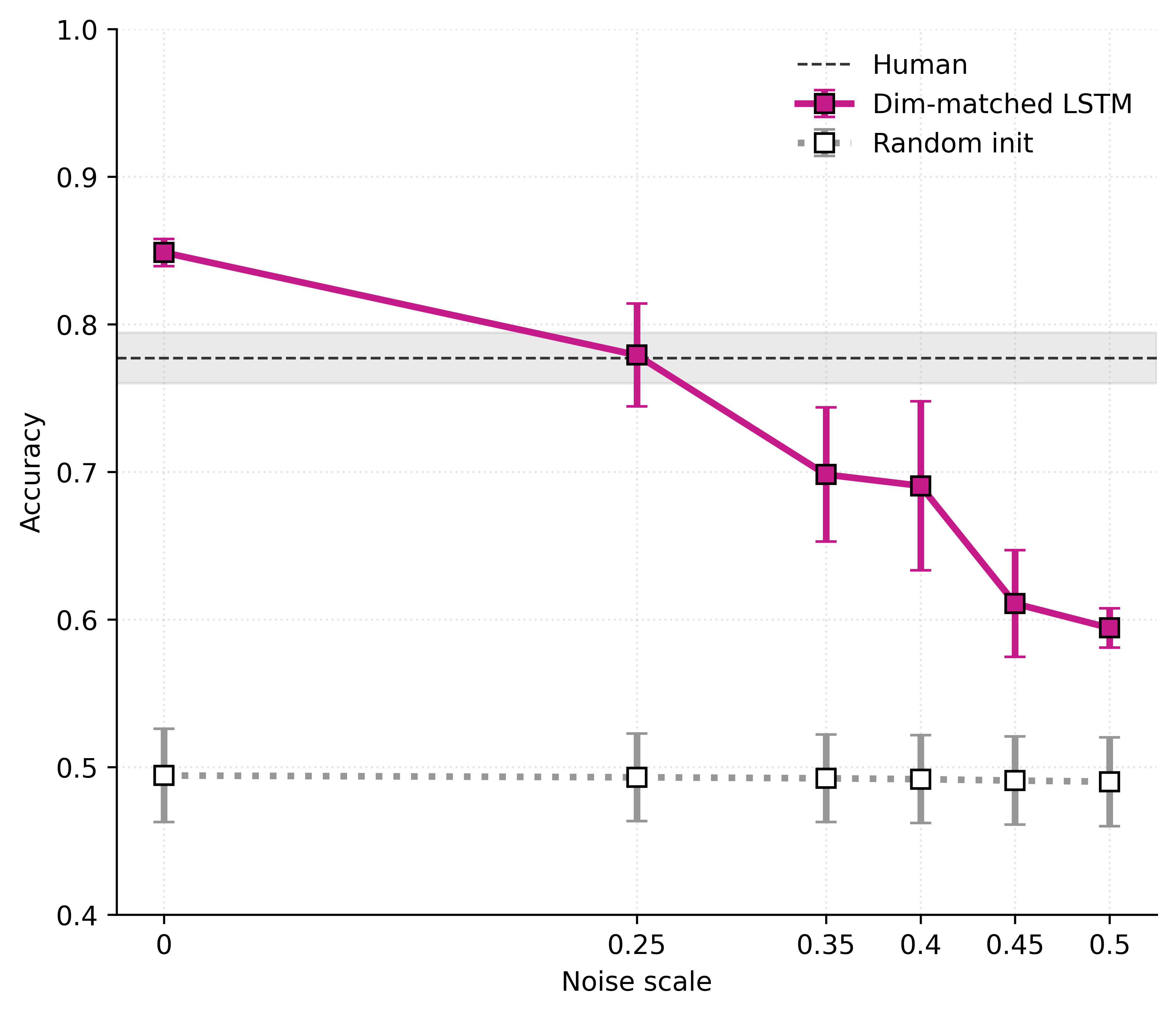}
        \caption{Accuracy}\label{fig:dimmatch_dose:acc}
    \end{subfigure}
\caption{\textbf{Comparison of \texttt{Abstract} models with dimensionality and perceptual uncertainty matching.}
(a) Correlation with human behavior decreases as more noise is injected. (b) Task accuracy also decreases as more noise is injected. Each level shows 6 seeds at their final checkpoints, including seeds that did not learn the task. Responses are averaged over 20 noise draws.
}\label{fig:dimmatch_dose}
\end{figure}

At comparable validation accuracy (about 0.75), projecting the one-hot input to the natural dimensionality lowers human consistency from $\rho(m,h)=0.183$ (one-hot) to $0.115$ (dimensionality-matched), and adding noise ($s=0.35$, final validation accuracy 0.76) lowers human consistency to $0$, against $0.452$ for the \texttt{Natural} LSTM.
Across noise levels (Fig.~\ref{fig:dimmatch_dose}), $\rho(m,h)$ falls from $0.151$ without noise to $0.050$ at $s=0.25$, where MTurk accuracy matches humans (0.779 vs.\ 0.777). At higher noise levels, some seeds never leave a plateau near 0.62 validation accuracy, so the two highest noise levels mostly reflect failed training.
\begin{figure}[ht] 
    \centering 
    \begin{minipage}{0.8\textwidth} 
    \begin{subfigure}{0.48\textwidth}
    \includegraphics[width=\linewidth]{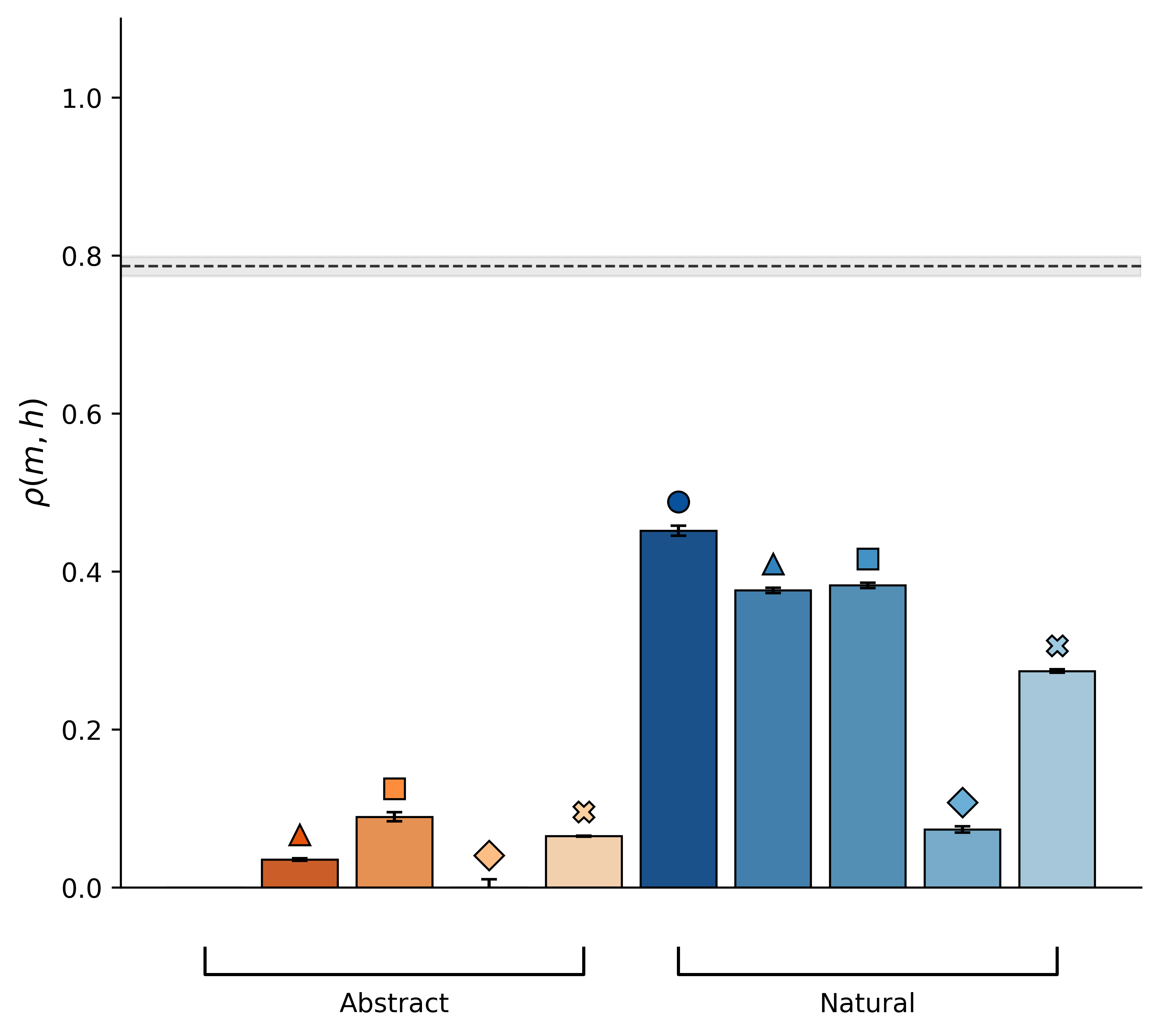}
    \caption{Correlation}\label{fig:transplant_dm_global:corr}
    \end{subfigure} 
    \hfill 
    \begin{subfigure}{0.48\textwidth}
    \includegraphics[width=\linewidth]{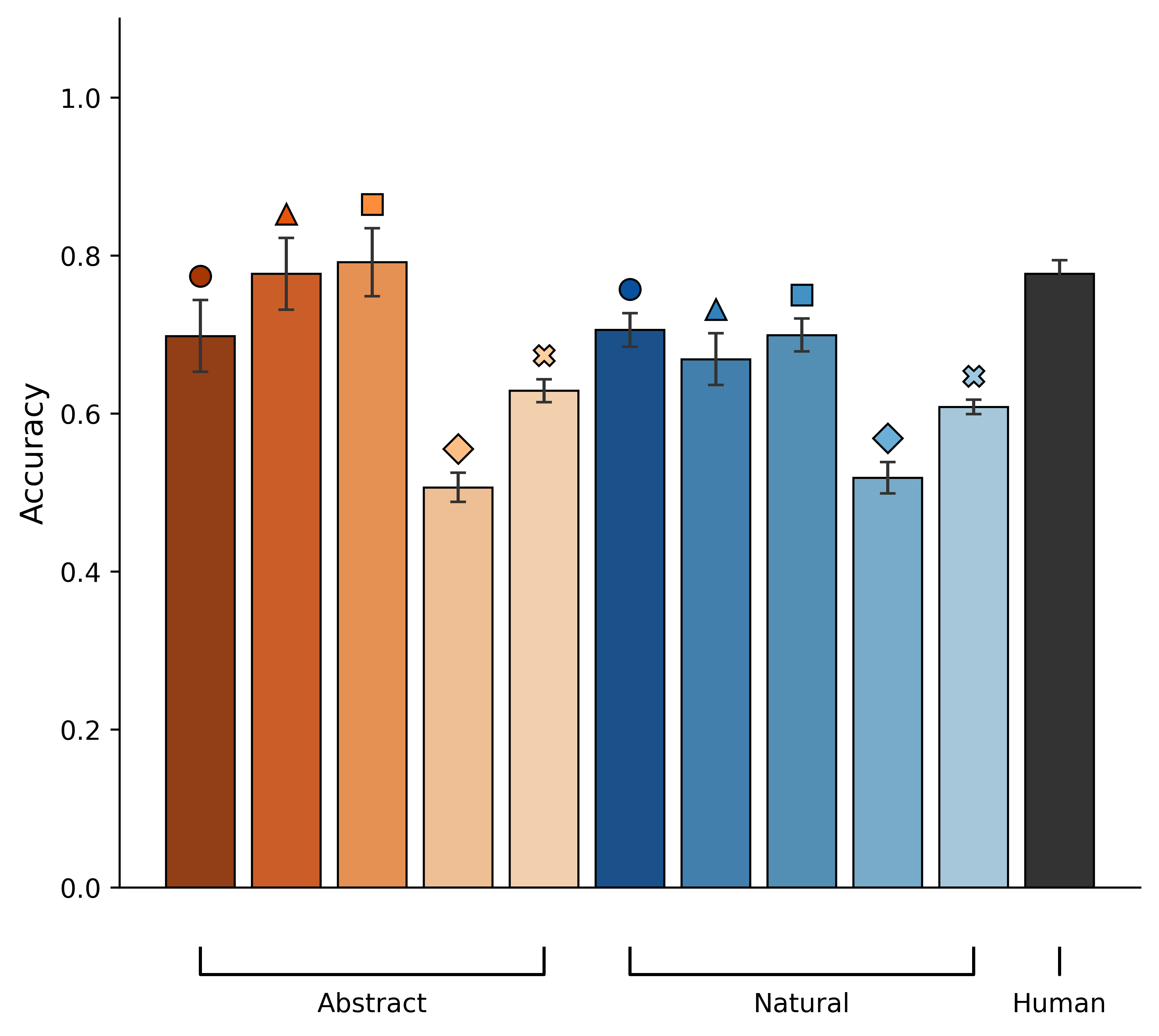}
    \caption{Accuracy}\label{fig:transplant_dm_global:acc}
    \end{subfigure}
    \end{minipage} 
    \begin{minipage}{0.15\textwidth}
    \centering
    \includegraphics[scale=0.5]{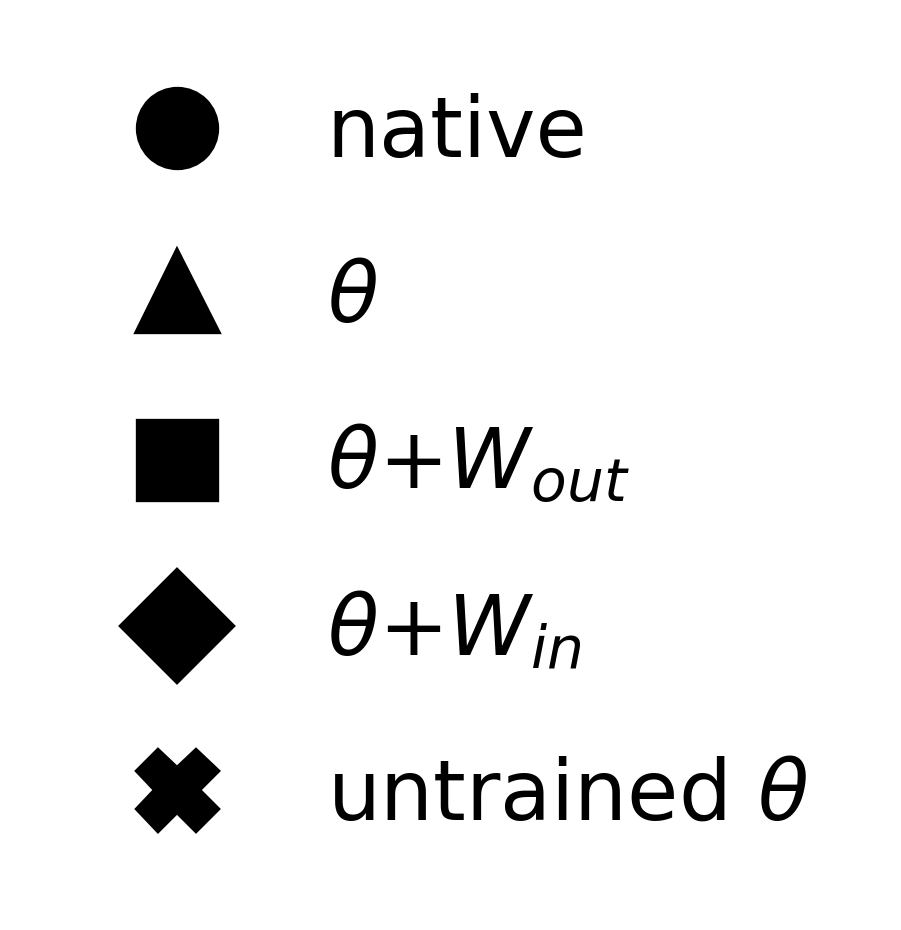}
    \end{minipage}
    \caption{\textbf{Behavioral metrics of dimensionality-matched \texttt{Abstract} and \texttt{Natural} models with substituted components.} Same conditions and colors as Fig.~\ref{fig:transplant_final_global}, but with dimensionality-matched \texttt{Abstract} models and $\theta+\mathbf{W}_\mathrm{in}$ substitution (diamond markers). Bars below zero are not drawn. }\label{fig:transplant_dm_global}
\end{figure}

\subsubsection{Frozen, randomly initialized vision encoder}\label{sec:appendix_random_encoder}
\begin{table}[ht]
\centering
\caption{\texttt{Natural} LSTM with a frozen pretrained or a frozen randomly initialized ResNet-CLIP encoder (6 seeds). Accuracy: mean $\pm$ SD over seeds; $\rho(m,h)$: mean over 20 split-halves.}
\label{tab:random_encoder}
\small
\begin{tabular}{lcccccc}
\toprule
 & & & \multicolumn{4}{c}{$\rho(m,h)$} \\
\cmidrule(lr){4-7}
Encoder & Val.\ acc. & MTurk acc. & All & Category & Identity & Position \\
\midrule
Pretrained & $0.771 \pm 0.029$ & $0.706 \pm 0.021$ & $0.452$ & $0.329$ & $0.534$ & $0.381$ \\
Random & $0.596 \pm 0.009$ & $0.516 \pm 0.012$ & $0.025$ & $0.000$ & $0.000$ & $0.101$ \\
\bottomrule
\end{tabular}
\end{table}
To test how much the \texttt{Natural} models rely on pretrained visual features, we replaced the pretrained ResNet-CLIP encoder of \texttt{Natural} LSTM with its default \texttt{timm} random initialization (shared by all six seeds) and kept it frozen.
All other modules were trained with the same recipe (Table~\ref{tab:native-training-config}).
These models barely learn the tasks (Table~\ref{tab:random_encoder}). Their final validation accuracy is $0.596 \pm 0.009$, against $0.771 \pm 0.029$ with the pretrained encoder, and MTurk accuracy is $0.516 \pm 0.012$.
Their human consistency is near zero overall ($\rho(m,h)=0.025$): 0 on category and identity trials and $0.101$ on position trials.
Frozen random features are therefore insufficient for the downstream modules to learn the tasks.

\subsubsection{Effect of architecture on behavioral consistency}\label{sec:appendix_vision_type}
\begin{figure}
\centering
\begin{minipage}{0.7\textwidth}
\begin{subfigure}{0.45\textwidth}
    \includegraphics[width=\linewidth]{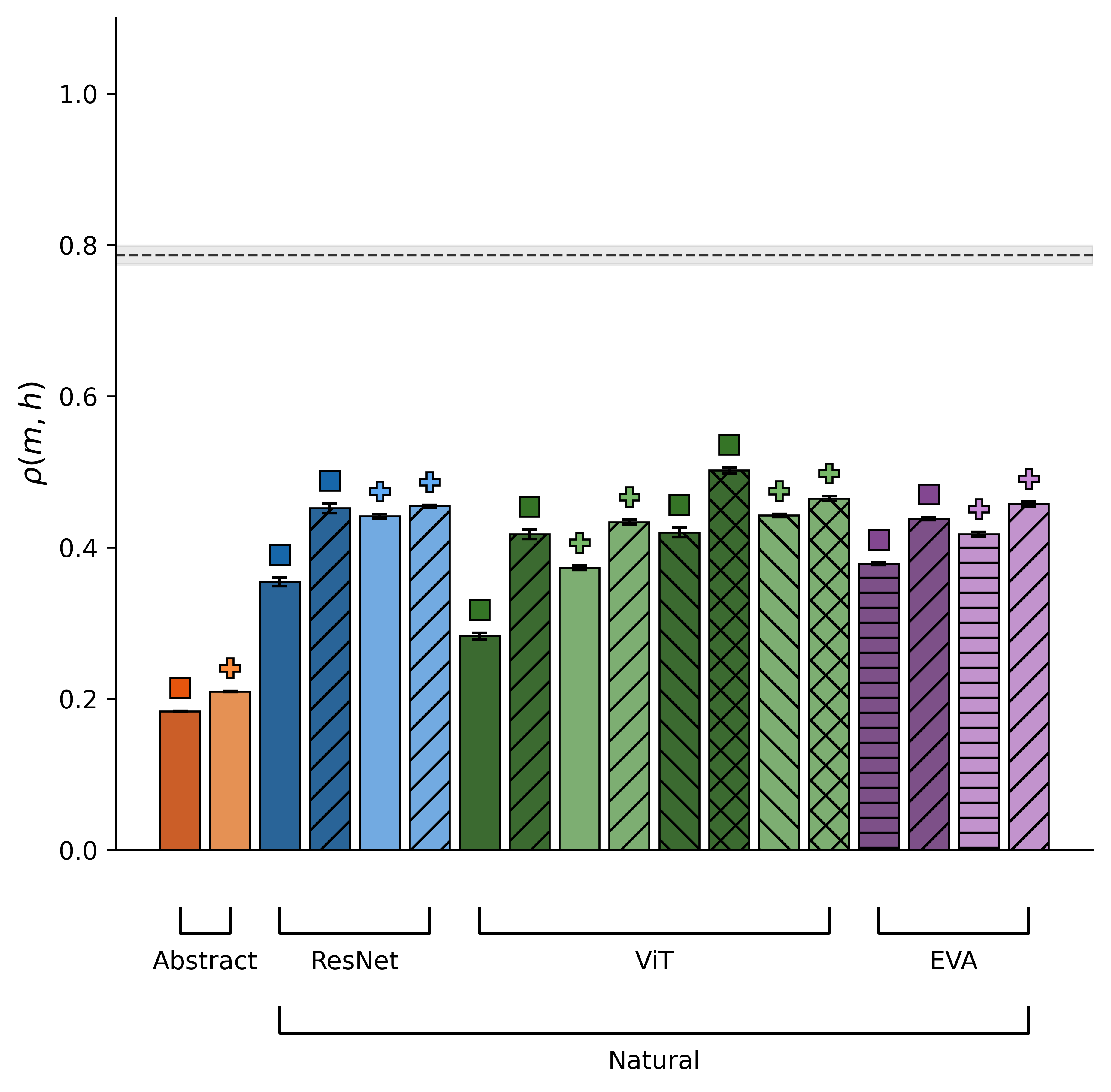}
    \caption{}\label{fig:vision_type_global:corr}
\end{subfigure}
\hfill
\begin{subfigure}{0.45\textwidth}
    \includegraphics[width=\linewidth]{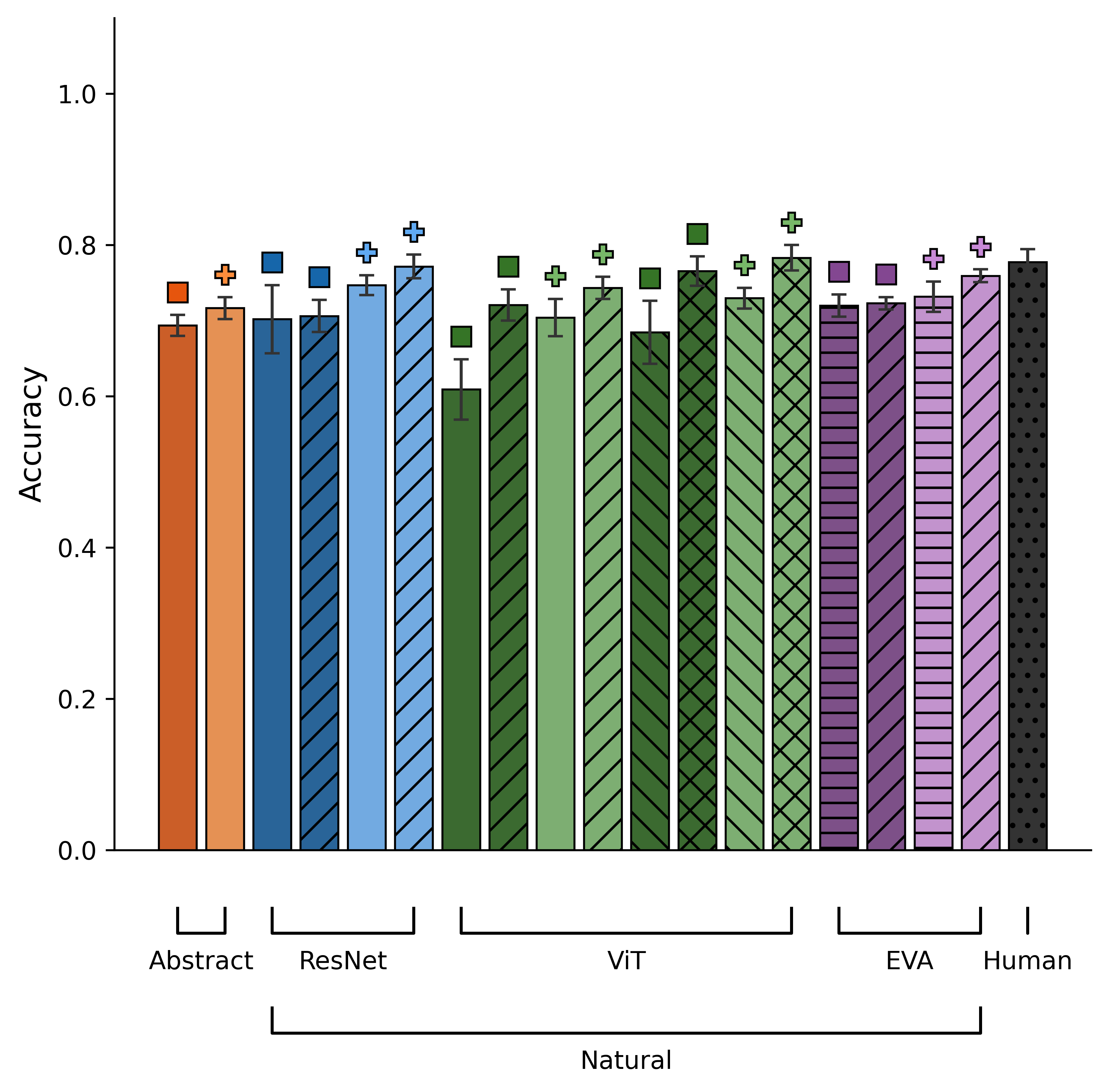}
    \caption{}\label{fig:vision_type_global:acc}
\end{subfigure}
\end{minipage}
\begin{minipage}{0.15\textwidth}
    \centering
    \includegraphics[scale=0.5]{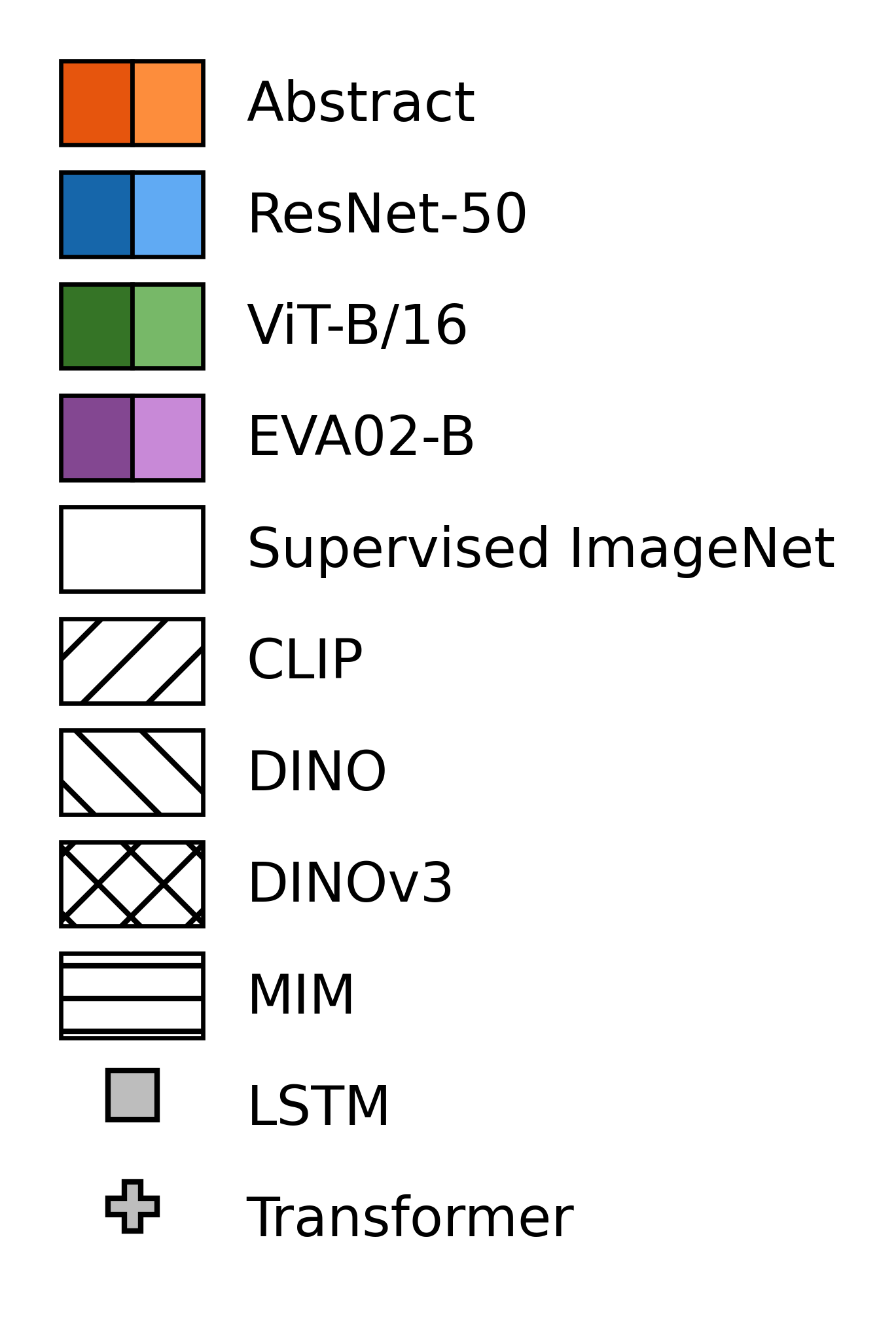}
\end{minipage}
\vspace{0.5em}

\begin{minipage}{0.7\textwidth}
\begin{subfigure}{0.45\textwidth}
    \includegraphics[width=\linewidth]{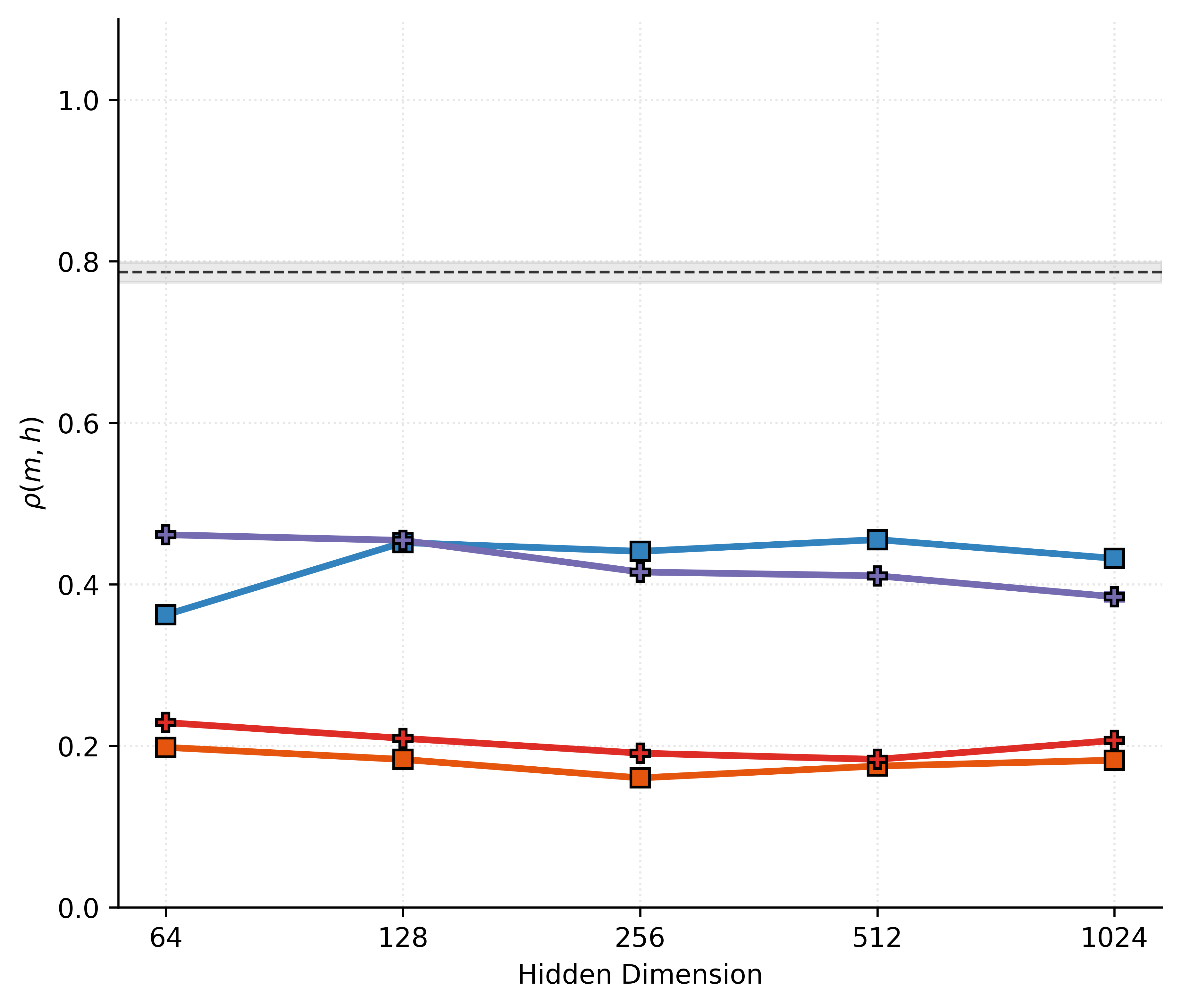}
    \caption{}\label{fig:seq_size_global:corr}
\end{subfigure}
\hfill
\begin{subfigure}{0.45\textwidth}
    \includegraphics[width=\linewidth]{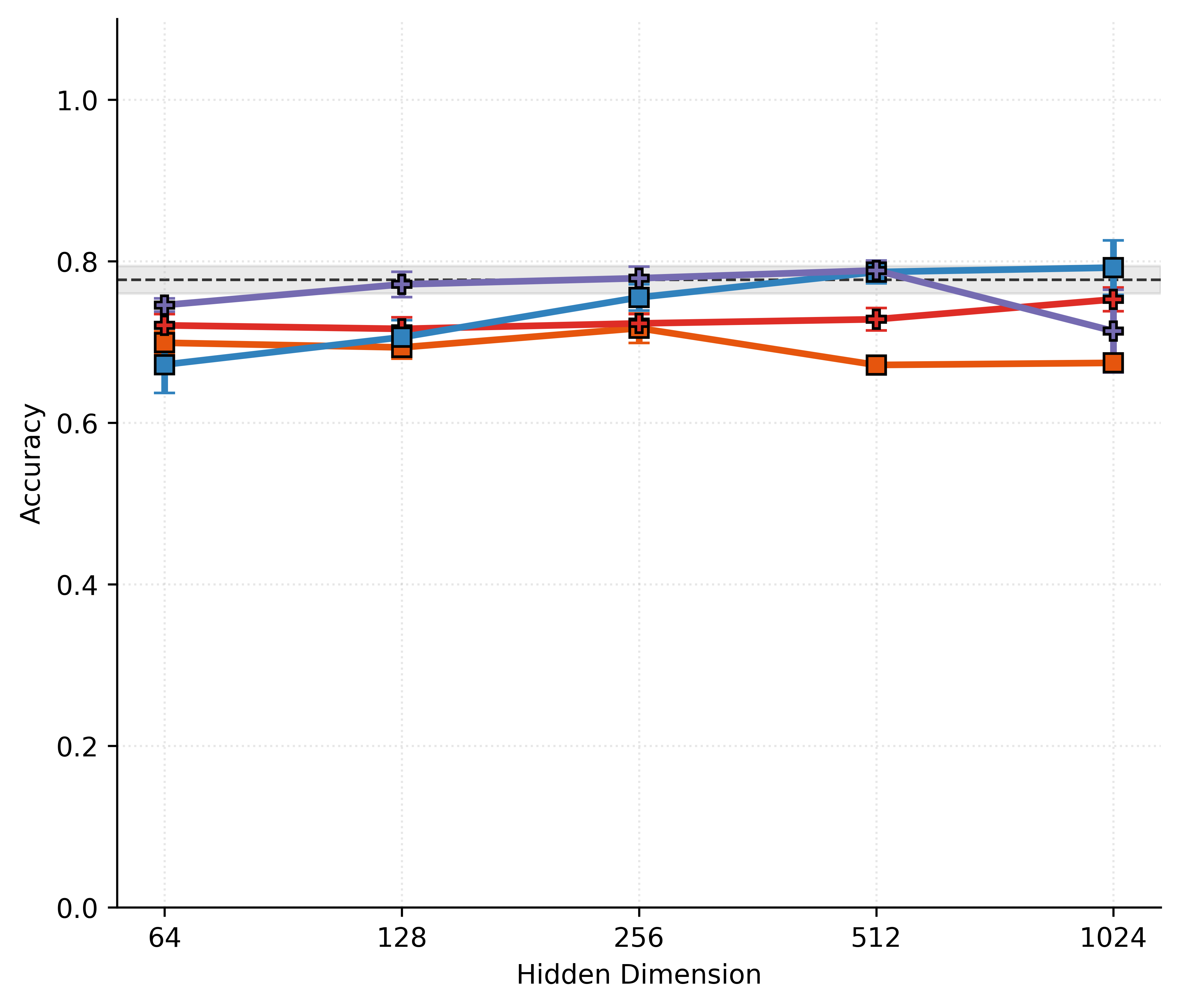}
    \caption{}\label{fig:seq_size_global:acc}
\end{subfigure}
\end{minipage}
\begin{minipage}{0.15\textwidth}
    \centering
    \includegraphics[scale=0.5]{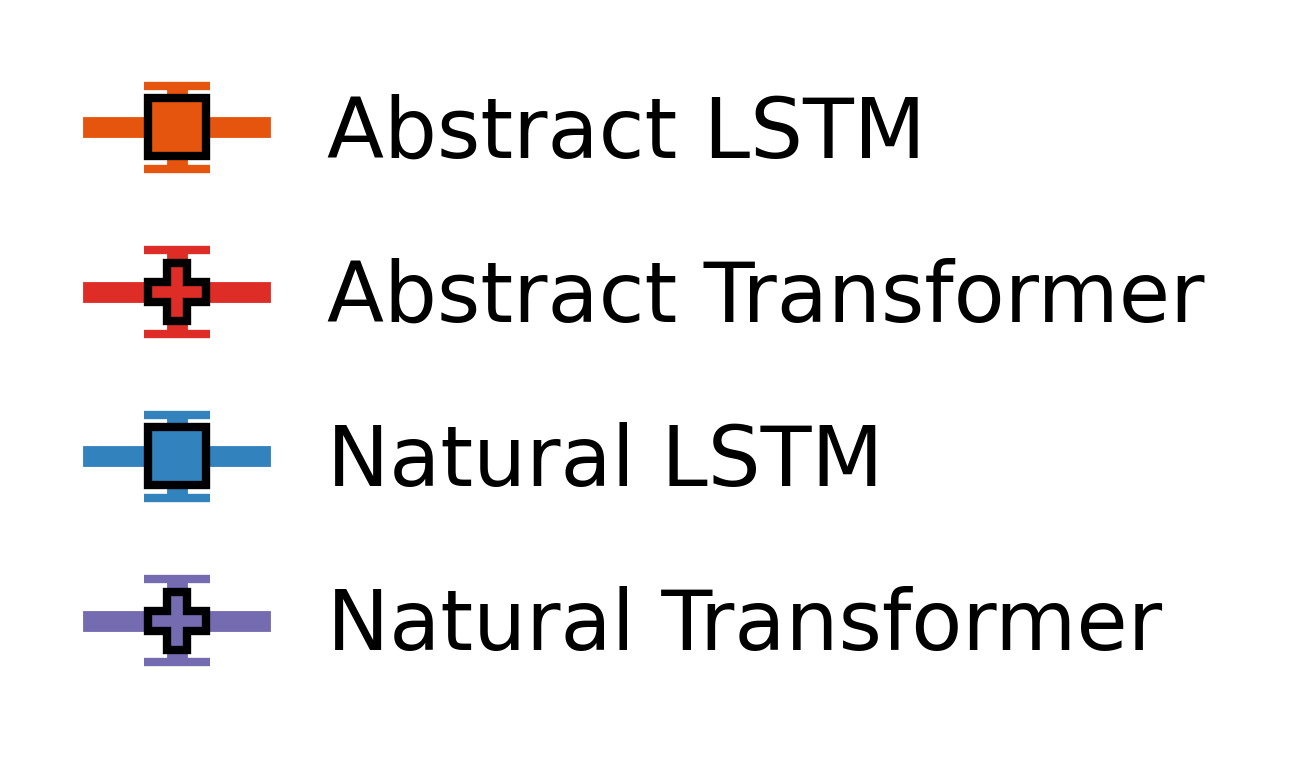}
\end{minipage}
\vspace{0.5em}

\begin{minipage}{0.7\textwidth}
\begin{subfigure}{0.45\textwidth}
    \includegraphics[width=\linewidth]{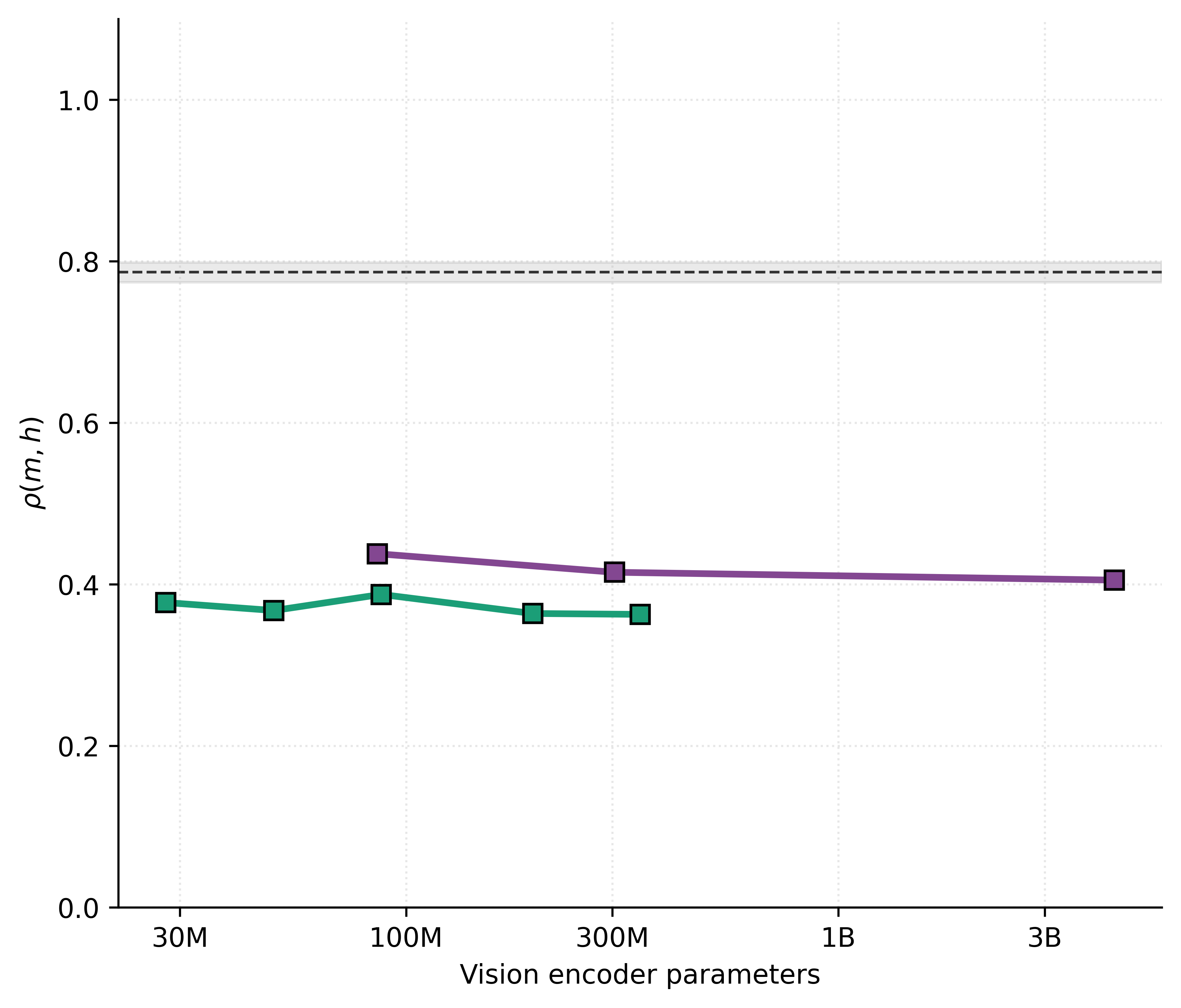}
    \caption{}\label{fig:encoder_size_global:corr}
\end{subfigure}
\hfill
\begin{subfigure}{0.45\textwidth}
    \includegraphics[width=\linewidth]{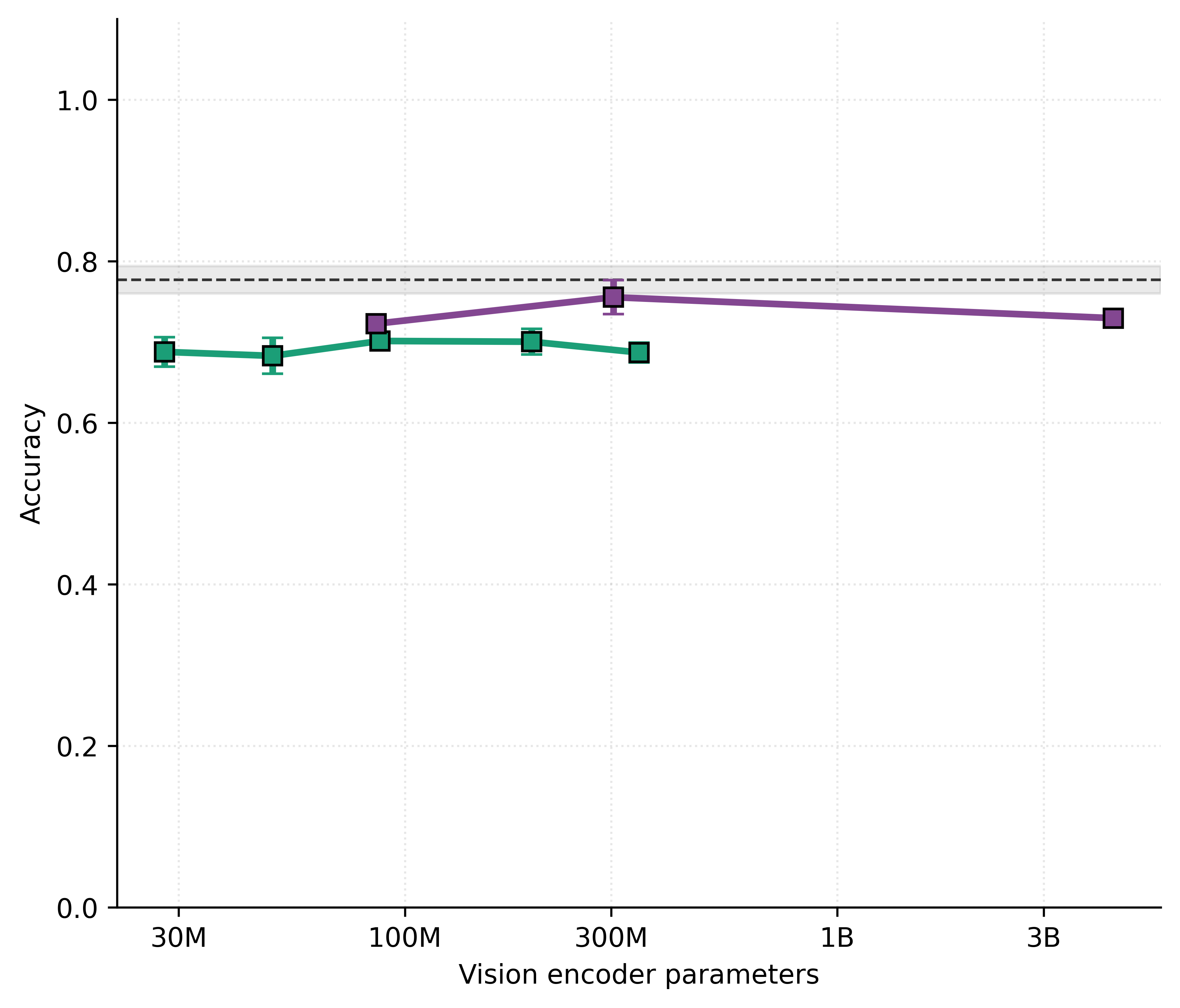}
    \caption{}\label{fig:encoder_size_global:acc}
\end{subfigure}
\end{minipage}
\begin{minipage}{0.15\textwidth}
    \centering
    \includegraphics[scale=0.5]{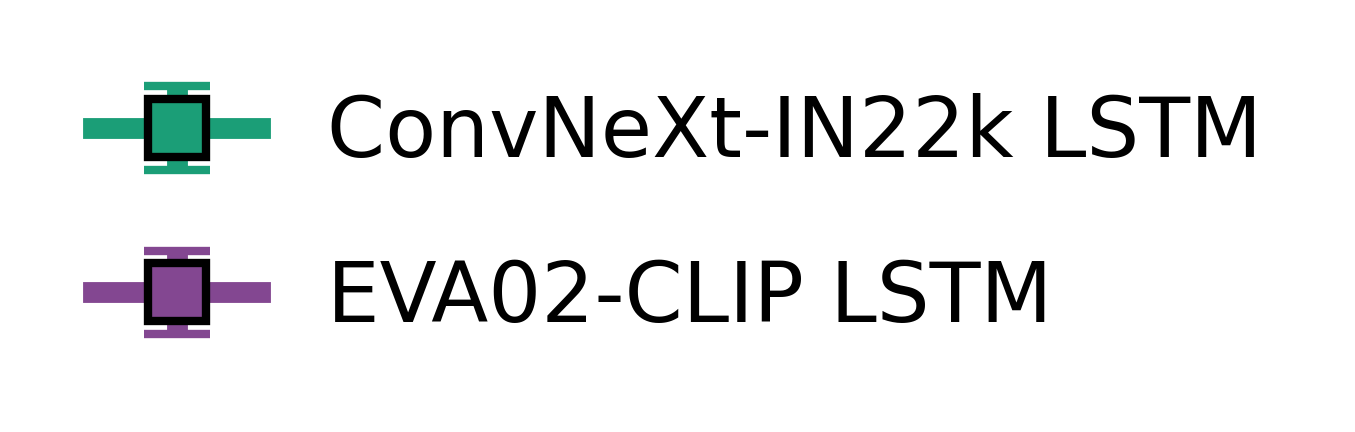}
\end{minipage}
\caption{
\textbf{Comparison of behavioral alignment across vision encoders, sequential architectures (LSTM vs. Transformer), hidden dimension sizes, and vision encoder sizes.} Hatch patterns indicate pretraining method: empty bars for supervised ImageNet \citep{krizhevsky2012imagenet}, forward slash for CLIP, backslash for DINOv1 \citep{caron2021emerging}, crossed diagonal lines for DINOv3 \citep{simeoni2025dinov3}, and horizontal lines for masked image modeling (MIM) \citep{bao2022beit}.  (a) Correlation with human behavior. (b) Task accuracy. (c) Effect of sequential module hidden dimension size on human alignment. (d) Effect of hidden size on task accuracy. Increasing model representation size increases accuracy, but does not increase behavioral alignment. For natural transformers, the human alignment declines steadily. (e) Effect of vision encoder size (number of parameters, log scale) on human alignment, for \texttt{Natural} LSTMs with ConvNeXt (supervised ImageNet-22k; Tiny to XLarge) and EVA02-CLIP (Base, Large, Enormous) encoders. (f) Effect of vision encoder size on task accuracy. Neither behavioral alignment nor accuracy increases consistently with encoder size.
}\label{fig:vision_type_global}
\end{figure}

Using the \texttt{timm} package \citep{rw2019timm}, we train and evaluate LSTMs and Transformers using various frozen vision encoders. To test whether a larger vision encoder brings \texttt{Natural} models closer to human behavior, we trained \texttt{Natural} LSTMs ($h=128$, 6 seeds) on two families of frozen encoders of increasing size (Fig.~\ref{fig:encoder_size_global:corr}).
The first is ConvNeXt \citep{liu2022convnet} pretrained with supervision on ImageNet-22k \citep{deng2009imagenet}, in five sizes from 28M (Tiny) to 348M (XLarge) parameters.
All five share the pretraining data, recipe and input resolution ($224\times224$), so only the encoder size changes.
The second is EVA02-CLIP \citep{fang2024eva,sun2023eva} at Base (86M), Large (303M) and Enormous (4.3B) sizes. 

\vspace{-0.1in}

Human consistency does not increase with encoder size in either family. Across a 12-fold range of ConvNeXt sizes, $\rho(m,h)$ stays between 0.363 and 0.388 and MTurk accuracy between 0.683 and 0.701, with no consistent trend.
For EVA02-CLIP, $\rho(m,h)$ decreases slightly from 0.438 (Base) to 0.415 (Large) and 0.405 (Enormous), while MTurk accuracy peaks at Large (0.723, 0.756 and 0.730).
The encoder family matters more than its size.
At nearly equal size, EVA02-CLIP Base reaches $\rho(m,h)=0.438$ against 0.388 for ConvNeXt Base, and it exceeds every ConvNeXt, including XLarge.

The two families differ in both architecture and pretraining objective, so this comparison does not isolate either factor, although it agrees with the advantage of CLIP pretraining in Fig.~\ref{fig:vision_type_global:corr}.
All eight encoders remain far below the human consistency ceiling ($\rho(h,h)=0.787$), and none exceeds the ResNet-CLIP LSTM used in the main text ($\rho(m,h)=0.452$).

\begin{figure}
\centering

\begin{subfigure}{0.4\textwidth}
    \includegraphics[width=\linewidth]{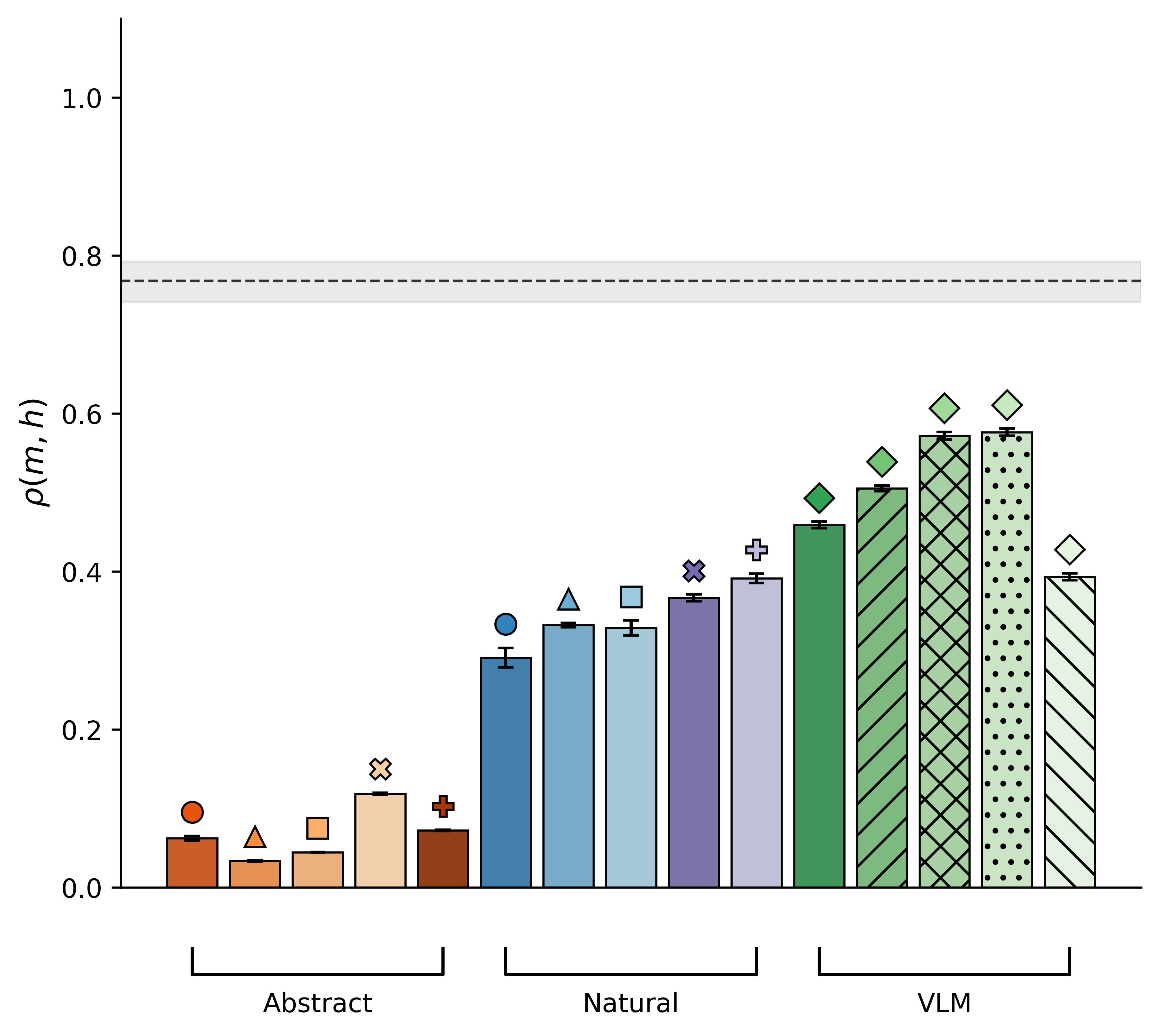}
    \caption{Category -- Correlation}
\end{subfigure}
\begin{subfigure}{0.4\textwidth}
    \includegraphics[width=\linewidth]{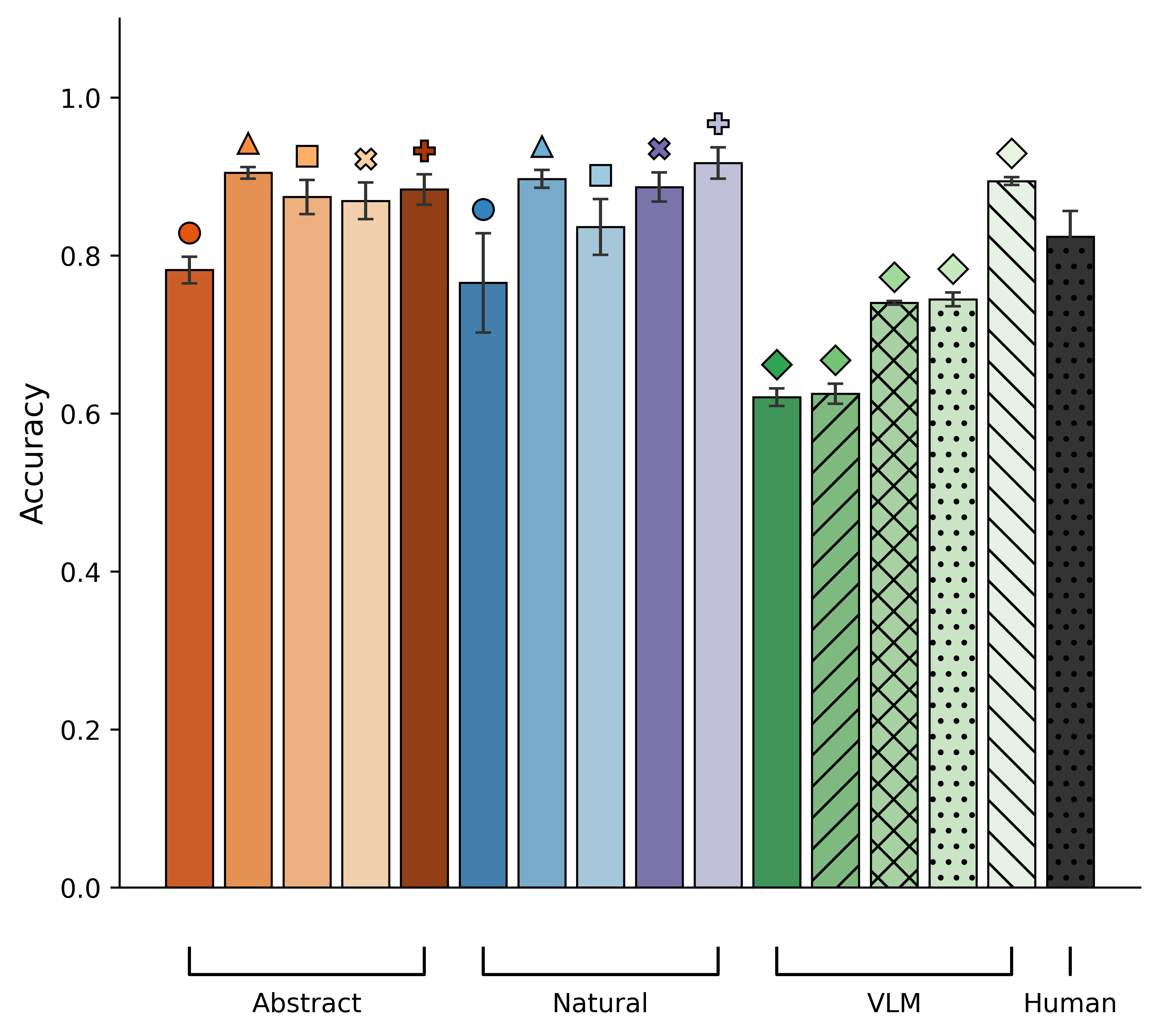}
    \caption{Category -- Accuracy}
\end{subfigure}

\vspace{0.5em}

\begin{subfigure}{0.4\textwidth}
    \includegraphics[width=\linewidth]{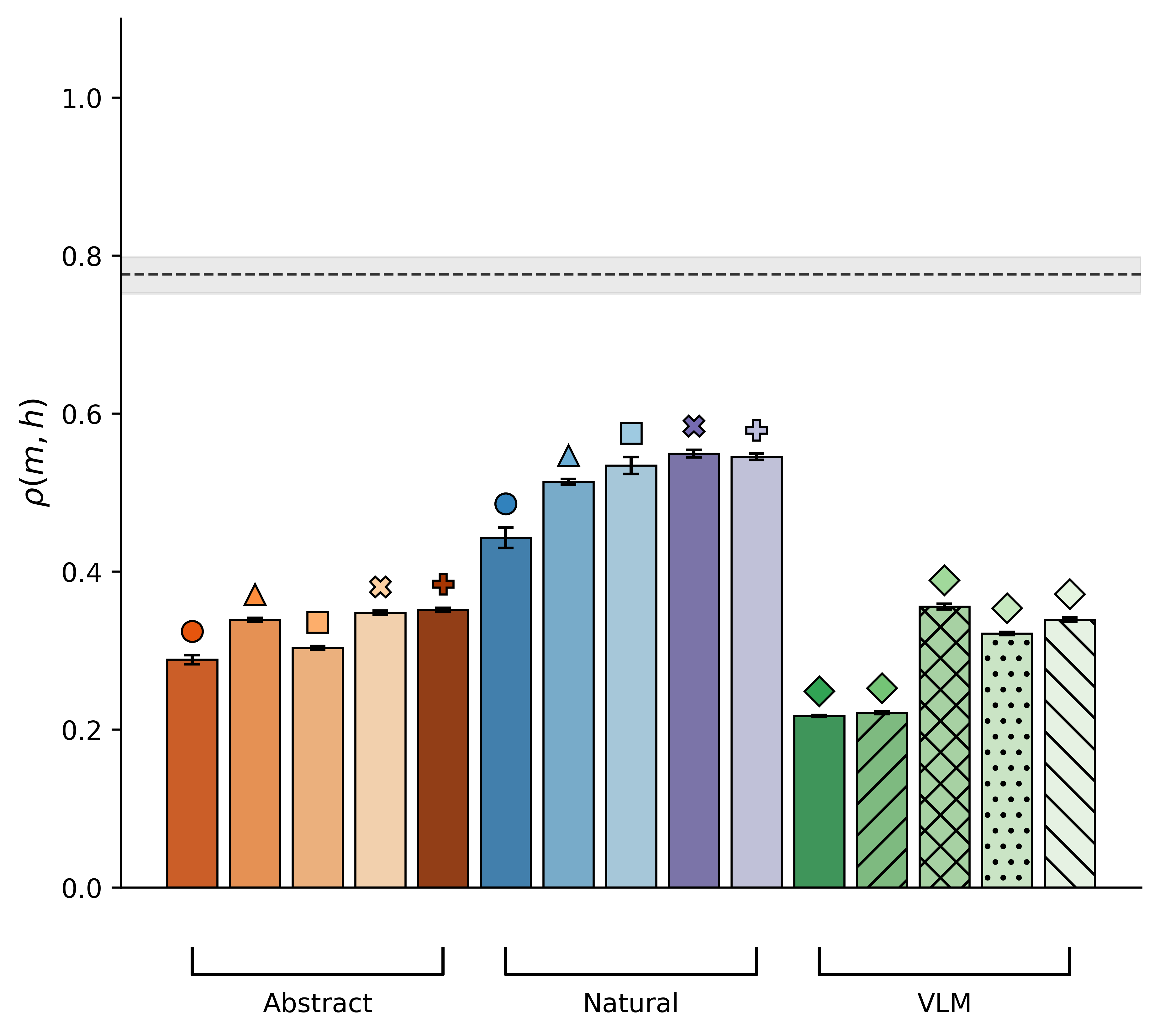}
    \caption{Identity -- Correlation}
\end{subfigure}
\begin{subfigure}{0.4\textwidth}
    \includegraphics[width=\linewidth]{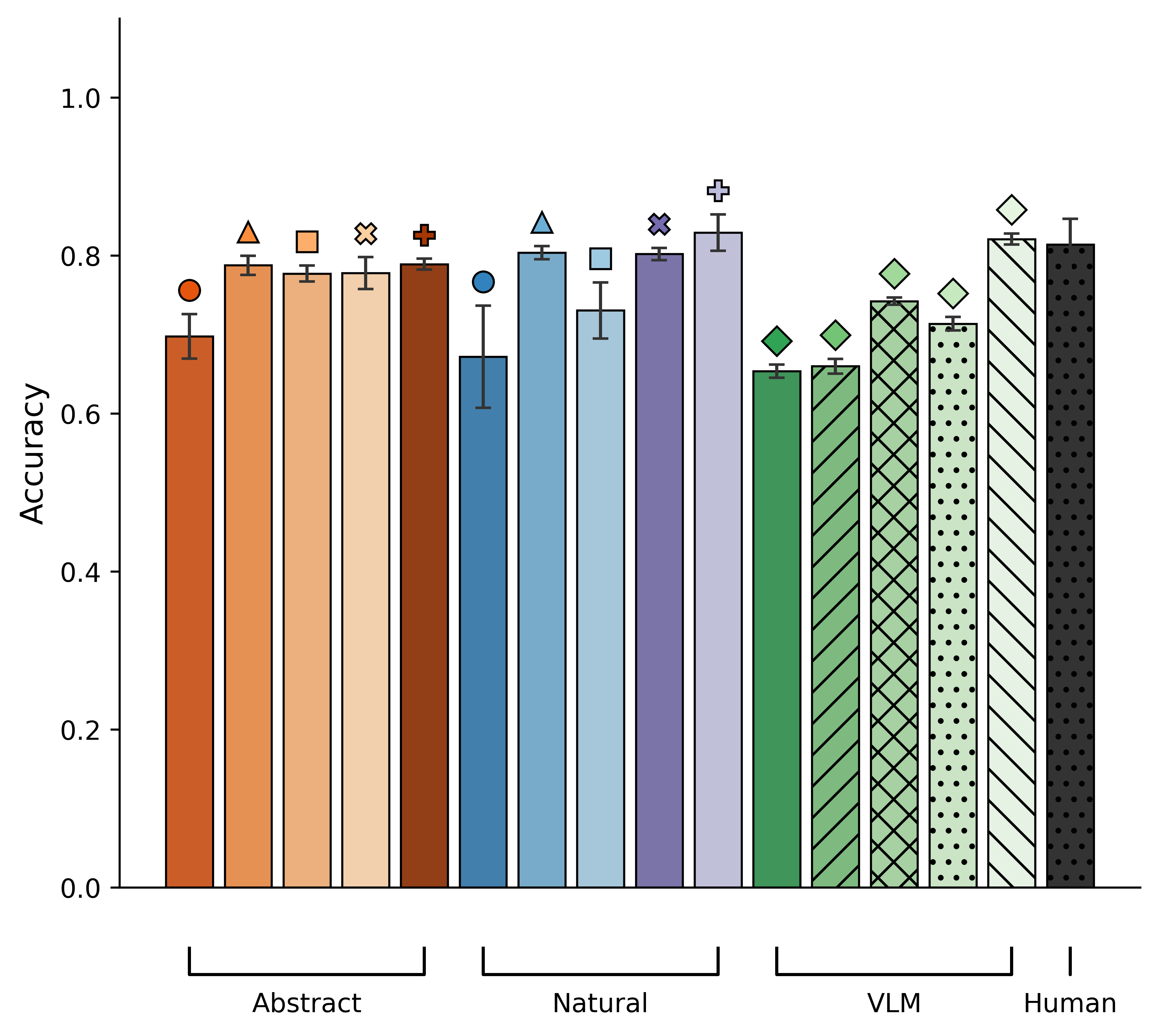}
    \caption{Identity -- Accuracy}
\end{subfigure}

\vspace{0.5em}

\begin{subfigure}{0.4\textwidth}
    \includegraphics[width=\linewidth]{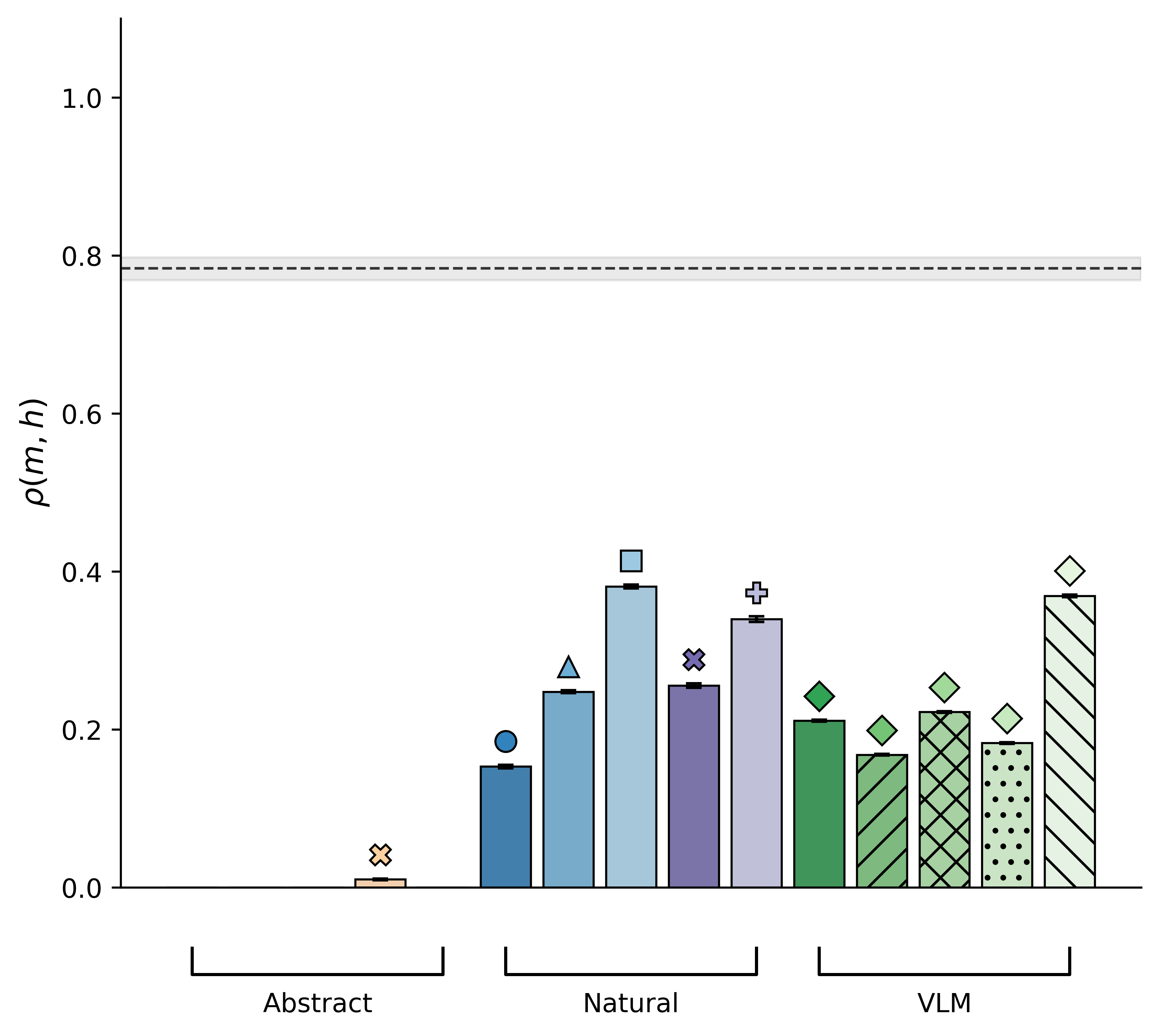}
    \caption{Position -- Correlation}
\end{subfigure}
\begin{subfigure}{0.4\textwidth}
    \includegraphics[width=\linewidth]{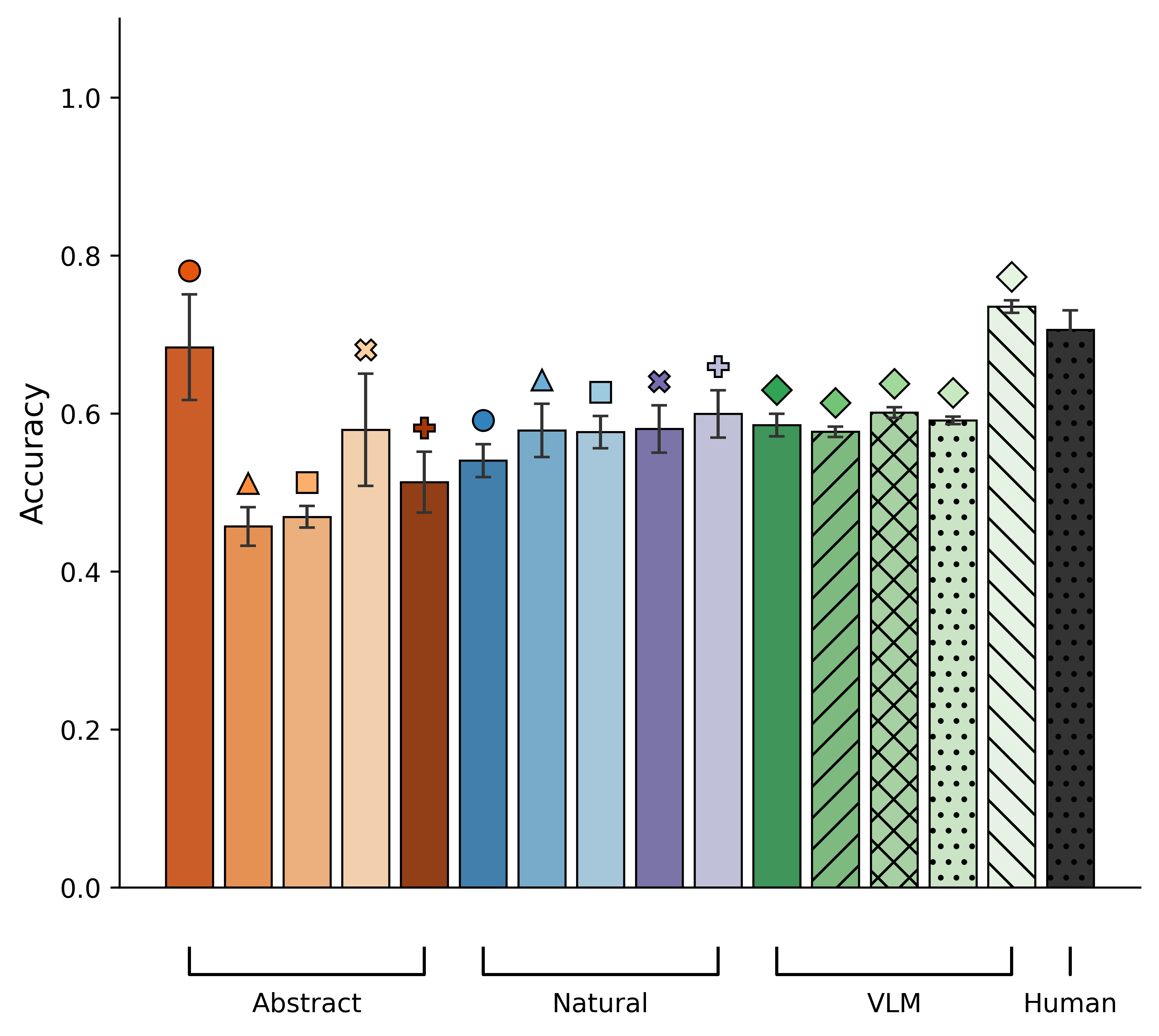}
    \caption{Position -- Accuracy}
\end{subfigure}

\caption{\textbf{Feature-level human consistency and accuracy of trained task models and VLMs.}
For every architecture, the \texttt{Natural} model is more human-consistent than the \texttt{Abstract} on every feature, and both model types are most human-consistent on identity trials.
Among \texttt{Natural} models, BERT and the Transformer lead on identity, the Transformer on category, and the LSTM on position.
VLMs are the most human-consistent models on category trials (0.39--0.58) but fall below every \texttt{Natural} model on identity trials.
On position trials, the human consistency of \texttt{Abstract} models is $0$ for every architecture except BERT ($0.01$).
We hypothesize that \texttt{Abstract} models solve position tasks from the one-hot position code, whereas \texttt{Natural} models, VLMs and humans judge position from the object's location over an ambiguous $3\times3$ grid on a naturalistic background.
\texttt{Abstract} models are evaluated at the 80\%-matched checkpoint; markers, error bars and the human reference follow Fig.~\ref{fig:seq_type_global}.}
\label{fig:seqsize_feature}
\end{figure}
\begin{figure}
\centering
\begin{subfigure}{0.22\textwidth}
    \includegraphics[width=\linewidth]{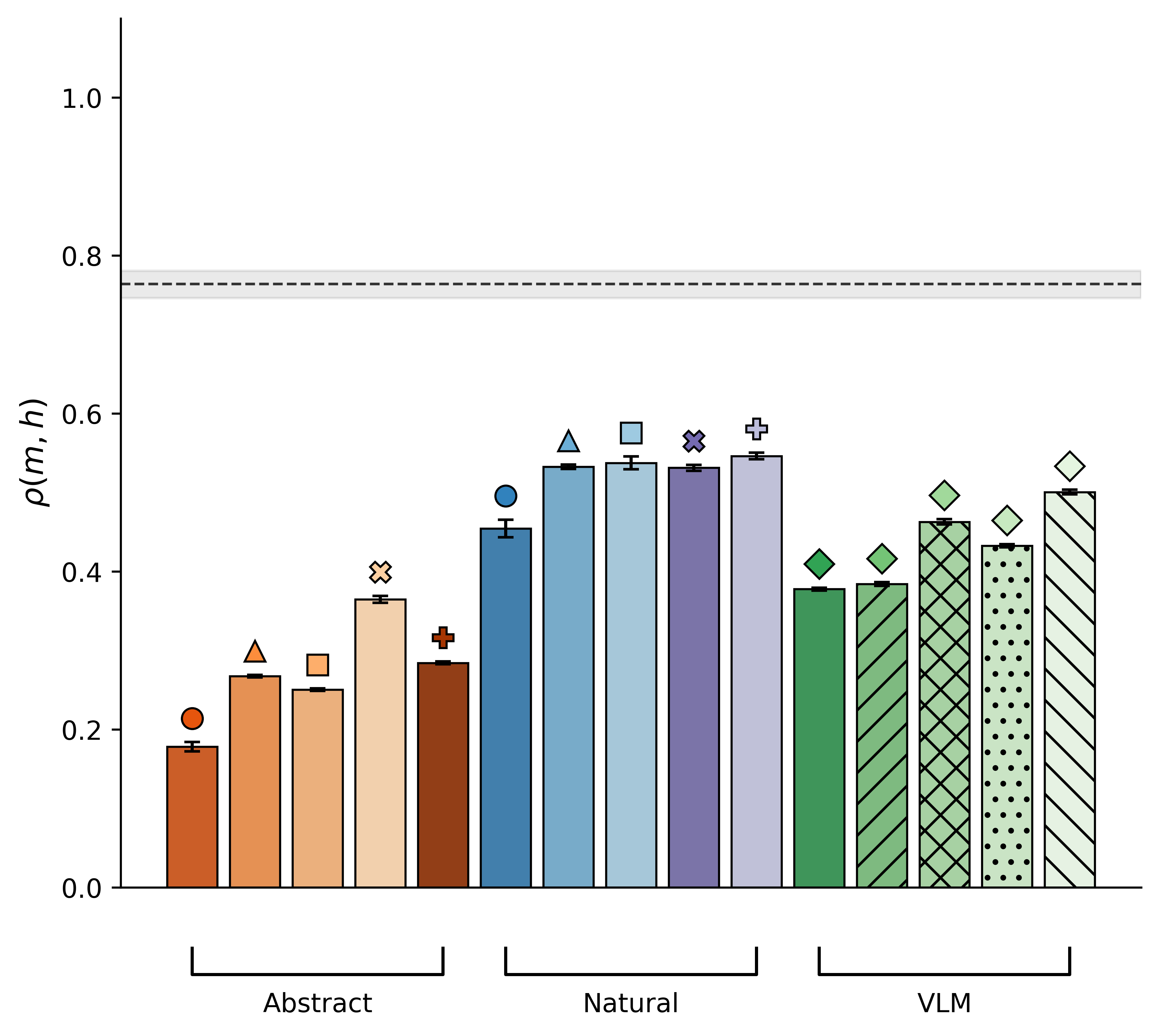}
    \caption{}
\end{subfigure}
\hfill
\begin{subfigure}{0.22\textwidth}
    \includegraphics[width=\linewidth]{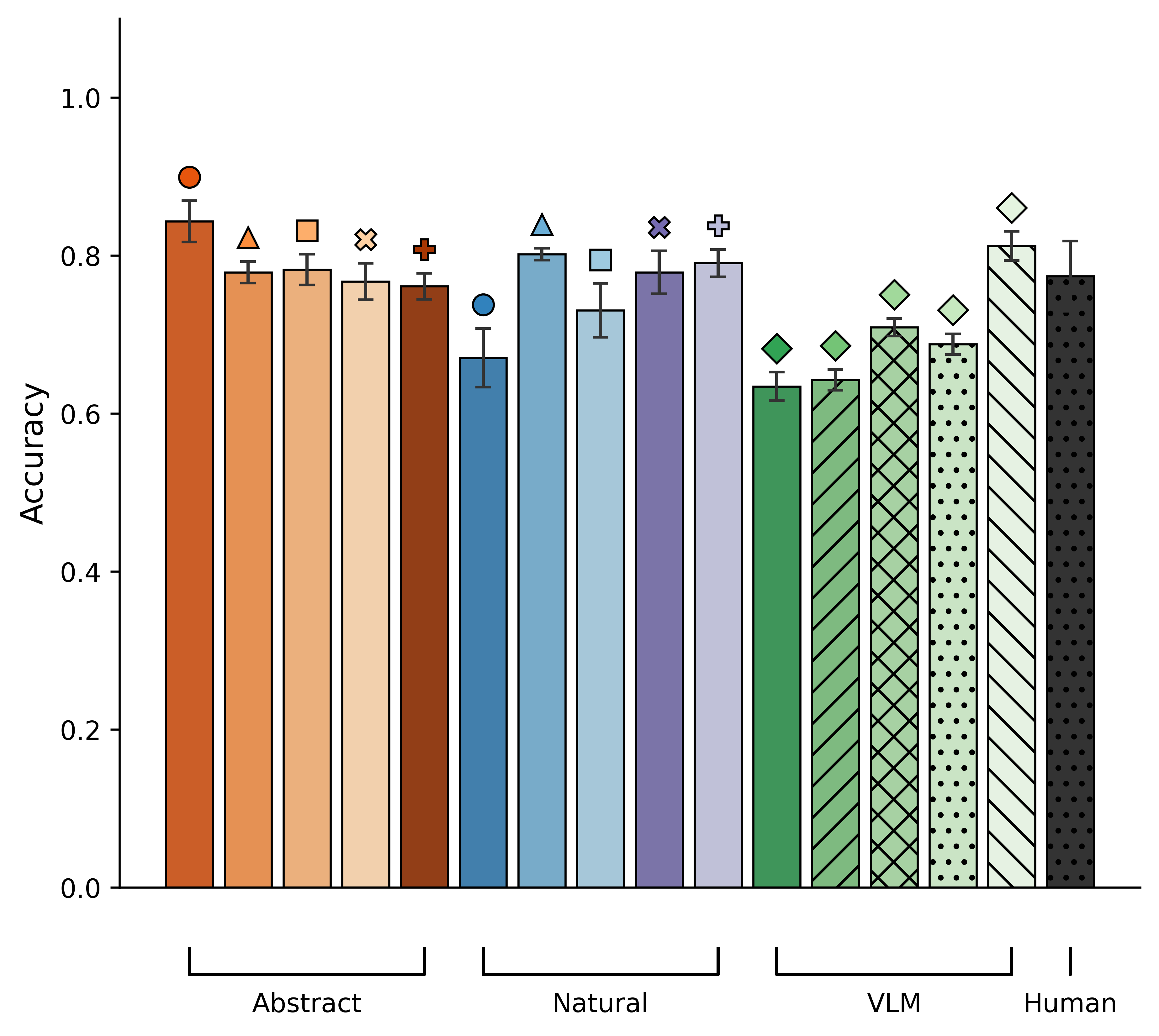}
    \caption{}
\end{subfigure}
\hfill
\begin{subfigure}{0.22\textwidth}
    \includegraphics[width=\linewidth]{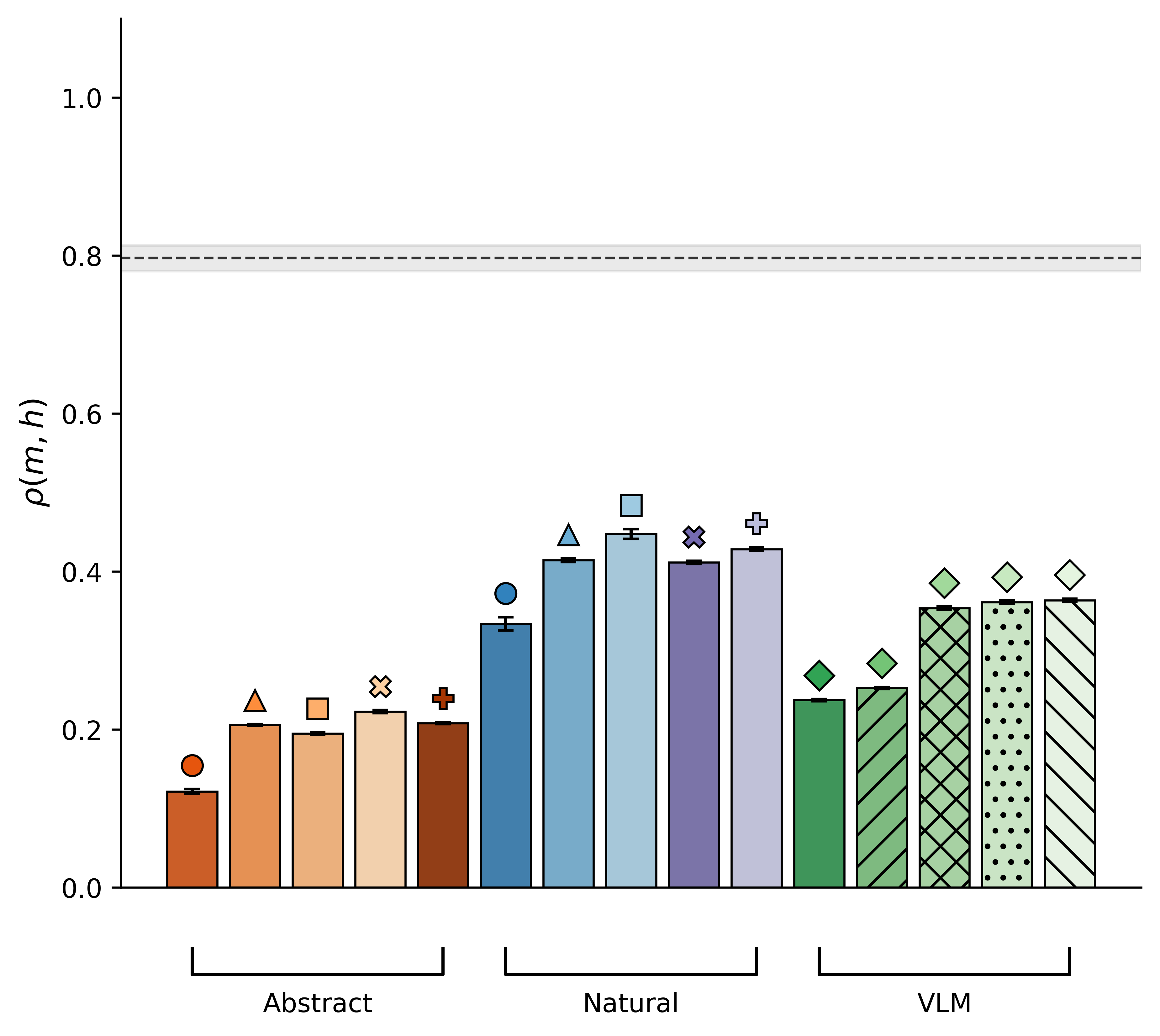}
    \caption{}
\end{subfigure}
\hfill
\begin{subfigure}{0.22\textwidth}
    \includegraphics[width=\linewidth]{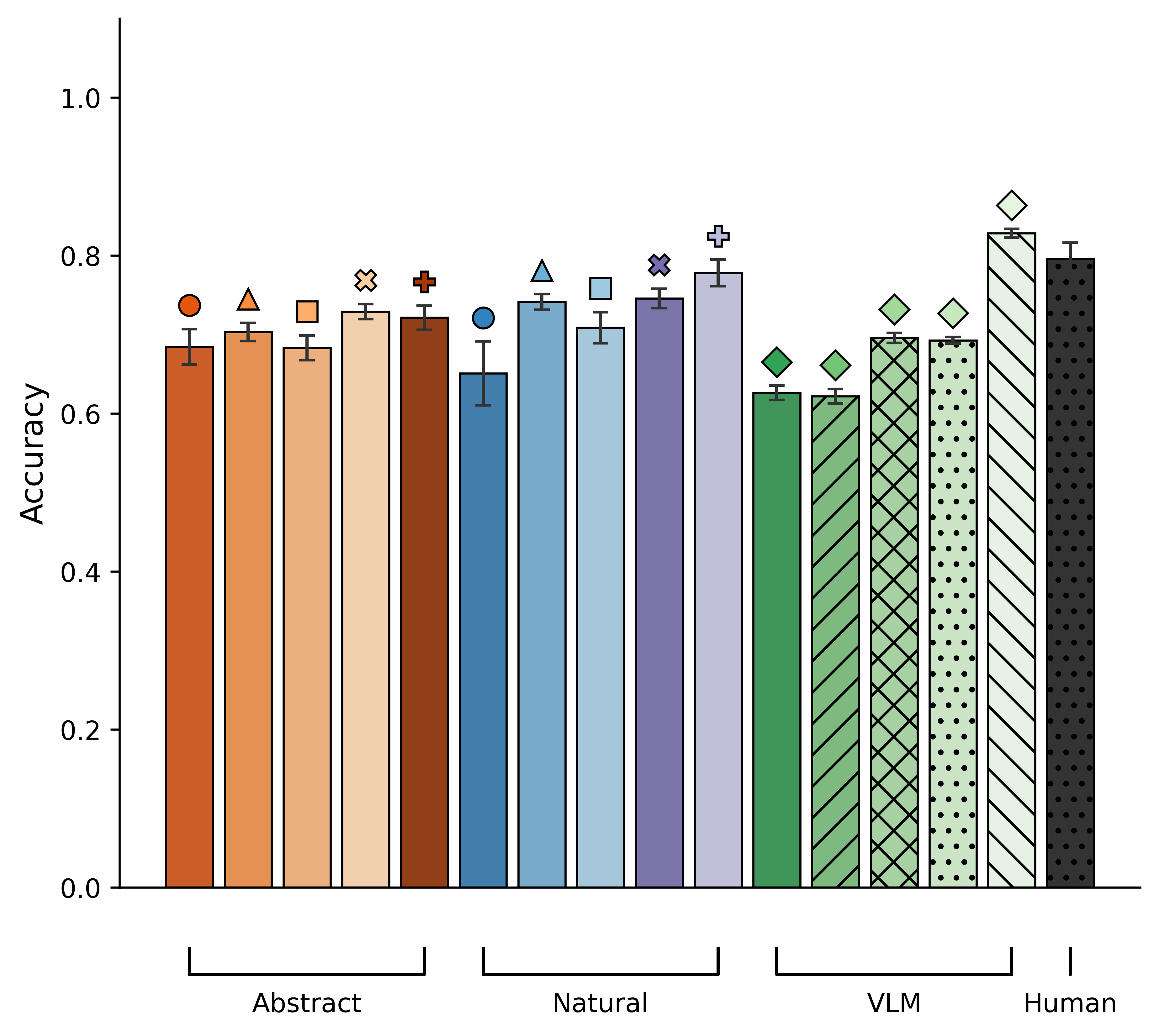}
    \caption{}
\end{subfigure}

\vspace{0.8em}

\begin{subfigure}{0.22\textwidth}
    \includegraphics[width=\linewidth]{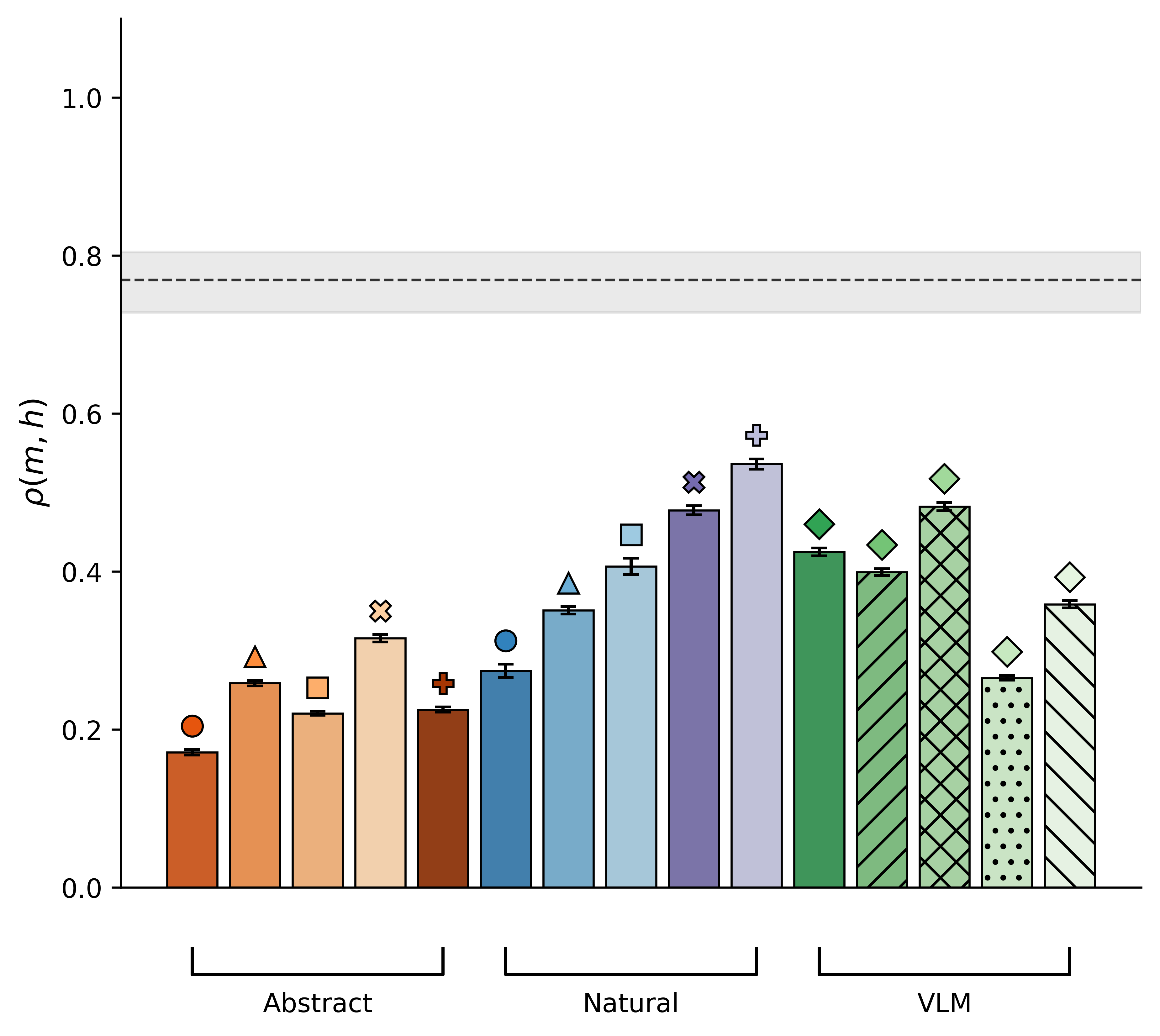}
    \caption{}
\end{subfigure}
\hfill
\begin{subfigure}{0.22\textwidth}
    \includegraphics[width=\linewidth]{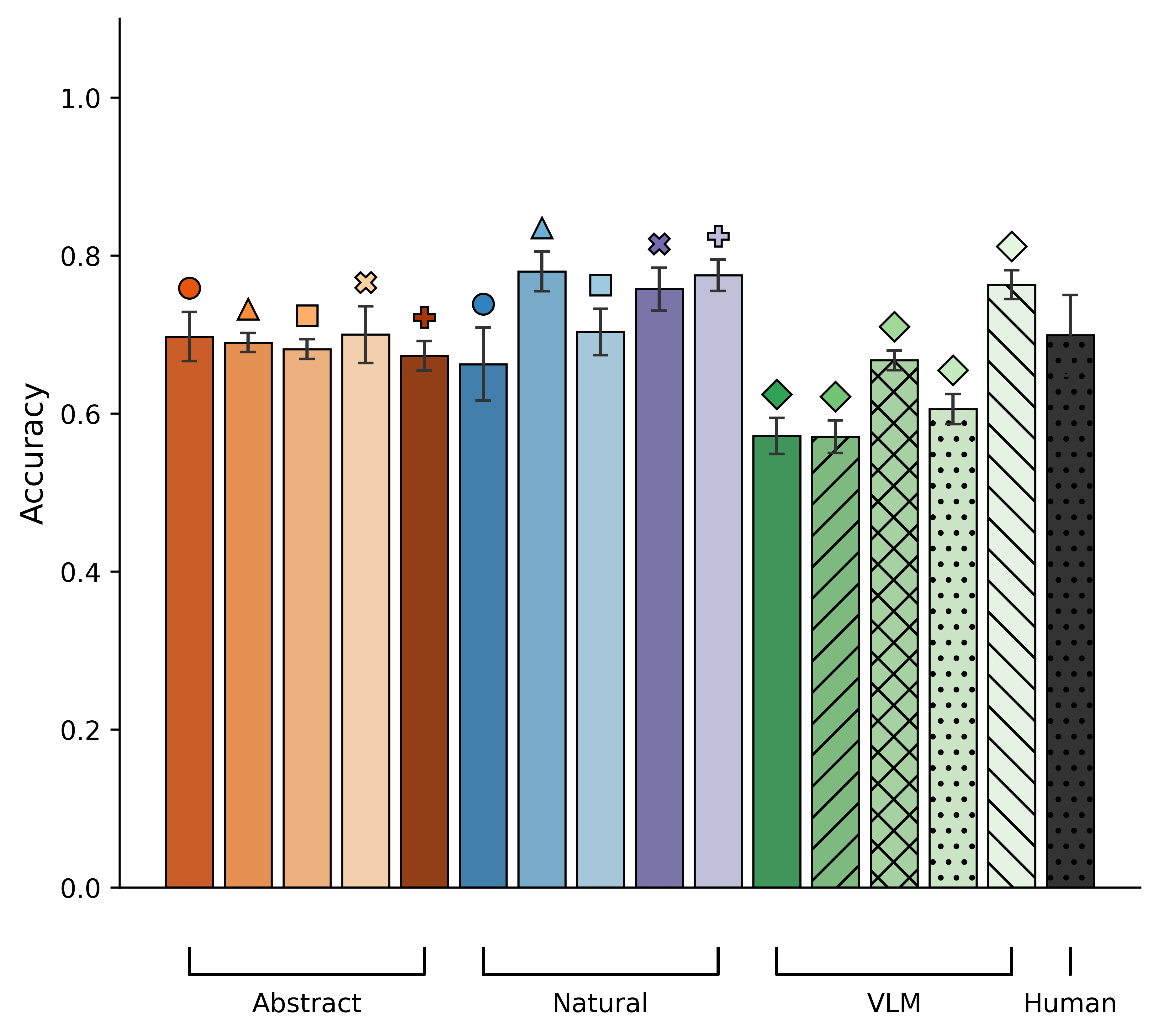}
    \caption{}
\end{subfigure}
\begin{subfigure}{0.22\textwidth}
    \includegraphics[width=\linewidth]{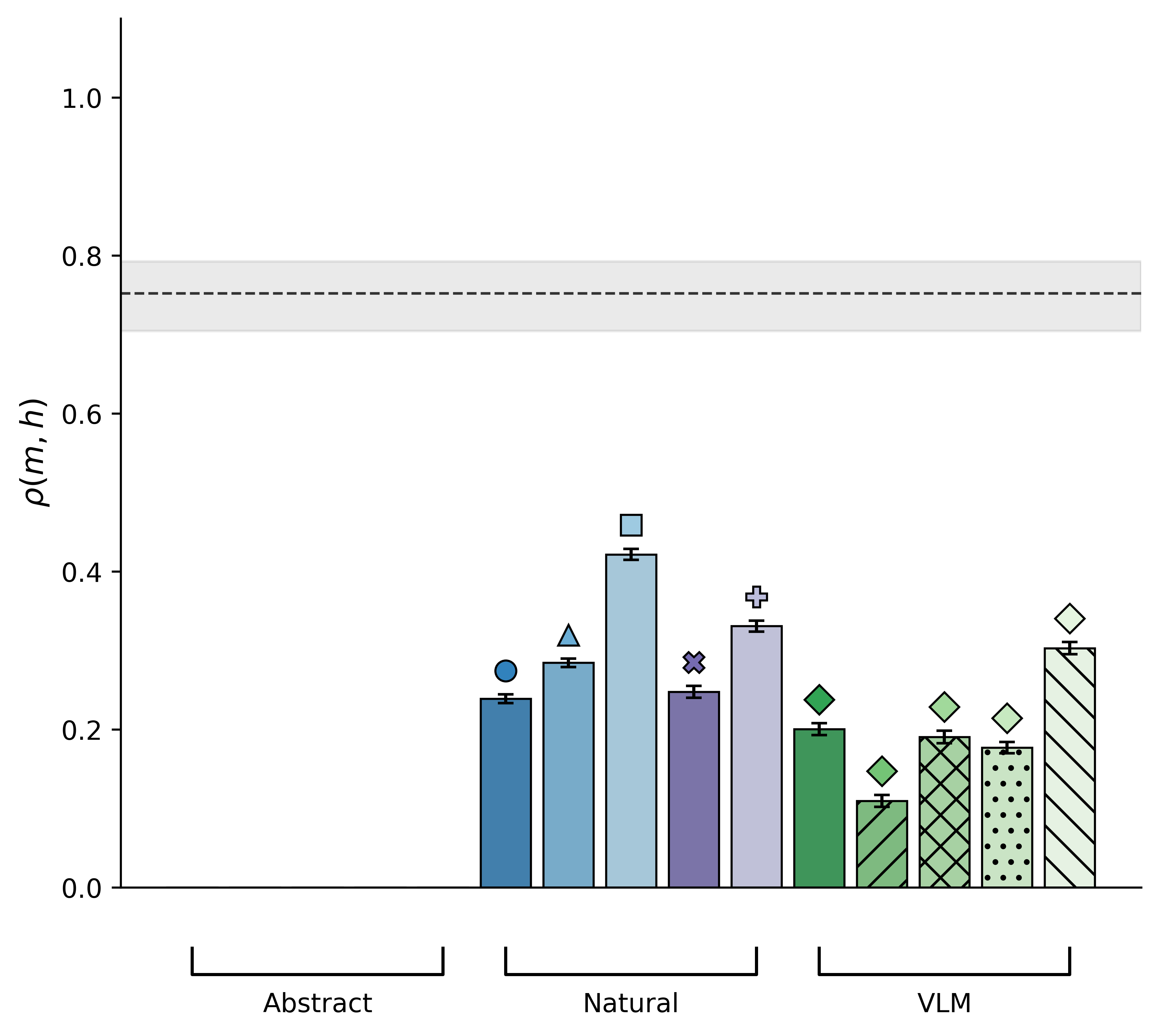}
    \caption{}
\end{subfigure}
\hfill
\begin{subfigure}{0.22\textwidth}
    \includegraphics[width=\linewidth]{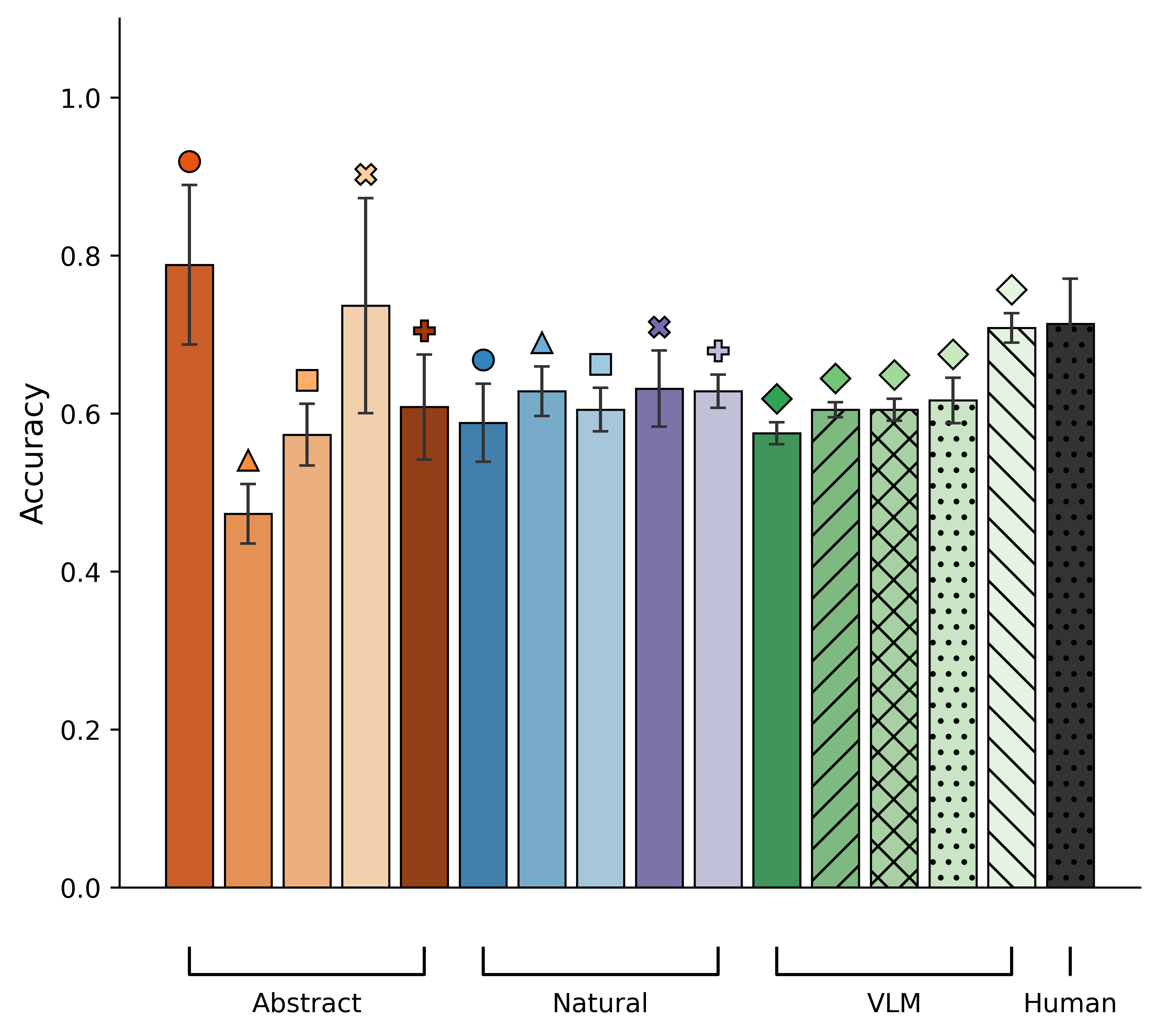}
    \caption{}
\end{subfigure}
\caption{\textbf{Task-family human consistency and accuracy by sequence architecture.}
(a, b) NBack, (c, d) InterDMS, (e, f) CtxDMS and (g, h) DMS; left of each pair $\rho(m,h)$, right MTurk accuracy.
For every architecture, the \texttt{Natural} model is more human-consistent than the \texttt{Abstract} in every task family.
Among \texttt{Natural} models, the LSTM leads on InterDMS and DMS, and the Transformer on CtxDMS, while four architectures are within 0.015 of each other on NBack.
Among \texttt{Abstract} models, BERT leads in every family. On DMS trials (the DMS-position task only) every \texttt{Abstract} model has zero correlation with humans.
\texttt{Abstract} models are evaluated at the 80\%-matched checkpoint; markers, error bars and the human reference follow Fig.~\ref{fig:seq_type_global}.}
\label{fig:seq_type_task_type}
\end{figure}
\subsubsection{Visual discriminability does not explain the behavioral gap}\label{sec:appendix_residualization}
\texttt{Natural} models and humans see the same images, so their agreement could reflect shared stimulus difficulty rather than shared task computation.
To test this, we regressed an image-based estimate of each trial's visual discriminability out of every human and model response vector and recomputed $\rho(m,h)$.
For trial $i$, let $s_i$ be the cosine similarity between the two stimuli in the feature space of a vision encoder.
A similar pair is easy to call a match but hard to call a non-match, so within each of the 18 task conditions we sign $s_i$ by the ground truth and center it within trial type, $d_i = (2m_i - 1)(s_i - \bar{s}_{m_i})$, where $m_i \in \{0,1\}$ is the correct answer and $\bar{s}_{m_i}$ is the mean similarity of trials of that type in the condition.
Centering makes $d$ orthogonal to the match/non-match label, so the regression removes differences in visual uncertainty between stimulus pairs but keeps the task structure differences between match and non-match trials.
Within each condition, each human subject's response vector $\mathbf{y}$ and each model seed's response vector are replaced by its residual $(I - XX^{+})\mathbf{y}$, with $X = [\mathbf{1}, \mathbf{d}]$.
Since $X$ depends only on the stimuli, humans and models are projected identically, while each responder's slope is fitted separately.
Removing the per-condition intercept changes $\rho(m,h)$ by itself, so we compare against a null in which $s$ is permuted within each condition before the same residualization (10 permutations), and report the effect of discriminability as the residualized gap minus the null gap.

\begin{table}[ht]
\centering
\caption{\texttt{Natural} minus \texttt{Abstract} difference in $\rho(m,h)$ (mean over the five sequence models; \texttt{Abstract} models at the 80\%-matched checkpoint) before and after regressing out visual discriminability. Null: mean $\pm$ SD over 10 within-condition permutations of the covariate. Effect: residualized minus null.}
\label{tab:residualization}
\small
\begin{tabular}{lccccc}
\toprule
Covariate & $R^2$ & Unregressed & Residualized & Null & Effect \\
\midrule
ResNet-CLIP & 0.029 & 0.229 & 0.249 & $0.252 \pm 0.001$ & $-0.003$ \\
DINOv3 & 0.042 & 0.229 & 0.233 & $0.252 \pm 0.002$ & $-0.019$ \\
\bottomrule
\end{tabular}
\end{table}
We computed $s$ with two encoders: ResNet-CLIP and DINOv3.
The covariate explains $R^2=0.029$ (ResNet-CLIP) and $0.042$ (DINOv3) of the variance in per-trial human accuracy (mean over conditions), and the two similarity estimates correlate at $r=0.44$ across trials.
Table~\ref{tab:residualization} reports the difference between the mean $\rho(m,h)$ of the five \texttt{Natural} and the five \texttt{Abstract} sequence models before residualization.
Relative to the null, ResNet-CLIP discriminability changes this gap by $-0.003$ and DINOv3 discriminability by $-0.019$ (8\% of the gap); per feature, the largest change is $-0.043$ on category trials. With the final \texttt{Abstract} checkpoints, the gap (0.291) changes by $-0.001$ and $-0.018$.
Visual discriminability, as estimated by either encoder, thus accounts for at most a small part of the behavioral gap.

\subsubsection{RSA}\label{sec:appendix/rsa}
Following the same split-half procedure for human consistency, for each random split-half, we average each partitioned group's response, resulting in a pair of vectors of shape $(18\times 100)$. We then calculate the average pairwise correlation either over the task space, or over the stimulus space, producing $\mathbf{C}_{\mathrm{task}}\in \left[ -1, 1\right]^{18\times 18}, \mathbf{C}_{\mathrm{stimulus}}\in \left[ -1, 1\right]^{100\times 100}$. We can then measure alignment with humans by taking the Pearson correlation between the upper triangles of the human $\mathbf{C}$ and model $\mathbf{C}$. Since we use the same 100 stimulus pairs for evaluation, given a stimulus pair, a human-like model would produce similar logits in \texttt{InterDMS-AA-Position} and \texttt{1-Back-Position} tasks, but dissimilar logits in \texttt{InterDMS-AA-Position} and \texttt{1-Back-Category} tasks.

\begin{figure}
\centering
    \begin{minipage}{0.78\textwidth} 

\begin{subfigure}{0.48\textwidth}
    \includegraphics[width=\linewidth]{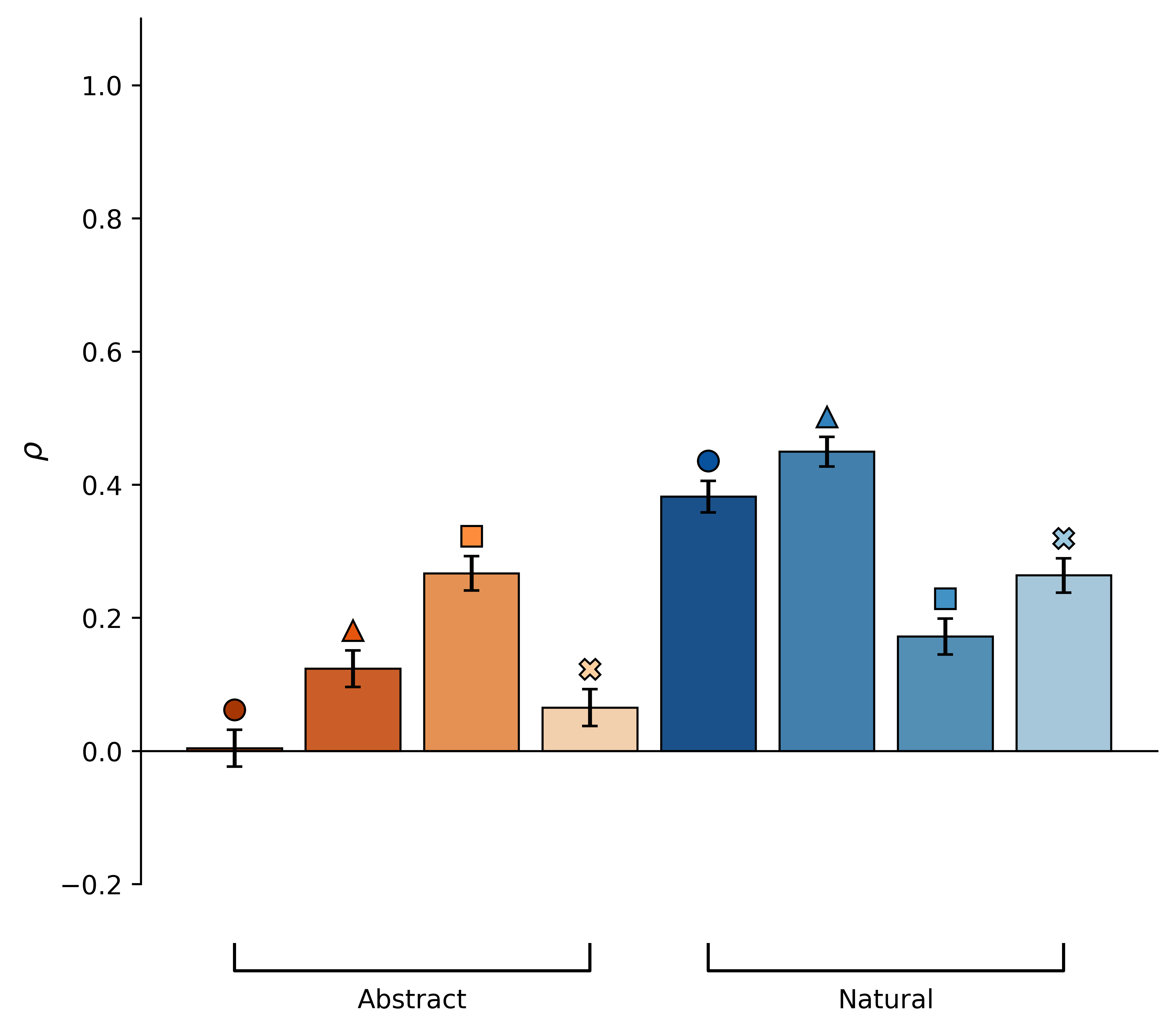}
    \caption{Cross-stimulus RSA}
\end{subfigure}
\hfill
\begin{subfigure}{0.48\textwidth}
    \includegraphics[width=\linewidth]{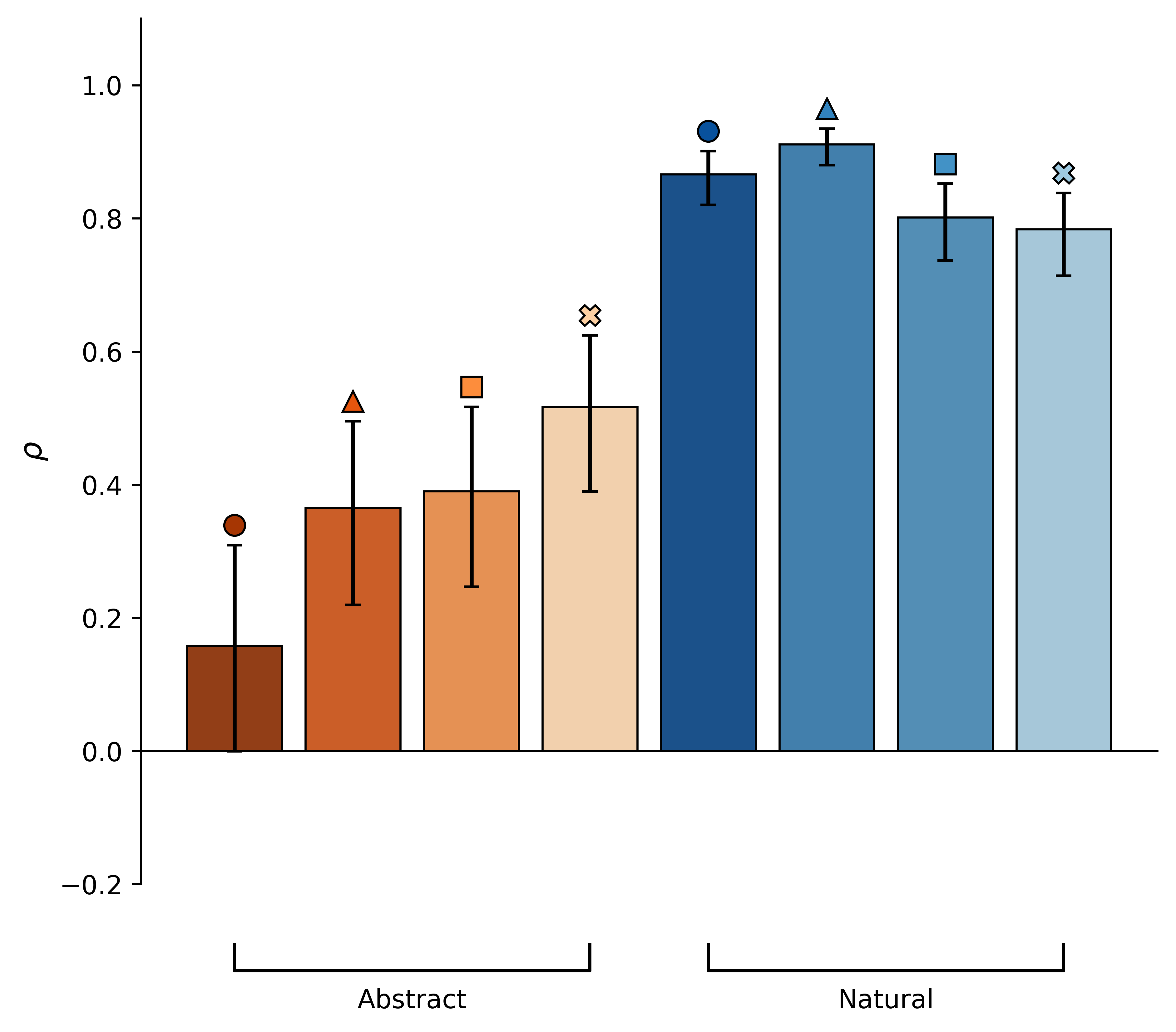}
    \caption{Cross-task RSA}\label{fig:new_transplant_rsa:ct}
\end{subfigure}
\end{minipage}
\begin{minipage}{0.15\textwidth}
    \centering
    \includegraphics[scale=0.5]{plots/transplant/final/legend.png}
    \end{minipage}
\caption{\textbf{Representational similarity analysis (RSA) of substituted models.}
The y-axis shows the Spearman correlation between the upper-triangular entries of the model and human representational dissimilarity matrices (RDMs).}
\label{fig:new_transplant_rsa}
\end{figure}
\begin{figure}
\centering
\begin{subfigure}{0.62\linewidth}
    \centering
    \includegraphics[width=\linewidth]{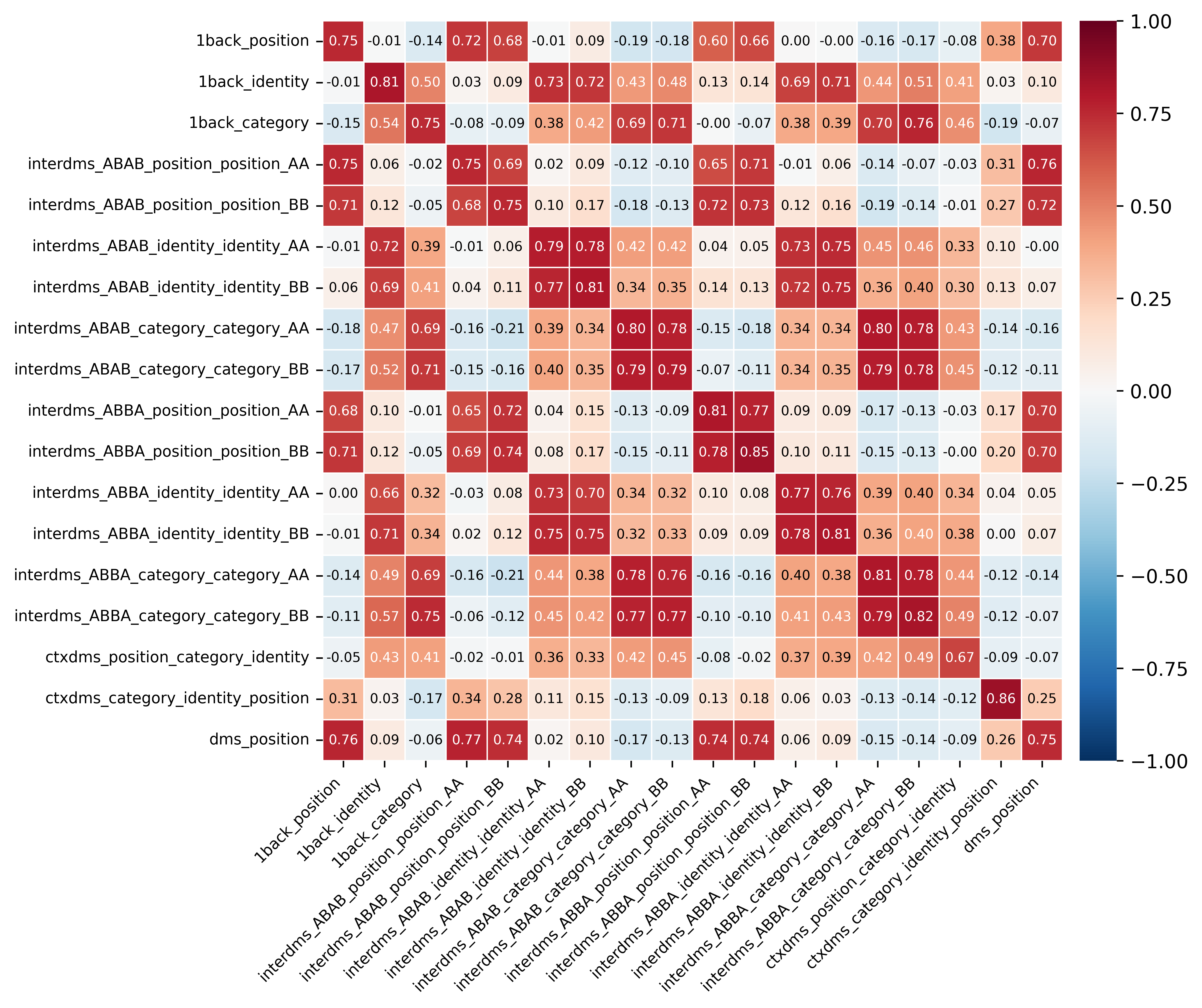}
    \caption{Human}\label{fig:rsa_ct_lstm:human}
\end{subfigure}

\vspace{0.5em}
\begin{minipage}{0.7\linewidth}
    \centering
    \begin{subfigure}{0.48\linewidth}
        \includegraphics[width=\linewidth]{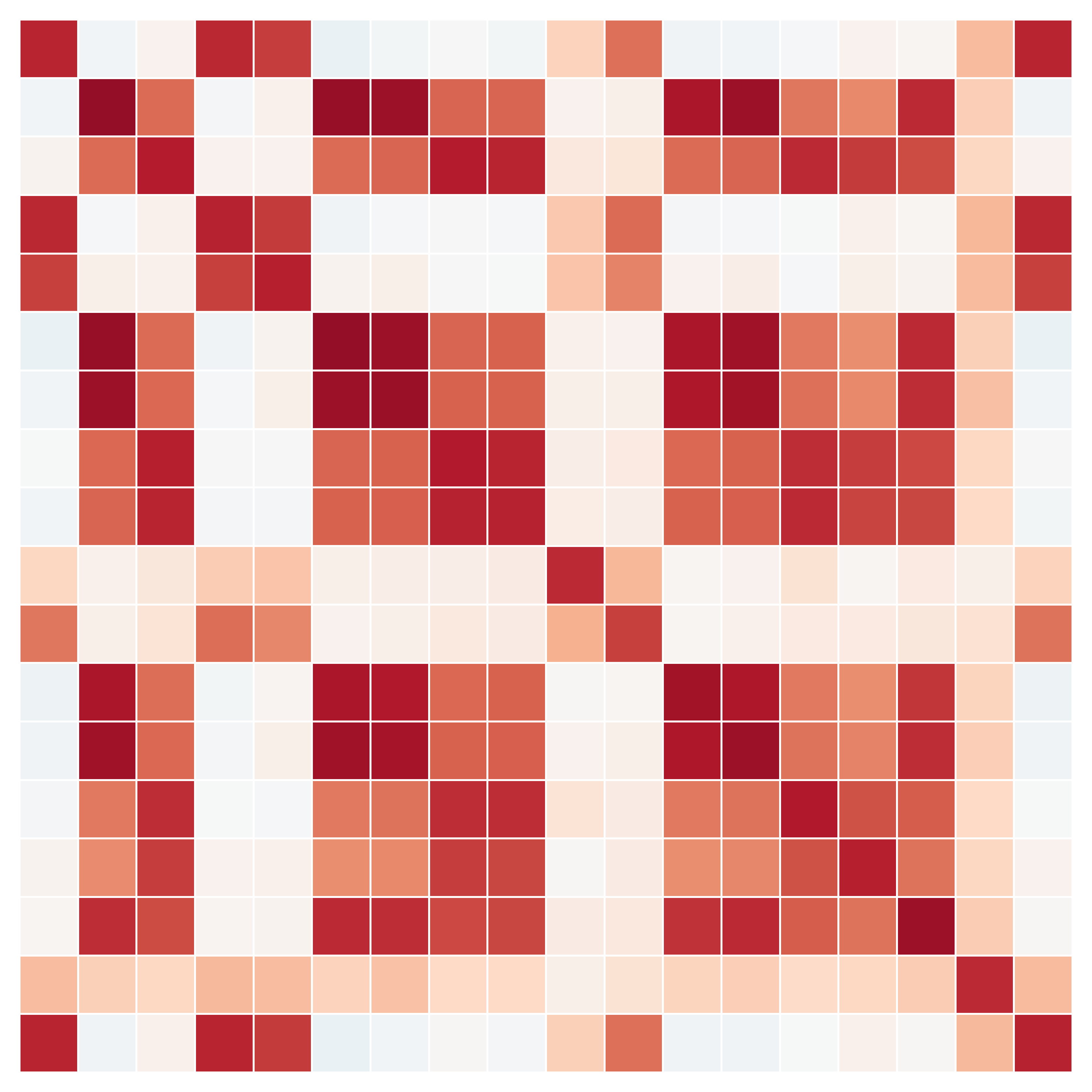}
        \caption{\texttt{Natural} LSTM}\label{fig:rsa_ct_lstm:nat}
    \end{subfigure}\hfill
    \begin{subfigure}[b]{0.48\linewidth}
        \includegraphics[width=\linewidth]{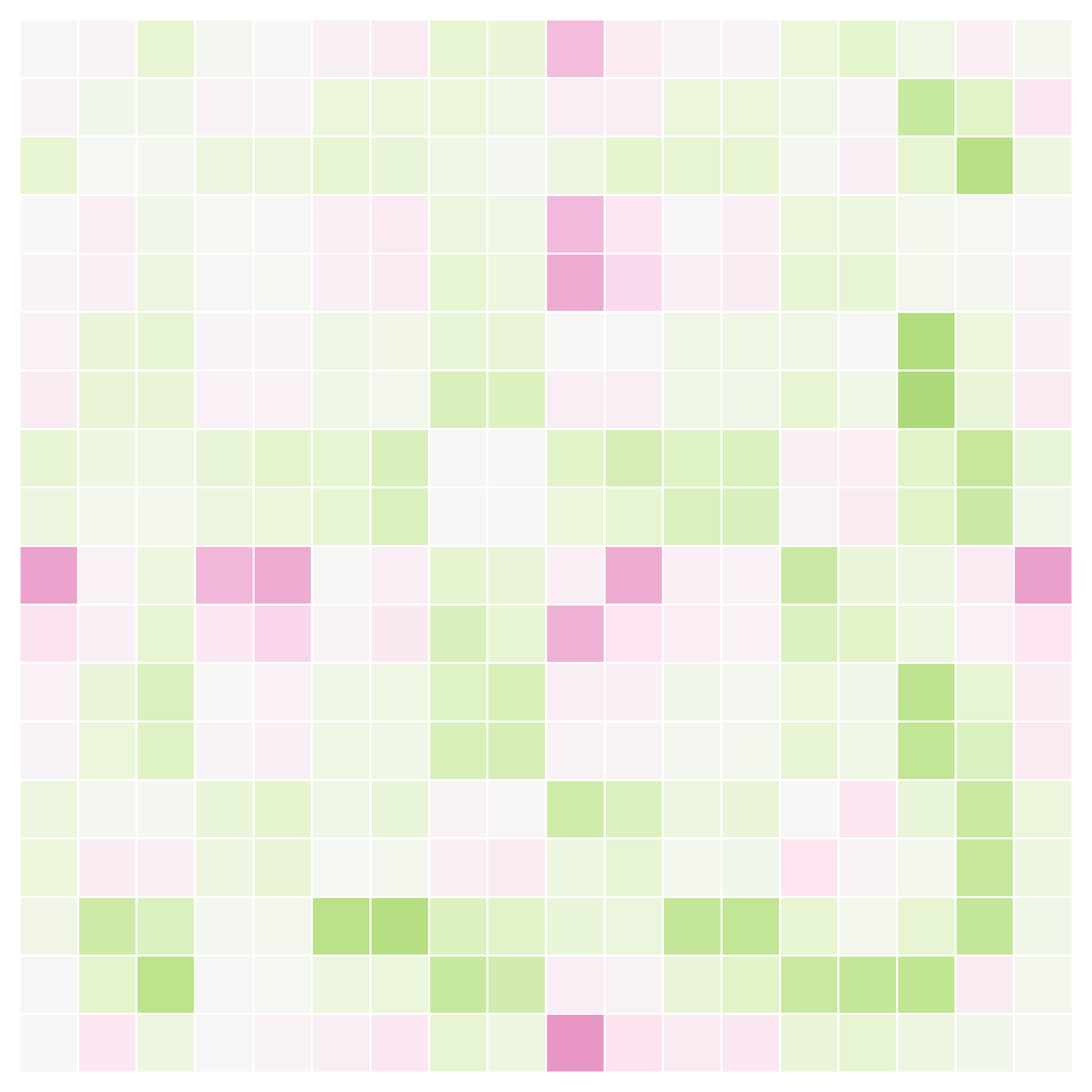}
        \caption{\texttt{Natural} LSTM $-$ human}\label{fig:rsa_ct_lstm:nat_diff}
    \end{subfigure}

    \vspace{0.5em}
    \begin{subfigure}{0.48\linewidth}
        \includegraphics[width=\linewidth]{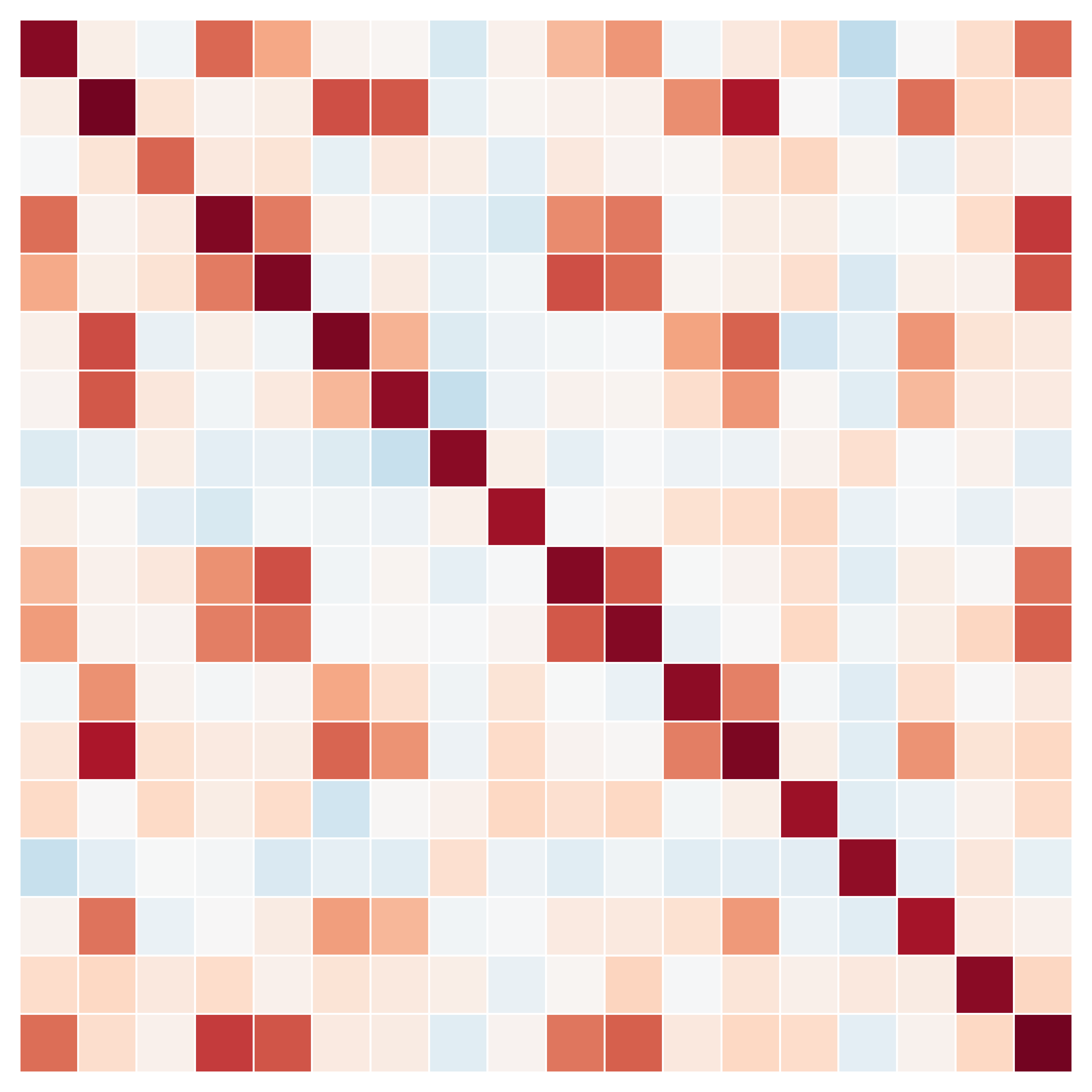}
        \caption{\texttt{Abstract} LSTM}\label{fig:rsa_ct_lstm:ab}
    \end{subfigure}\hfill
    \begin{subfigure}{0.48\linewidth}
        \includegraphics[width=\linewidth]{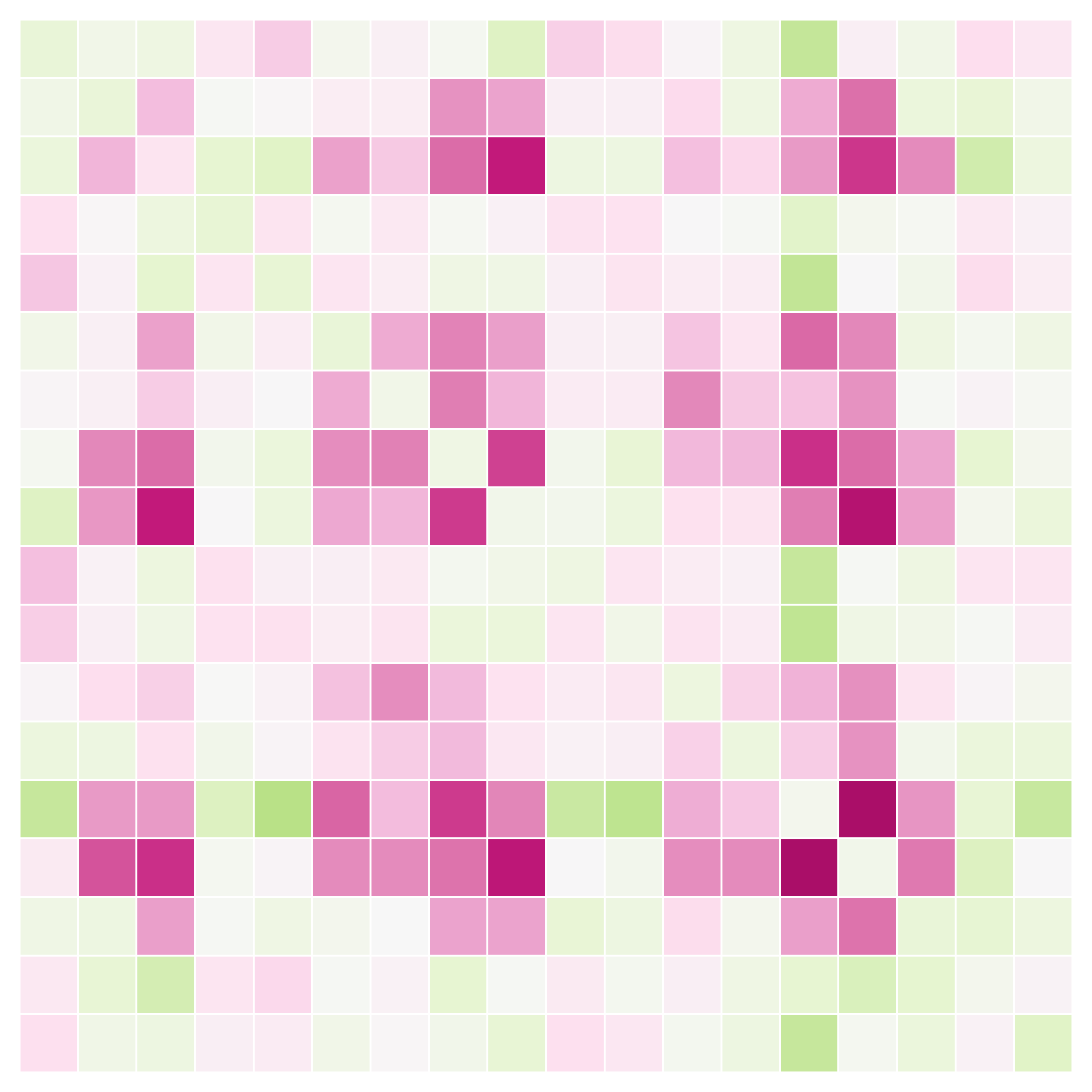}
        \caption{\texttt{Abstract} LSTM $-$ human}\label{fig:rsa_ct_lstm:ab_diff}
    \end{subfigure}
\end{minipage}\hspace{0.02\linewidth}
\begin{minipage}{0.2\linewidth}
    \centering
    \includegraphics[scale=0.5]{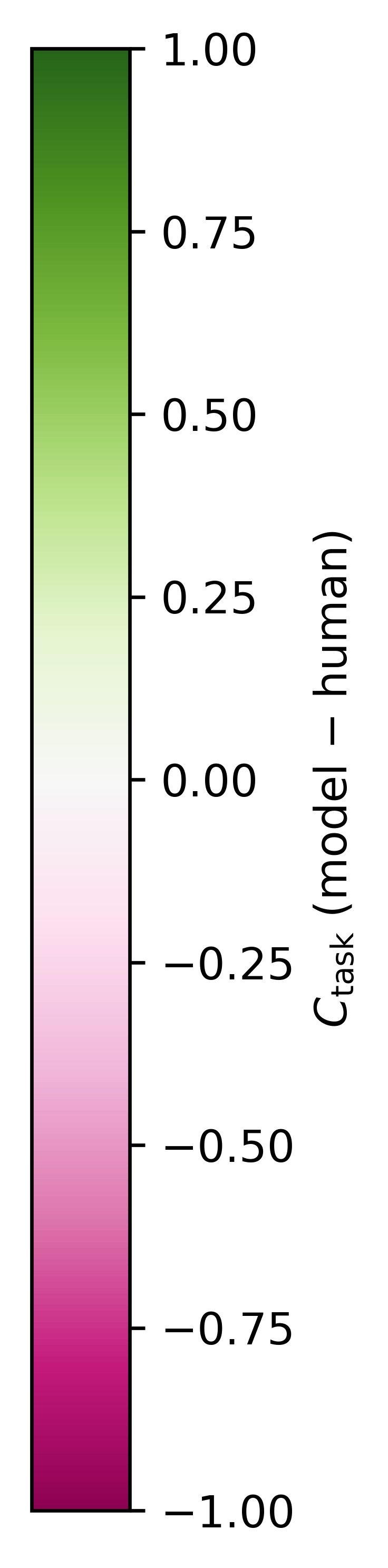}
\end{minipage}
\caption{\textbf{Cross-task response structure of humans and LSTM models.}
(a) Human cross-task matrix $\mathbf{C}_{\mathrm{task}}$ (Appendix~\ref{sec:appendix/rsa}); rows and columns are the 18 task conditions, in the same order in every panel.
(b, d) $\mathbf{C}_{\mathrm{task}}$ of the \texttt{Natural} LSTM (ResNet-CLIP) and of the \texttt{Abstract} LSTM at its 80\%-matched checkpoint, on the same scale as (a).
(c, e) Model minus human, on the shared scale at the right. Green cells are condition pairs the model correlates more strongly than humans do, pink cells less strongly.
The diagonal is the split-half reliability of each condition, not 1.}
\label{fig:rsa_ct_lstm}
\end{figure}

Fig.~\ref{fig:rsa_ct_lstm} shows the cross-task matrices $\mathbf{C}_{\mathrm{task}}$ of the human participants, the \texttt{Natural} LSTM and the \texttt{Abstract} LSTM, and the difference of each model's matrix from the human one.

\subsubsection{DSA}\label{sec:appendix/dsa}
For each task and model seed, we project the states onto their top $\mu$ principal components (time steps as samples, no whitening) and report results for $\mu \in \{2, 5, 10, 20, 50\}$. We then apply Hankel alternative view of Koopman (HAVOK) \citep{brunton2017chaos} with 2 delays and rank 3, selected by a preliminary grid search over delays $\{1,2\}$ and ranks $\{2,3\}$. Finally, we compute the angular Procrustes distance between the operators of every pair of model seeds after optimal orthogonal alignment (Adam, 200 iterations, learning rate $5\times10^{-3}$). Unlike the Wasserstein distance between eigenvalue spectra, which at rank 3 compares only three eigenvalues, it compares the full operators. We report the mean $\pm$ SEM over the 12 tasks and all seed pairs. Note that similar dynamics should not be read as similar recurrent weights, since different weights can produce similar dynamics and similar weights can produce different dynamics under different inputs.
\begin{figure}[ht]
  \centering
  \begin{minipage}{0.75\linewidth}
    \centering
    \begin{subfigure}{0.45\linewidth}
      \includegraphics[width=\linewidth]{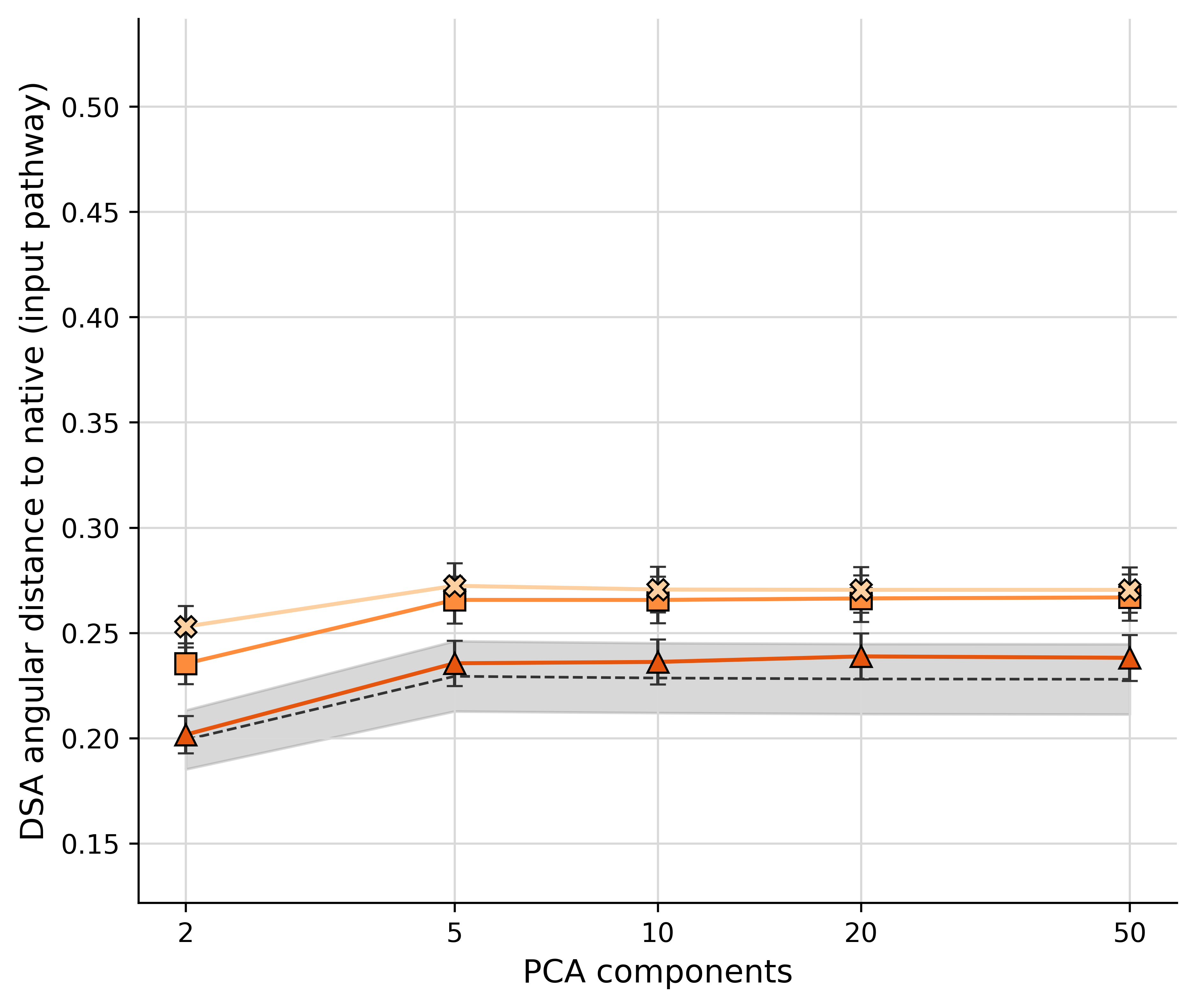}
      \caption{Abstract $\mathbf{z}_{\mathrm{proj}_t}$}\label{fig:dsa_pre:ab}
    \end{subfigure}\hfill
    \begin{subfigure}{0.45\linewidth}
      \includegraphics[width=\linewidth]{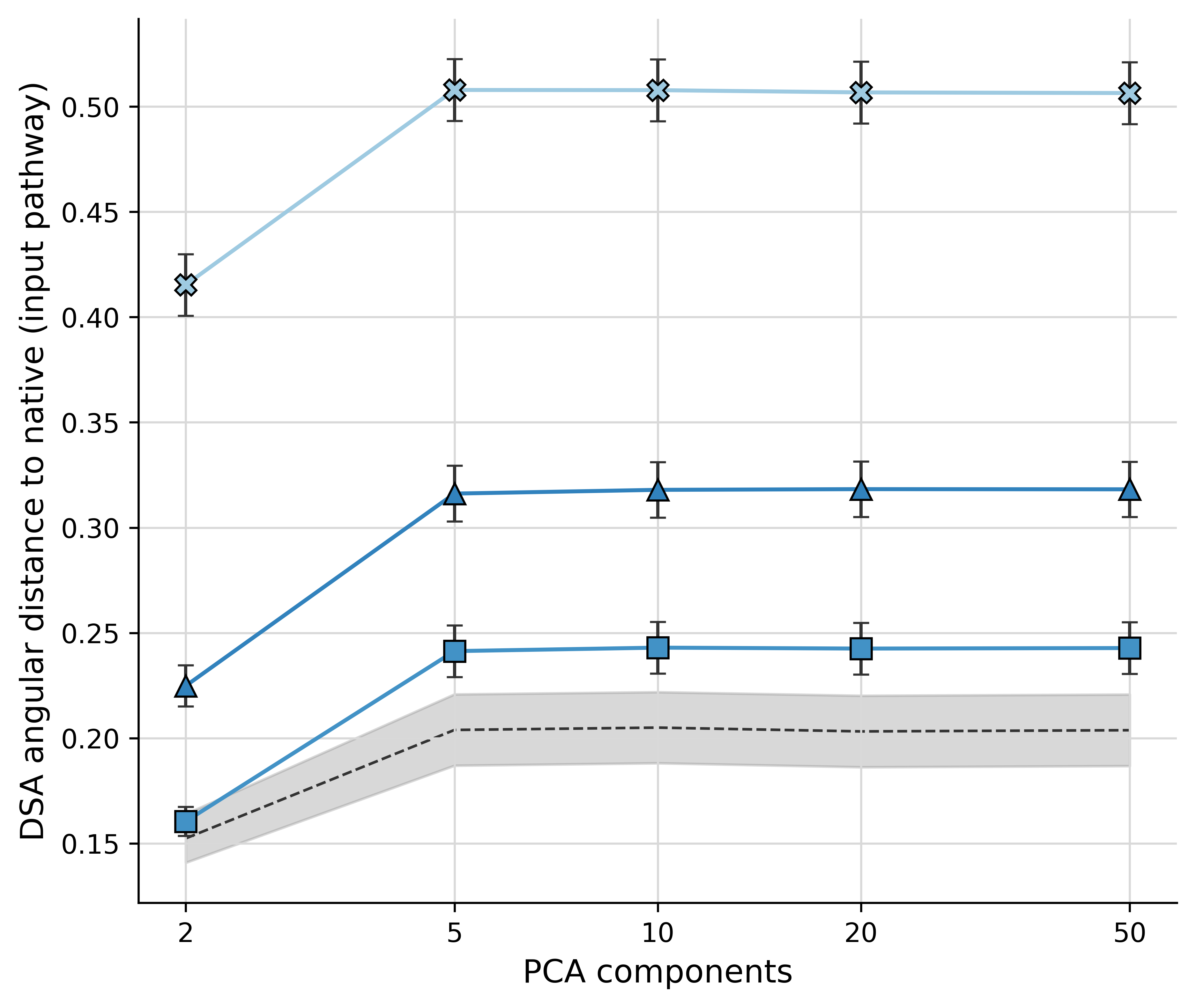}
    \caption{Natural $\mathbf{z}_{\mathrm{proj}_t}$}\label{fig:dsa_pre:nat}
    \end{subfigure}
  \end{minipage}
  \begin{minipage}{0.2\textwidth}
    \centering
    \includegraphics[scale=0.5]{plots/dsa/final/legend.png}
    \end{minipage}
    \caption{
  \textbf{Dynamic distances of retrained models to native models in the projected input.}
    Each panel shows the DSA angular distance between each retrained model and the native models of the same input modality, computed on the projected input $\mathbf{z}_{\mathrm{proj}_t}$, against the number of PCA components.
    (a). In models that receive abstract input,
    $\theta$ is at the noise floor at every dimension, while $\theta+\mathbf{W}_{\mathrm{out}}$ and the untrained $\theta$ lie just above it.
    (b). In models that receive ResNet embedding input,
    $\theta+\mathbf{W}_{\mathrm{out}}$ lies at or just above the noise floor, $\theta$ is well above it, and the untrained $\theta$ is farthest.
    Closeness in $\mathbf{z}_{\mathrm{proj}_t}$ does not predict closeness in the hidden state $\mathbf{h}_t$ (Fig.~\ref{fig:dsa}). With abstract input, $\theta$ is at the noise floor here but not in $\mathbf{h}_t$. With ResNet input, $\theta$ is far from native here but at the noise floor in $\mathbf{h}_t$, while $\theta+\mathbf{W}_{\mathrm{out}}$ shows the reverse.}\label{fig:dsa_pre}
\end{figure}
\begin{figure}[ht]
  \centering
  \begin{minipage}{0.75\linewidth}
    \centering
    \begin{subfigure}{0.45\linewidth}
      \includegraphics[width=\linewidth]{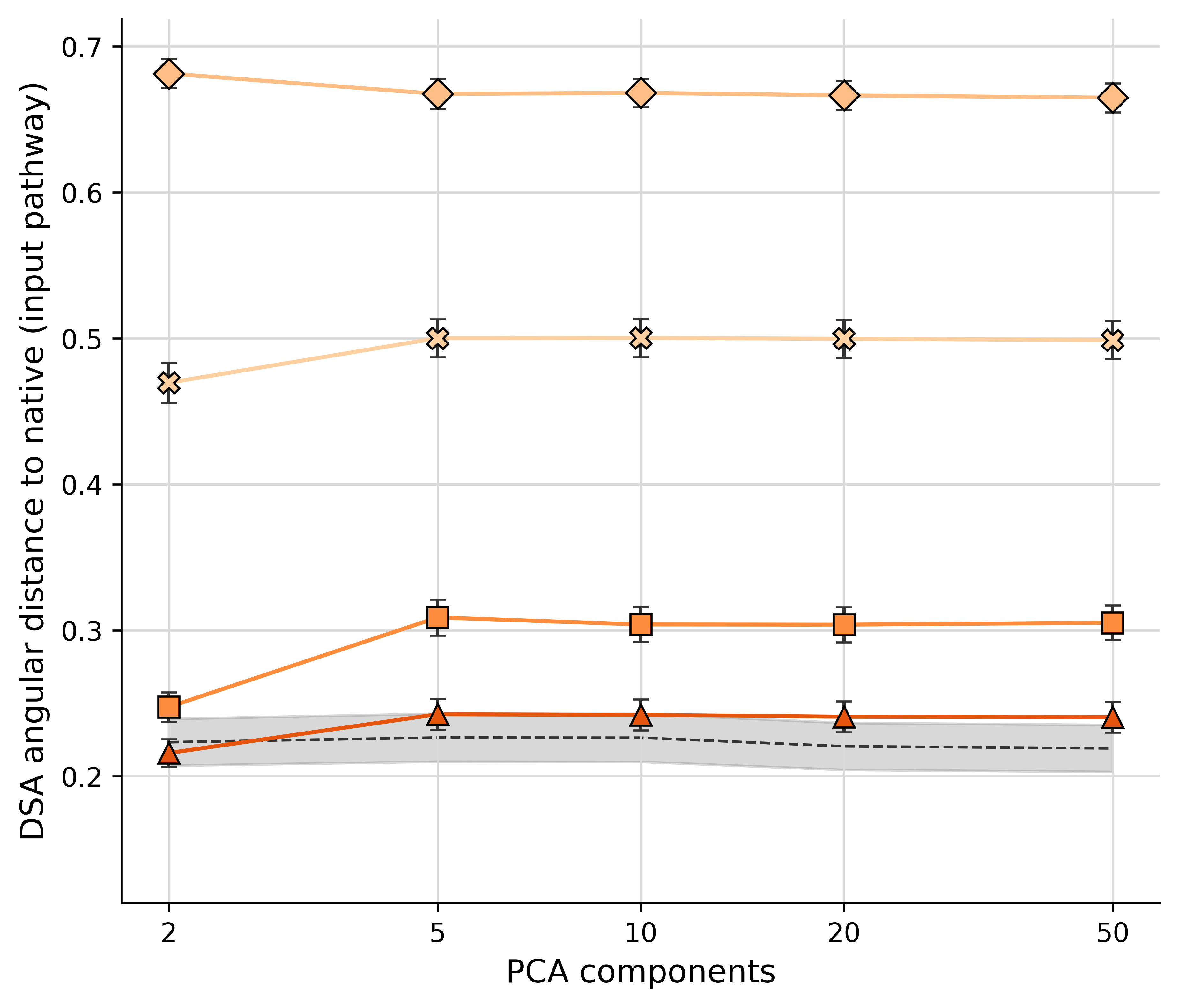}
      \caption{Abstract $\mathbf{z}_{\mathrm{proj}_t}$}\label{fig:dsa_dm:ab_pre}
    \end{subfigure}\hfill
    \begin{subfigure}{0.45\linewidth}
      \includegraphics[width=\linewidth]{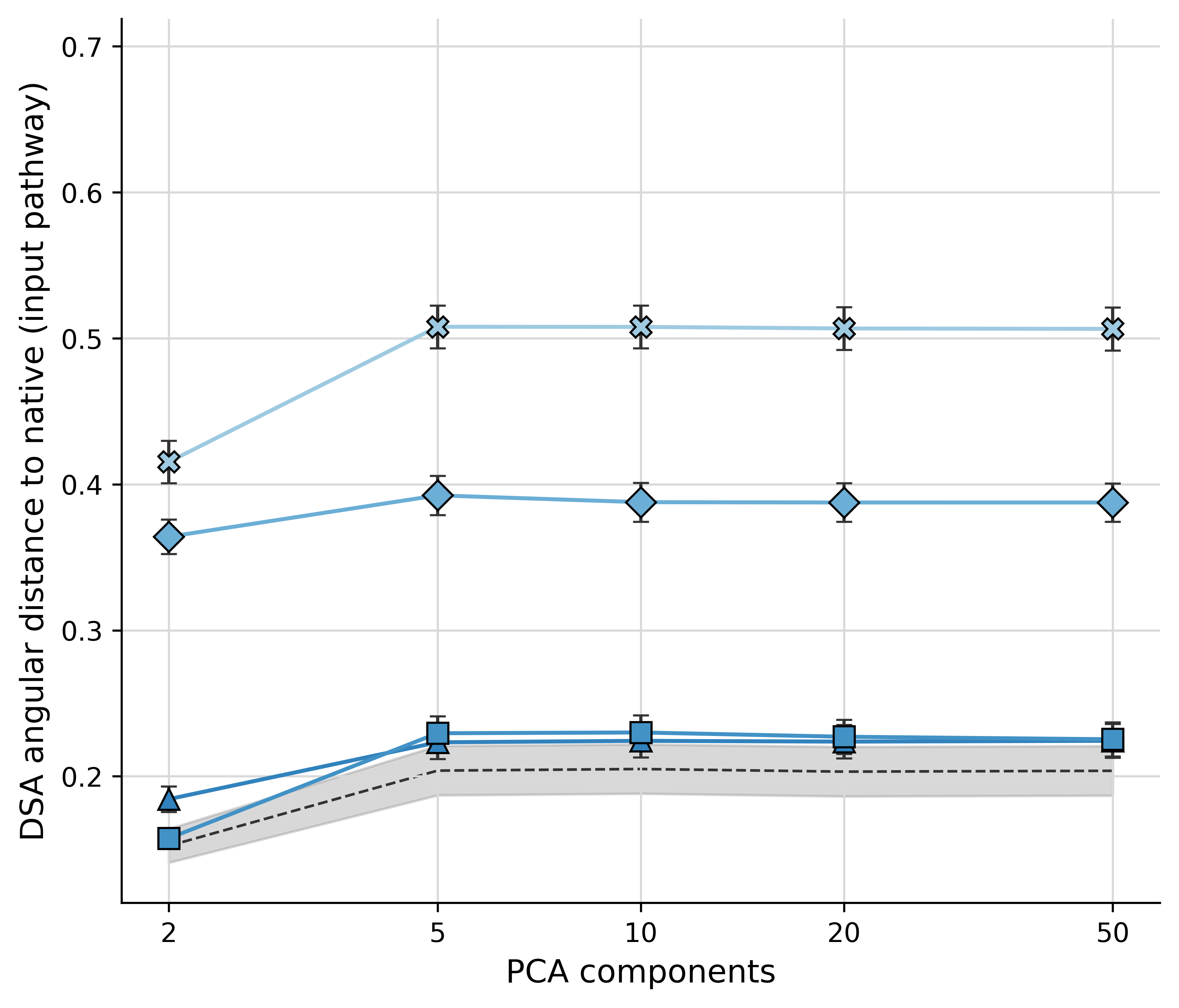}
    \caption{Natural $\mathbf{z}_{\mathrm{proj}_t}$}\label{fig:dsa_dm:nat_pre}
    \end{subfigure}
    \par\vspace{0.5em}
    \begin{subfigure}{0.45\linewidth}
      \includegraphics[width=\linewidth]{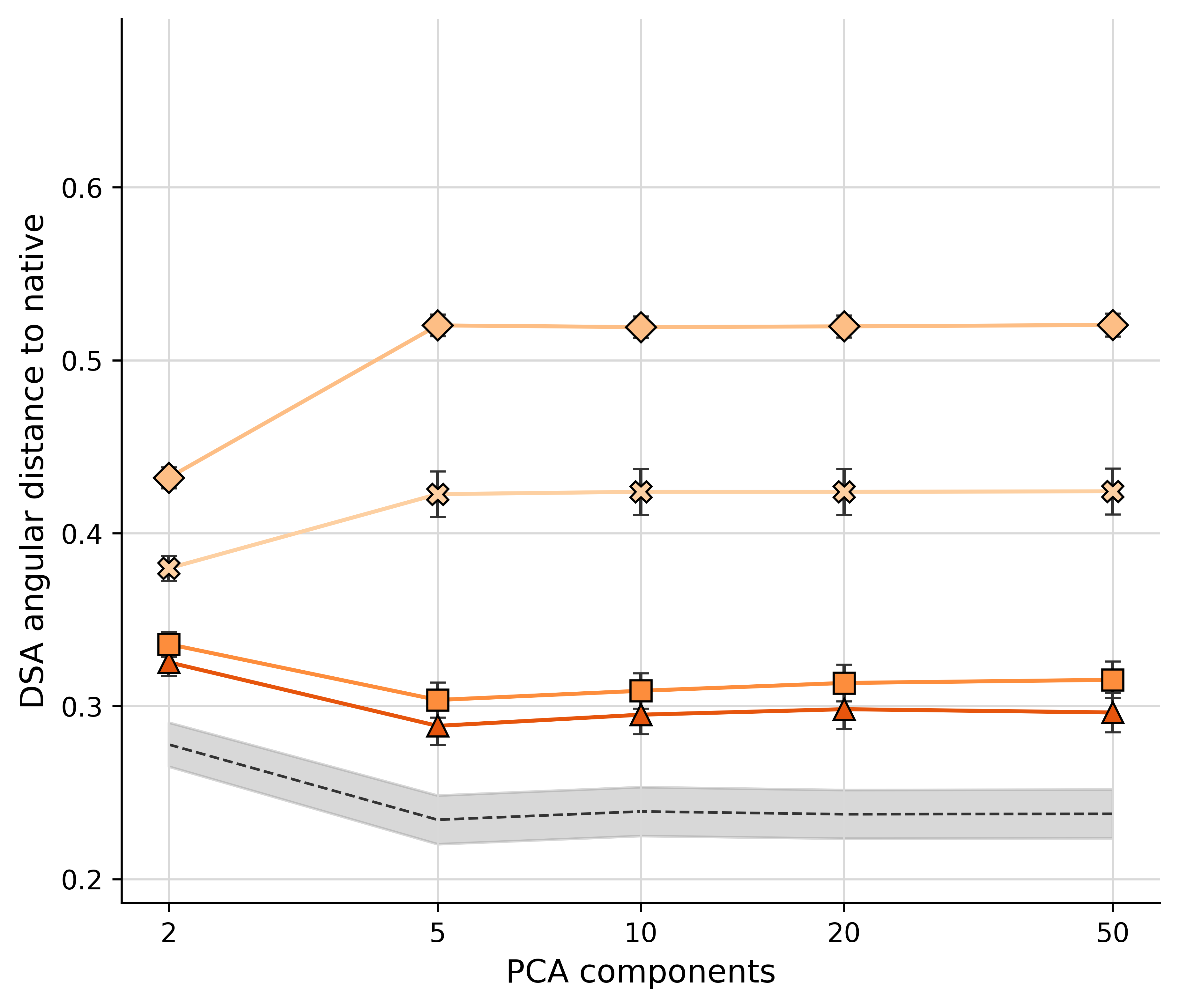}
      \caption{Abstract $\mathbf{h}_t$}\label{fig:dsa_dm:ab_seq}
    \end{subfigure}\hfill
    \begin{subfigure}{0.45\linewidth}
      \includegraphics[width=\linewidth]{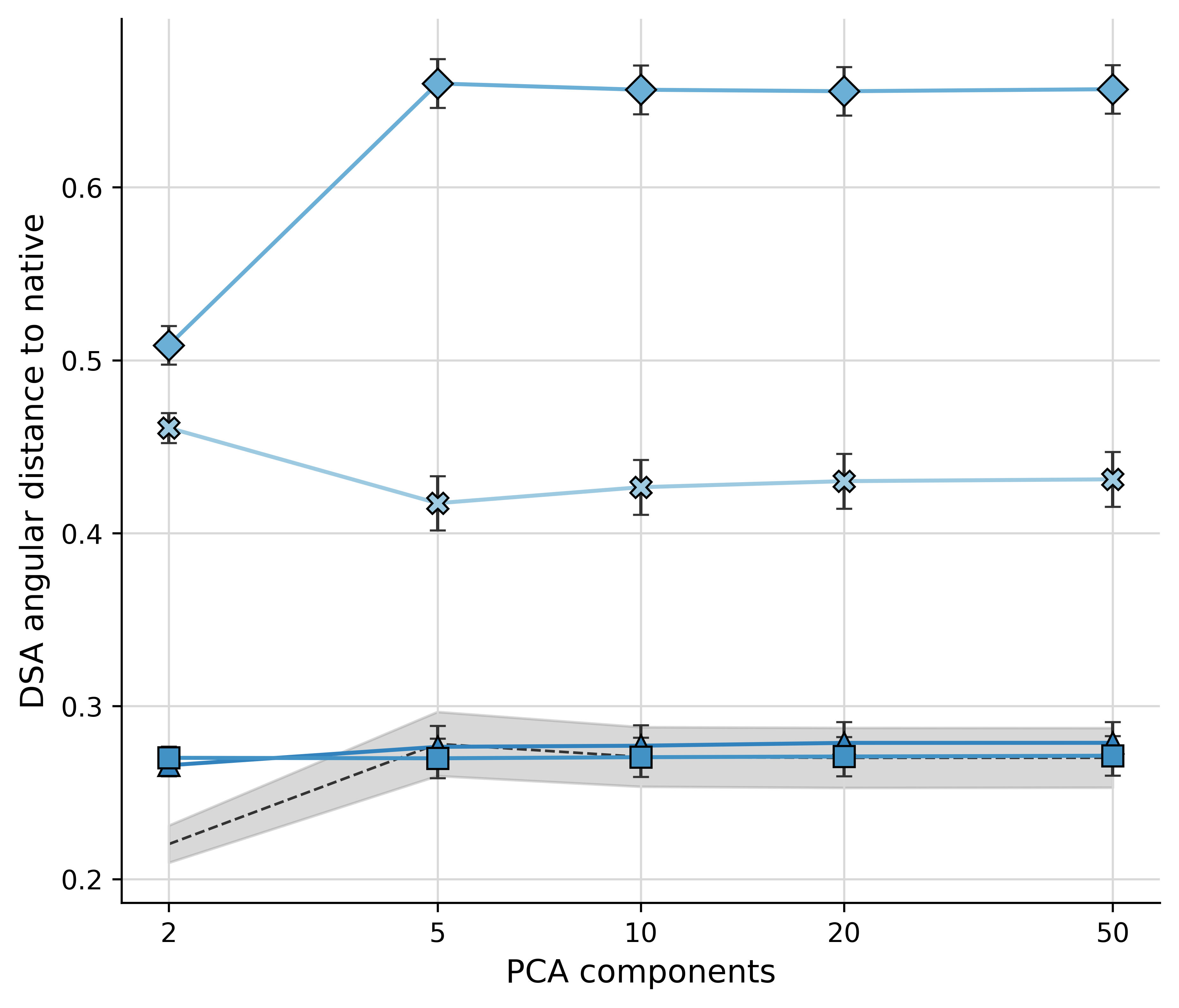}
    \caption{Natural $\mathbf{h}_t$}\label{fig:dsa_dm:nat_seq}
    \end{subfigure}
  \end{minipage}
  \begin{minipage}{0.2\textwidth}
    \centering
    \includegraphics[scale=0.5]{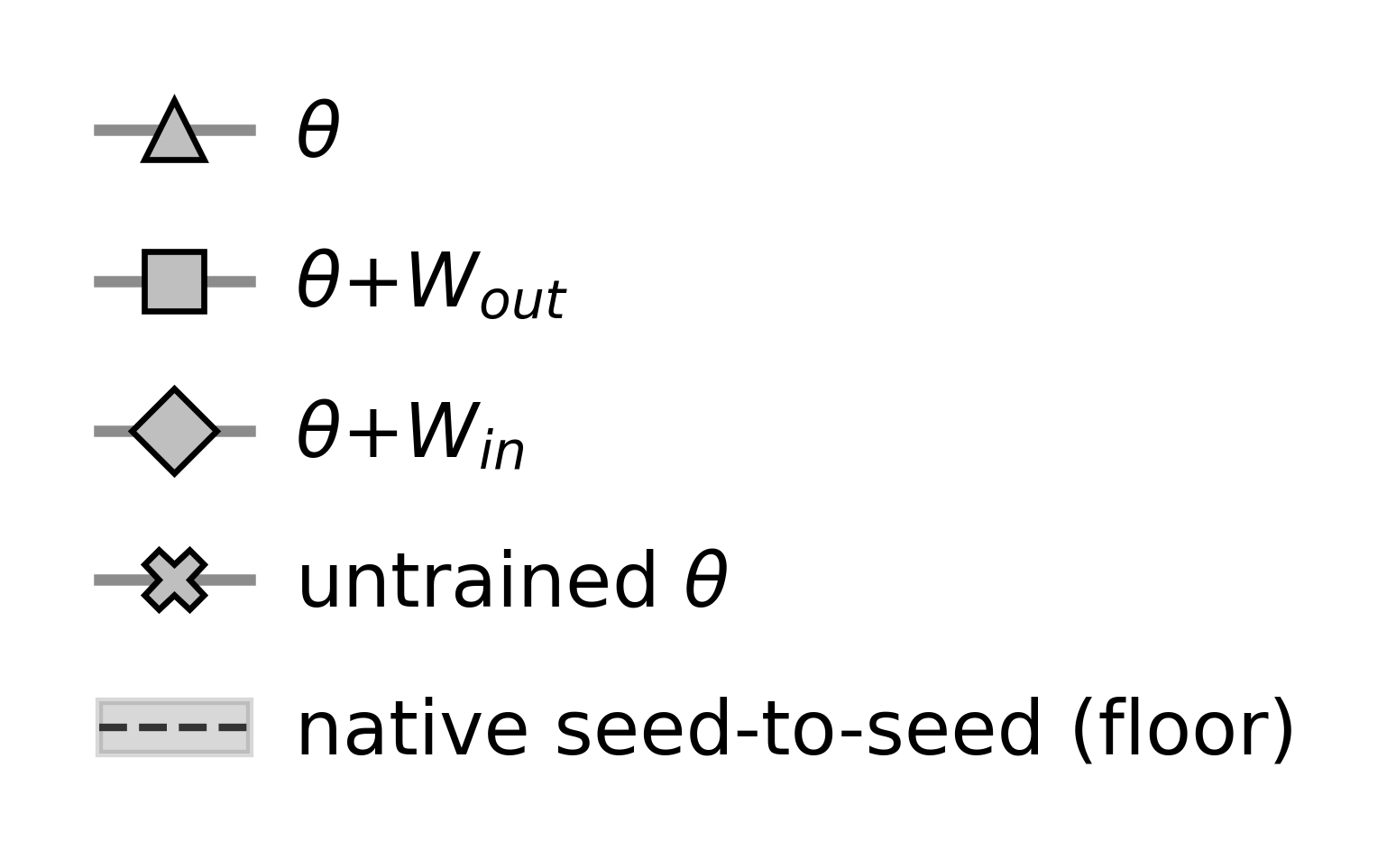}
    \end{minipage}
  \caption{
\textbf{Dynamic distances of retrained dimensionality-matched models to native models.}
    Each panel shows the DSA angular distance between each retrained model and the native models of the same input modality, against the number of PCA components, computed on the projected input $\mathbf{z}_{\mathrm{proj}_t}$ (a, b) or the hidden state $\mathbf{h}_t$ (c, d).
    (a). In models that receive dimensionality-matched abstract input,
    $\theta$ is at the noise floor in $\mathbf{z}_{\mathrm{proj}_t}$ at every dimension and $\theta+\mathbf{W}_{\mathrm{out}}$ lies above it, while the untrained $\theta$ and $\theta+\mathbf{W}_{\mathrm{in}}$ are far above it, with $\theta+\mathbf{W}_{\mathrm{in}}$ farthest.
    (b). In models that receive ResNet embedding input,
    $\theta$ and $\theta+\mathbf{W}_{\mathrm{out}}$ are at the noise floor in $\mathbf{z}_{\mathrm{proj}_t}$ from 5 dimensions on, while $\theta+\mathbf{W}_{\mathrm{in}}$ and the untrained $\theta$ are far above it, with the untrained $\theta$ farthest.
    (c). In $\mathbf{h}_t$ of models that receive abstract input,
    $\theta$ and $\theta+\mathbf{W}_{\mathrm{out}}$ lie above the noise floor at every dimension but well below the untrained $\theta$, and $\theta+\mathbf{W}_{\mathrm{in}}$ is farthest.
    (d). In $\mathbf{h}_t$ of models that receive ResNet embedding input,
    $\theta$ and $\theta+\mathbf{W}_{\mathrm{out}}$ are at the noise floor from 5 dimensions on, while the untrained $\theta$ sits far above it and $\theta+\mathbf{W}_{\mathrm{in}}$ is farthest.
  }\label{fig:dsa_dm}
\end{figure}
\begin{figure}
\centering

\begin{subfigure}{0.48\textwidth}
    \includegraphics[width=\linewidth]{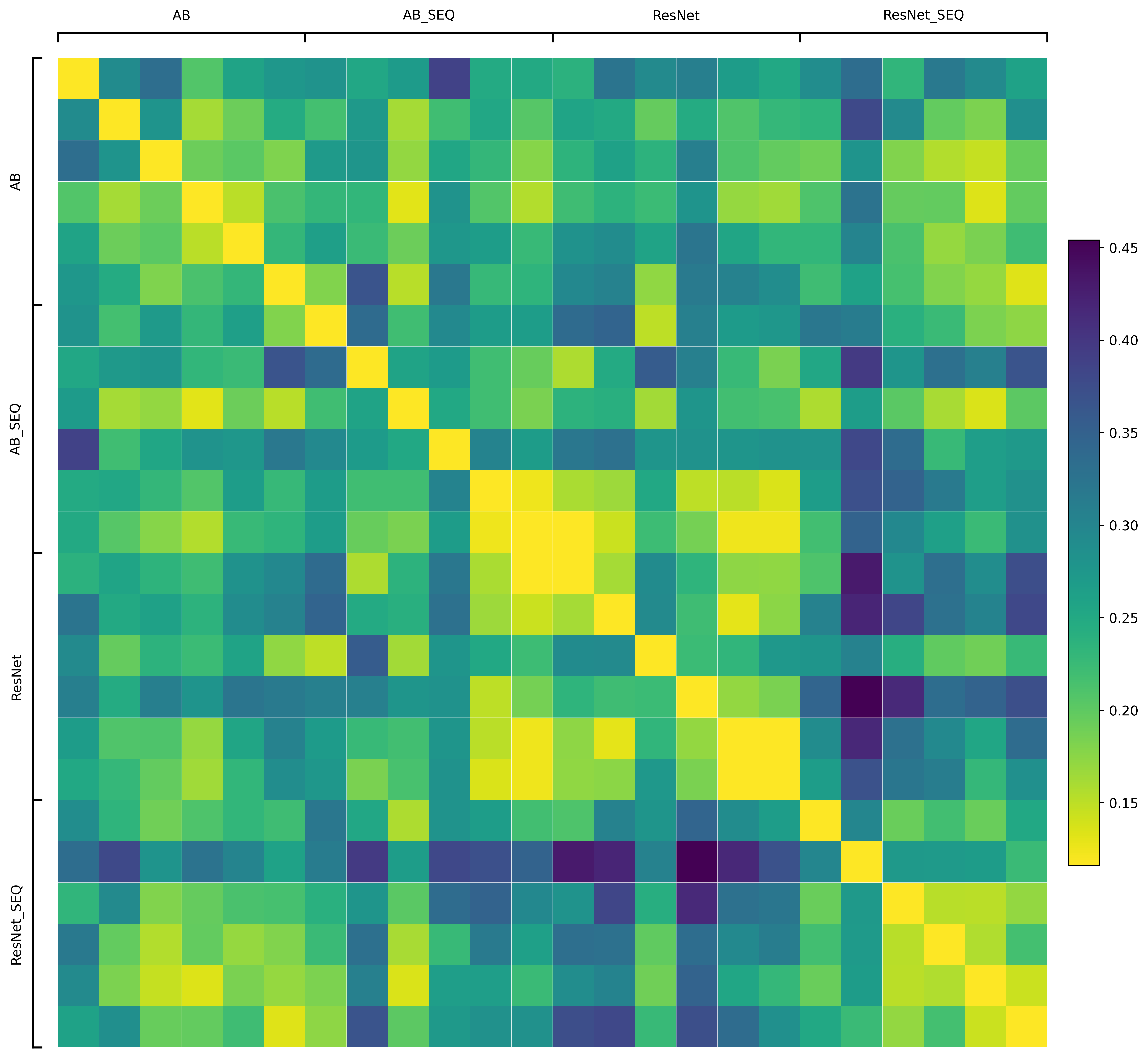}
    \caption{}
\end{subfigure}
\hfill
\begin{subfigure}{0.48\textwidth}
    \includegraphics[width=\linewidth]{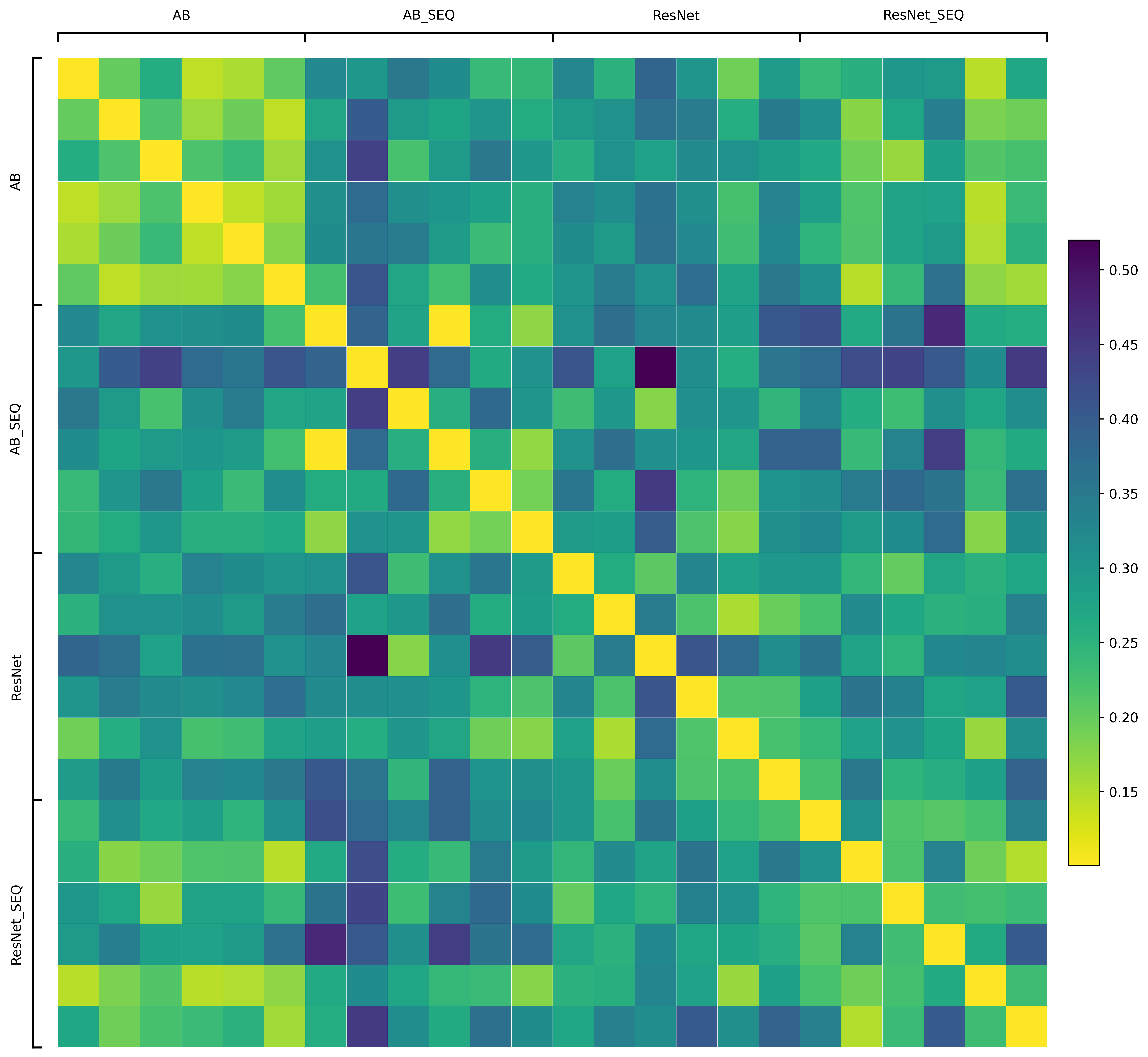}
    \caption{}
\end{subfigure}

\vspace{0.5em}

\begin{subfigure}{0.48\textwidth}
    \includegraphics[width=\linewidth]{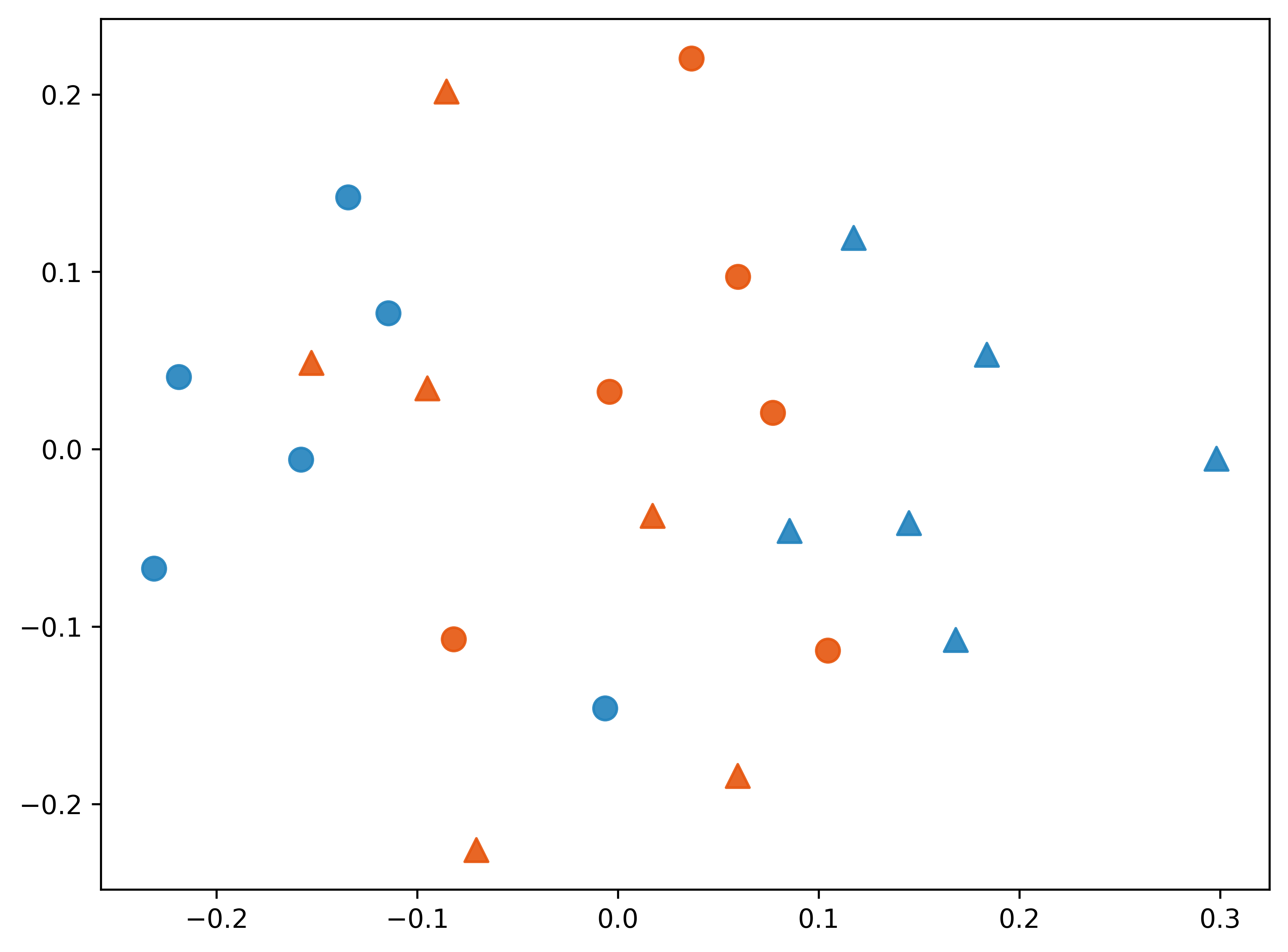}
    \caption{}
\end{subfigure}
\hfill
\begin{subfigure}{0.48\textwidth}
    \includegraphics[width=\linewidth]{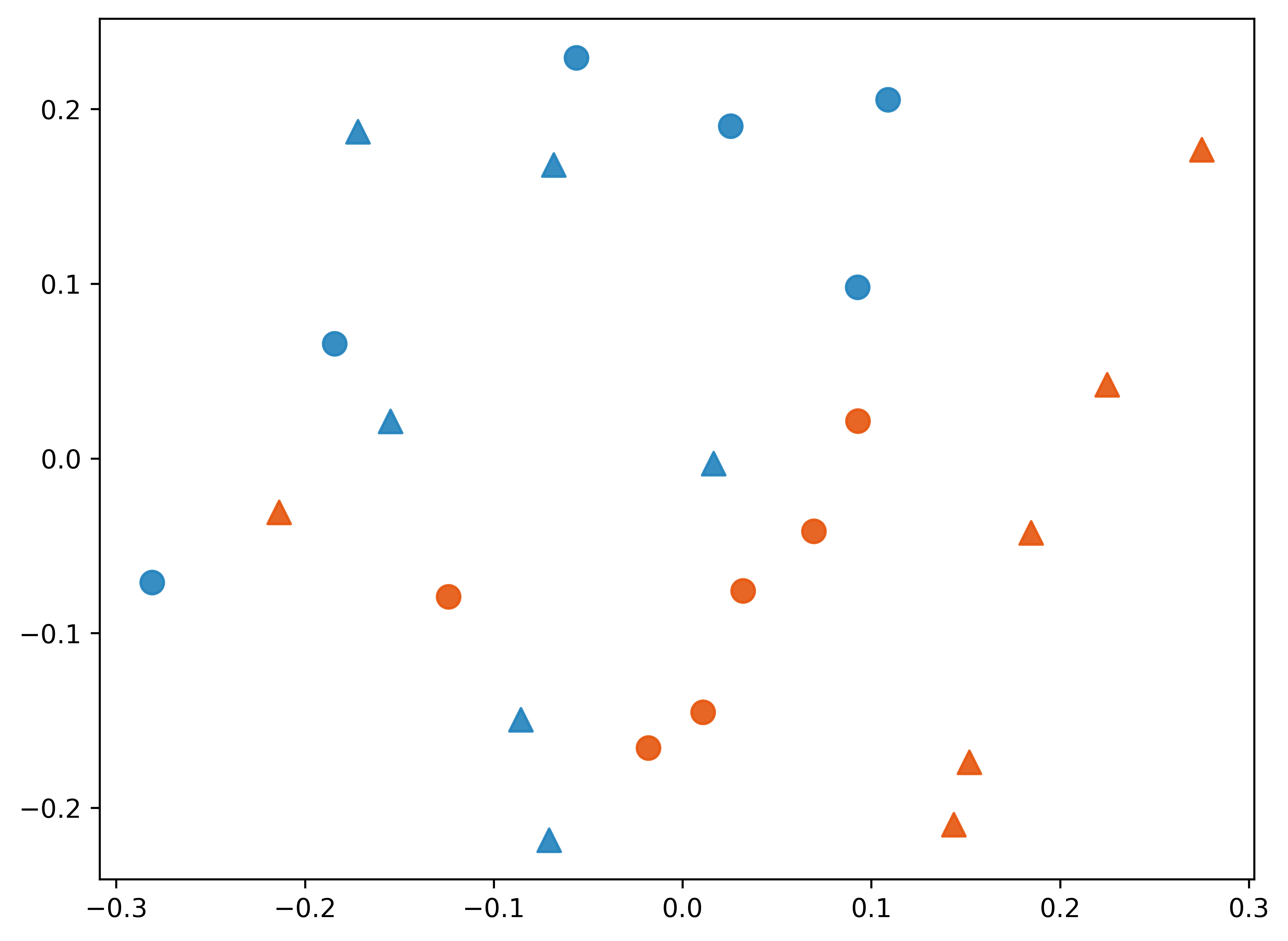}
    \caption{}
\end{subfigure}

\caption{\textbf{Pairwise DSA distances between native and substituted models.} (a) DSA pairwise distance heatmap on $\mathbf{z}_{\mathrm{proj}_t}$. (b) DSA pairwise distance heatmap on $\mathbf{h}_t$. (c) MDS \citep{kruskal1964multidimensional} applied to the pre-computed pairwise distance matrix of $\mathbf{z}_{\mathrm{proj}_t}$. (d) MDS applied to the pairwise distance matrix of $\mathbf{h}_t$.}
\label{fig:dsa_big_picture}
\end{figure}

Fig.~\ref{fig:dsa_big_picture} shows the pairwise DSA distances between the native and substituted models of Sec.~\ref{sec:results3} (untrained $\theta$ omitted), averaged over the 12 tasks.

\paragraph{Sensitivity to the DSA settings.}
We repeated the analysis of Sec.~\ref{sec:results3} in 19 configurations, varying one setting at a time around the default above with $\mu=10$: $\mu \in \{2, 5, 10, 20, 50\}$; 1 or 3 delays; rank 2 or 4; the Euclidean Procrustes distance or the Wasserstein distance between eigenvalue spectra instead of the angular distance; 500 or 1000 iterations at learning rate $5\times10^{-3}$, or 1000 iterations at $10^{-3}$ or $10^{-2}$; and random seeds 0, 7, 42, and 2025 instead of 2026.
We only kept configurations that our trials support: with $d$ delays, a 6-frame trial gives $6-d$ one-step transitions, so the rank cannot exceed $6-d$, which rules out ranks above 4 with 2 delays and above 3 with 3 delays.
We did not vary PCA whitening.
HAVOK fits the operator on the leading right singular vectors of the Hankel matrix, which form an orthonormal basis regardless of whether the input components are whitened, so whitening changes only which rank-3 subspace is kept.
Because whitening gives every component unit variance, that subspace no longer follows the dominant directions of the trajectories, so whitened DSA measures a different quantity, and one that depends on $\mu$.

For each comparison and configuration, we test the per-task distances with a one-sided paired Wilcoxon signed-rank test over the 12 tasks ($p<0.05$), where the distance between two models is the mean over all pairs of their seeds and a model's seed floor is the mean over pairs of its own seeds.
A model is within a floor when its distance is not significantly above it, which is a failure to reject rather than evidence of equivalence.
The optimizer settings and random seeds change every task-averaged distance by at most 0.012 and leave every comparison in Table~\ref{tab:dsa_sensitivity} unchanged relative to the default setting, so the table and the counts below cover the other 11 configurations.
\begin{table}[ht]
\centering
\caption{\textbf{DSA sensitivity, number of configurations in which each comparison holds.} Each column varies one setting around the default setting of the main text ($\mu=10$, 2 delays, rank 3, angular distance), which is counted once, in the $\mu$ column. The floor is the mean distance between seeds of the native model of the input modality; ``within'' means not significantly above it (one-sided paired Wilcoxon signed-rank test over the 12 tasks, $p<0.05$). The 8 optimizer and seed configurations give the default outcome in every row and are not shown.}
\label{tab:dsa_sensitivity}
\small
\begin{tabular}{lcccccc}
\toprule
Comparison & Feature & $\mu$ & Delays & Rank & Distance & Total \\
 & & 2--50 & 1, 3 & 2, 4 & Eucl./Wass. & \\
\midrule
\multicolumn{7}{l}{\emph{Natural input}} \\
$\theta_{\mathrm{Abs}\to\mathrm{Nat}}$ within floor & $\mathbf{h}_t$ & 4/5 & 1/2 & 1/2 & 2/2 & 8/11 \\
Untrained $\theta$ farther than $\theta_{\mathrm{Abs}\to\mathrm{Nat}}$ & $\mathbf{h}_t$ & 5/5 & 2/2 & 1/2 & 1/2 & 9/11 \\
$\theta_{\mathrm{Abs}\to\mathrm{Nat}}$ above floor & $\mathbf{z}_{\mathrm{proj}_t}$ & 5/5 & 2/2 & 0/2 & 2/2 & 9/11 \\
$(\theta+\mathbf{W}_{\mathrm{out}})_{\mathrm{Abs}\to\mathrm{Nat}}$ within floor & $\mathbf{z}_{\mathrm{proj}_t}$ & 3/5 & 1/2 & 1/2 & 1/2 & 6/11 \\
Dim.-matched $\theta$, $\theta+\mathbf{W}_{\mathrm{out}}$ within floor & $\mathbf{h}_t$ & 4/5 & 1/2 & 1/2 & 2/2 & 8/11 \\
\midrule \multicolumn{7}{l}{\emph{Abstract input}} \\
$\theta_{\mathrm{Nat}\to\mathrm{Abs}}$ within floor & $\mathbf{z}_{\mathrm{proj}_t}$ & 5/5 & 2/2 & 1/2 & 2/2 & 10/11 \\
\bottomrule
\end{tabular}
\end{table}

Under natural input, $\theta_{\mathrm{Abs}\to\mathrm{Nat}}$ stays within the \texttt{Natural} floor in 8 of the 11 configurations and leaves it with $\mu=2$, 3 delays or rank 2, where its distance above the floor is 21--60\% of the untrained $\theta$'s (11\% at the default setting); the dimensionality-matched $\theta$ and $\theta+\mathbf{W}_{\mathrm{out}}$ fail in the same three.
An untrained $\theta$ stays farther from the native models in 9 (all but rank 2 and the Wasserstein distance), and the projected input of $\theta_{\mathrm{Abs}\to\mathrm{Nat}}$ stays above the floor in 9 (all but ranks 2 and 4).
$(\theta+\mathbf{W}_{\mathrm{out}})_{\mathrm{Abs}\to\mathrm{Nat}}$ is above the floor in all 11, but its projected input is within the floor in only 6.
The dimensionality-matched $\theta+\mathbf{W}_{\mathrm{in}}$ is above the floor in all 11 for both input modalities, but it is not consistently farther from the native models than an untrained $\theta$: with natural input the comparison is borderline ($p=0.055$ at the default setting, 5 of 11 configurations and 5 of the 8 optimizer and seed reruns), and with abstract input it holds in 2 of 11.
Under abstract input, $\theta_{\mathrm{Nat}\to\mathrm{Abs}}$ is above the \texttt{Abstract} floor in all 11, and its projected input is within that floor in 10 (all but rank 2).
It is closer to the native \texttt{Abstract} models than an untrained $\theta$ on average (0.300 vs.\ 0.350 at the default setting) but not consistently across tasks (1 of 11), whereas the dimensionality-matched $\theta_{\mathrm{Nat}\to\mathrm{Abs}}$ is closer in 8 of the 11 configurations.

\subsection{Training Settings}
We detail training configuration in Table \ref{tab:native-training-config}. For dimensionality matched and visual difficulty matched \texttt{Abstract} models, their input dimension is 171 and they are trained for 10 epochs; everything else stayed the same as \texttt{Natural} models. Substituted models use the training recipe of the native models of the modality they are retrained on (e.g., $\theta_{\mathrm{Nat}\to\mathrm{Abs}}$ uses the \texttt{Abstract} recipe).
\begin{table}
\centering
\caption{\textbf{Training configuration.}}
\label{tab:native-training-config}
\small
\begin{tabular}{p{0.42\linewidth}p{0.45\linewidth}}
\hline
\textbf{Item} & \textbf{Value} \\
\hline
Input dimension & $171$ for natural models (image embedding of size $128$ + one-hot task embedding of size $43$), $128$ for abstract models\\
Hidden dimension & RNN: $256$, All other models: $128$ \\
Layers & 1 \\
Learning rate & $D\in\{64, 128, 256\}$: 3e-4 for \texttt{Natural} models, 3e-3 for \texttt{Abstract} models\par
$D\in\{512, 1024\}$: 1e-4 for \texttt{Natural} models, 3e-4 for \texttt{Abstract} models  \\
Label smoothing & 0.2 for natural models, none for abstract \\
Attention head & 8 \\
Optimizer & AdamW \\
Loss function & Cross-entropy loss \\
Epochs & $10$ for \texttt{Abstract} models, $20$ for \texttt{Natural} models \\
Train size & 256000\\
Validation size & 64000 \\
Batch size & 128 \\
Weight decay & 1e-4 \\
Dropout & 0.1 for $\mathrm{FC_{in}}$ (\texttt{Natural} only), $0.3$ for $\mathrm{FC_{out}}$ \\
Gradient clipping & 2.0 \\
GPU & \texttt{1 x NVidia H100 SXM5 (80GB)} for \texttt{Natural}, None for \texttt{Abstract} \\
Execution time & 4 hours for \texttt{Natural}, 1 hour for \texttt{Abstract} \\
\hline
\end{tabular}
\end{table}

\subsection{VLM Evaluation Details}\label{sec:appendix/vlms}

We evaluated 4 state-of-the-art models: Gemini 3.1 Pro Preview \citep{google2026gemini31pro}, Qwen3 VL 235B A22B Thinking, Qwen3 VL 30B A3B Thinking \citep{qwen3technicalreport}, and Nemotron Nano 12B 2 VL \citep{nvidia2025nvidianemotronnanov2}. All models were accessed via OpenRouter API. All 4 models were tested in reasoning mode, while Nemotron was additionally evaluated without reasoning by adding \texttt{no-thinking} to the system prompt for comparison against the reasoning baseline.

To ensure a fair and rigorous comparison between human participants and Vision-Language Models (VLMs), we designed an evaluation pipeline that closely emulated the interactive, sequential nature of the behavioral experiment. This section details the prompting structure and trial emulation methods used for all evaluated VLMs.

VLMs were provided with the same task instructions as human participants. For all the trials, images and instructions from previous steps were kept in the context of the VLMs. To ensure that the models strictly adhered to the two-alternative forced-choice format, we implemented an iterative response filtering and retry mechanism: any trial where the VLM generated a response at an action frame that was not exactly ``x'' or ``b'' was removed and rerun until the model successfully produced a correctly formatted response.

\subsubsection{Sequential Trial Emulation}

Unlike standard VLM benchmarks where an entire task is presented in a single prompt, our pipeline presented each trial sequentially. For each trial, the VLM maintained a continuous context window across the sequence of frames. 

At the start of every trial, the model was initialized with a System Prompt that contained the global task instructions. Then, for each frame in the sequence ($t=1, \dots, T$):
\begin{enumerate}
    \item The model was provided with the visual stimulus (the image for frame $t$, encoded in base64).
    \item The model was provided with a Per-Frame Instruction (text prompt) corresponding to the current step in the trial logic. 
\end{enumerate}

These text prompts were sourced identically from the instructions provided to human participants during their interactive learning sessions. At action frames (where a response was required), the model was forced to output its decision based on the history of frames observed so far.

\subsubsection{Prompting Structure}

\paragraph{System Prompts}
The system prompt served as the global task instruction, mirroring the ``Welcome Screen'' text shown to human participants before beginning the experiment. To format this for VLMs, we made minor adjustments (e.g. replacing ``Press X'' with ``Respond 'x'''). The model was explicitly constrained to output only the target characters.

For example, the system prompt for the Contextual Delayed Match-to-Sample (Category $\to$ Object/Position) task was:

{\footnotesize
\begin{verbatim}
• You will perform a Contextual Delayed Match-to-Sample task.

• Each trial consists of a sequence of 3 images. You will make two 
  comparisons, the second comparison depends on the result of the first:
  1) Category match:
     - Compare the 2nd frame with the 1st frame.
     - Remember if they are from the same category (for example, both planes).
  2) Identity/Position match:
     - Compare the 3rd frame with the 2nd frame.
     - If first comparison was a match, then compare their IDENTITY:
     - Otherwise, compare their POSITION:
     - If they match according to the relevant feature, respond 'x'.
     - Otherwise, respond 'b'.

• The categories include Cars, Planes, Boats, Fruits, Faces, Animals, 
  Chairs, and Tables.

Only output 'x' or 'b'.
\end{verbatim}
}
\textit{(A similar structure was used across all 12 tasks, accurately reflecting their specific task rules and contingencies).}

\paragraph{Per-Frame Instructions}
To emulate the learning phase, the VLM was given explicit, step-by-step instructions alongside each image frame. These per-frame instructions guided the model on whether it should encode the current stimulus into working memory or execute a comparison. 

For instance, in an Interleaved Delayed Match-to-Sample ABAB Position--Position trial, the 4 frames were presented sequentially with the following instructions:

\begin{itemize}
    \item Frame 1 (A):
    \begin{itemize}
        \item \textit{Image:} [Stimulus Image 1]
        \item \textit{Instruction:} \texttt{Remember the POSITION of this object.}
    \end{itemize}
    \item Frame 2 (B):
    \begin{itemize}
        \item \textit{Image:} [Stimulus Image 2]
        \item \textit{Instruction:} \texttt{Remember the POSITION of this object.}
    \end{itemize}
    \item Frame 3 (A - Action Frame):
    \begin{itemize}
        \item \textit{Image:} [Stimulus Image 3]
        \item \textit{Instruction:} \texttt{Compare POSITION with the FIRST image. Respond 'x' if same, 'b' if different.}
    \end{itemize}
    \item Frame 4 (B - Action Frame):
    \begin{itemize}
        \item \textit{Image:} [Stimulus Image 4]
        \item \textit{Instruction:} \texttt{Compare POSITION with the SECOND image. Respond 'x' if same, 'b' if different.}
    \end{itemize}
\end{itemize}

By maintaining this strict sequential history and providing per-frame behavioral anchors, we ensured the VLMs experienced the temporal and cognitive demands of the tasks as close to human participants as possible.
\begin{table}
\centering
\small
\caption{\textbf{Complete listing of all 43 tasks.} Each task is defined by a
family-specific parameter (memory lag $n$ for NBack, interleaving pattern
for InterDMS) and its comparison dimension(s); for CtxDMS the arrows give
the feature ordering. Raw frames are the natural sequence length; all tasks
are padded to 6 frames in multi-task training. Action frames are the frames
that require a response (1-based).}
\label{tab:all_tasks}
\begin{tabular}{@{}c c c c c l@{}}
\toprule
\textbf{Family} & \textbf{\# of frames} & \textbf{Param.} & \textbf{Action frame} & \textbf{ID} & \textbf{Comparison dimension} \\
\midrule
\multirow{9}{*}{NBack} & \multirow{9}{*}{6}
  & \multirow{3}{*}{$n=1$} & \multirow{3}{*}{2--6} & 1 & position \\
& & & & 2 & identity \\
& & & & 3 & category \\
\cmidrule(l){3-6}
& & \multirow{3}{*}{$n=2$} & \multirow{3}{*}{3--6} & 4 & position \\
& & & & 5 & identity \\
& & & & 6 & category \\
\cmidrule(l){3-6}
& & \multirow{3}{*}{$n=3$} & \multirow{3}{*}{4--6} & 7 & position \\
& & & & 8 & identity \\
& & & & 9 & category \\
\midrule
\multirow{27}{*}{InterDMS} & \multirow{27}{*}{4}
  & \multirow{9}{*}{AABB} & \multirow{9}{*}{2, 4} & 10 & position, position \\
& & & & 11 & position, identity \\
& & & & 12 & position, category \\
& & & & 13 & identity, position \\
& & & & 14 & identity, identity \\
& & & & 15 & identity, category \\
& & & & 16 & category, position \\
& & & & 17 & category, identity \\
& & & & 18 & category, category \\
\cmidrule(l){3-6}
& & \multirow{9}{*}{ABAB} & \multirow{9}{*}{3, 4} & 19 & position, position \\
& & & & 20 & position, identity \\
& & & & 21 & position, category \\
& & & & 22 & identity, position \\
& & & & 23 & identity, identity \\
& & & & 24 & identity, category \\
& & & & 25 & category, position \\
& & & & 26 & category, identity \\
& & & & 27 & category, category \\
\cmidrule(l){3-6}
& & \multirow{9}{*}{ABBA} & \multirow{9}{*}{3, 4} & 28 & position, position \\
& & & & 29 & position, identity \\
& & & & 30 & position, category \\
& & & & 31 & identity, position \\
& & & & 32 & identity, identity \\
& & & & 33 & identity, category \\
& & & & 34 & category, position \\
& & & & 35 & category, identity \\
& & & & 36 & category, category \\
\midrule
\multirow{4}{*}{CtxDMS} & \multirow{4}{*}{3}
  & \multirow{4}{*}{--} & \multirow{4}{*}{3} & 37 & position $\to$ category $\to$ identity \\
& & & & 38 & position $\to$ identity $\to$ category \\
& & & & 39 & identity $\to$ position $\to$ category \\
& & & & 40 & category $\to$ identity $\to$ position \\
\midrule
\multirow{3}{*}{DMS} & \multirow{3}{*}{2}
  & \multirow{3}{*}{--} & \multirow{3}{*}{2} & 41 & position \\
& & & & 42 & identity \\
& & & & 43 & category \\
\bottomrule
\end{tabular}
\end{table}

\newpage
\section*{NeurIPS Paper Checklist}

\begin{enumerate}

\item {\bf Claims}
    \item[] Question: Do the main claims made in the abstract and introduction accurately reflect the paper's contributions and scope?
    \item[] Answer: \answerYes{} 
    \item[] Justification: The abstract and introduction explicitly state the paper's main contributions and clearly delineate the scope of the work. 

\item {\bf Limitations}
    \item[] Question: Does the paper discuss the limitations of the work performed by the authors?
    \item[] Answer: \answerYes{} 
    \item[] Justification: We acknowledge that the analysis does not exhaustively cover all possible model variants, and the experiments are limited by available computational resources, which may restrict the breadth of evaluations. 

\item {\bf Theory assumptions and proofs}
    \item[] Question: For each theoretical result, does the paper provide the full set of assumptions and a complete (and correct) proof?
    \item[] Answer: \answerNA{}. 
    \item[] Justification: The paper does not include theoretical results.

    \item {\bf Experimental result reproducibility}
    \item[] Question: Does the paper fully disclose all the information needed to reproduce the main experimental results of the paper to the extent that it affects the main claims and/or conclusions of the paper (regardless of whether the code and data are provided or not)?
    \item[] Answer: \answerYes{} 
    \item[] Justification: The paper provides all necessary details to reproduce the main experimental results, including model architectures, training procedures, hyperparameters, and evaluation protocols. Additionally, we release the code and data to further facilitate reproducibility.

\item {\bf Open access to data and code}
    \item[] Question: Does the paper provide open access to the data and code, with sufficient instructions to faithfully reproduce the main experimental results, as described in supplemental material?
    \item[] Answer: \answerYes{} 
    \item[] Justification: We release the full codebase along with the data and accompanying instructions to reproduce all experimental results in the camera-ready version.

\item {\bf Experimental setting/details}
    \item[] Question: Does the paper specify all the training and test details (e.g., data splits, hyperparameters, how they were chosen, type of optimizer) necessary to understand the results?
    \item[] Answer: \answerYes{} 
    \item[] Justification: All training and evaluation details, including data splits, hyperparameters, optimization strategies, and model configurations, are specified in the methods section and further elaborated in the appendix, providing sufficient information to understand and interpret the reported results.

\item {\bf Experiment statistical significance}
    \item[] Question: Does the paper report error bars suitably and correctly defined or other appropriate information about the statistical significance of the experiments?
    \item[] Answer: \answerYes{} 
    \item[] Justification: Statistical analyses are performed wherever possible, with the methodology and sources of variability clearly described in the corresponding sections.

\item {\bf Experiments compute resources}
    \item[] Question: For each experiment, does the paper provide sufficient information on the computer resources (type of compute workers, memory, time of execution) needed to reproduce the experiments?
    \item[] Answer: \answerYes{} 
    \item[] Justification: The paper provides sufficient information on the computational resources used for each experiment.
    
\item {\bf Code of ethics}
    \item[] Question: Does the research conducted in the paper conform, in every respect, with the NeurIPS Code of Ethics \url{https://neurips.cc/public/EthicsGuidelines}?
    \item[] Answer: \answerYes{} 
    \item[] Justification: The authors have reviewed and strictly adhered to the NeurIPS Code of Ethics in all aspects of this research. MTurk workers were compensated fairly for their time.

\item {\bf Broader impacts}
    \item[] Question: Does the paper discuss both potential positive societal impacts and negative societal impacts of the work performed?
    \item[] Answer: \answerNA{} 
    \item[] Justification: There is no societal impact of the work performed.
    
\item {\bf Safeguards}
    \item[] Question: Does the paper describe safeguards that have been put in place for responsible release of data or models that have a high risk for misuse (e.g., pre-trained language models, image generators, or scraped datasets)?
    \item[] Answer: \answerNA{} 
    \item[] Justification: The paper poses no such risk.

\item {\bf Licenses for existing assets}
    \item[] Question: Are the creators or original owners of assets (e.g., code, data, models), used in the paper, properly credited and are the license and terms of use explicitly mentioned and properly respected?
    \item[] Answer: \answerYes{} 
    \item[] Justification: All datasets, code, and models used in this paper are properly credited with appropriate citations, and their respective licenses and terms of use are acknowledged and respected. We will use CC-BY 4.0.

\item {\bf New assets}
    \item[] Question: Are new assets introduced in the paper well documented and is the documentation provided alongside the assets?
    \item[] Answer: \answerNA{} 
    \item[] Justification: The paper does not release new assets.

\item {\bf Crowdsourcing and research with human subjects}
    \item[] Question: For crowdsourcing experiments and research with human subjects, does the paper include the full text of instructions given to participants and screenshots, if applicable, as well as details about compensation (if any)? 
    \item[] Answer: \answerYes{} 
    \item[] Justification: The paper includes full details of the platform, behavioral task instructions given to participants, compensation information, and relevant screenshots, all provided in accordance with the NeurIPS Code of Ethics.

\item {\bf Institutional review board (IRB) approvals or equivalent for research with human subjects}
    \item[] Question: Does the paper describe potential risks incurred by study participants, whether such risks were disclosed to the subjects, and whether Institutional Review Board (IRB) approvals (or an equivalent approval/review based on the requirements of your country or institution) were obtained?
    \item[] Answer: \answerYes{} 
    \item[] Justification: The paper does not pose potential risks, IRB approval has been obtained.

\item {\bf Declaration of LLM usage}
    \item[] Question: Does the paper describe the usage of LLMs if it is an important, original, or non-standard component of the core methods in this research? Note that if the LLM is used only for writing, editing, or formatting purposes and does \emph{not} impact the core methodology, scientific rigor, or originality of the research, declaration is not required.
    \item[] Answer: \answerNA{} 
    \item[] Justification: The core method development does not involve LLMs as an original components.

\end{enumerate}
\end{document}